\documentclass[12pt]{article}

\usepackage{amsthm,color}
\usepackage{amsfonts}
\usepackage{graphicx}
\usepackage{mathrsfs}
\usepackage{amsmath}
\usepackage{amssymb}
\usepackage{float}
\usepackage{natbib}
\usepackage{booktabs}
\usepackage{url}
\usepackage{multirow}
\usepackage{arydshln}
\usepackage{threeparttable}
\usepackage{courier}
\usepackage{authblk} 

\usepackage{multicol}
\usepackage{latexsym}
\usepackage{psfrag}
\usepackage[usenames,dvipsnames]{xcolor}

\usepackage{breakcites}
\usepackage{bbm}
\usepackage{epsfig,epstopdf}
\usepackage[colorlinks, linkcolor=black, citecolor=black]{hyperref}

\usepackage[utf8]{inputenc}
\usepackage[english]{babel}
\usepackage{caption}
\usepackage{setspace} 

\usepackage{amssymb,amsmath,epsfig,epsf,psfrag,graphicx,rotating,times,color}
\usepackage{multirow}
\usepackage{amssymb,amsmath,epsfig,epsf,psfrag,graphicx,rotating,times}
\usepackage{color,latexsym,amsfonts,amsthm,amscd}
\usepackage{multirow}
\usepackage{booktabs}
\usepackage{mathabx}

\usepackage{graphicx}

\usepackage{sectsty}
\sectionfont{\fontsize{12}{15}\selectfont}
\subsectionfont{\fontsize{12}{15}\selectfont}

\newtheorem{theorem}{Theorem}
\newtheorem{lemma}{Lemma}

\newtheorem{remark}{Remark}

\newtheorem{defi}{Definition}

\def\mathB{\mathcal{B}}

\def\mathF{\mathcal{F}}
\def\mathX{\mathcal{X}}
\def\mathY{\mathcal{Y}}
\def\mathH{\mathcal{H}}
\def\mathZ{\mathcal{Z}}

\def\mathC{\mathcal{C}}

\def\mathI{\mathcal{I}}

\def\mathT{\mathcal{T}}
\def\mathM{\mathcal{M}}

\def\FF{\mathbb{F}}
\def\UU{\mathbb{U}}
\def\VV{\mathbb{V}}
\def\PP{\mathbb{P}}
\def\GG{\mathbb{G}}
\def\MM{\mathbb{M}}
\def\RR{\mathbb{R}}
\def\QQ{\mathbb{Q}}
\def\WW{\mathbb{W}}

\def\bx{\boldsymbol{x}}
\def\bz{\boldsymbol{z}}
\def\bv{\boldsymbol{v}}

\def\proof{{\it Proof.\ }}

\def\epf{\quad $\blacksquare$}

\begin{document}

\title{\Large
 Maximum profile binomial likelihood estimation for the semiparametric Box--Cox power transformation model} \vspace{10mm}
\author{
Pengfei Li$^1$, Tao Yu$^2$, Baojiang Chen$^3$, and Jing Qin$^4$ }

\date{}
 \maketitle

\begin{center}
$^1$Pengfei Li is Professor,
Department of Statistics and Actuarial Sciences,
University of Waterloo,
Waterloo, ON, Canada, N2L 3G1\\
(Email: \emph{pengfei.li@uwaterloo.ca})\\
$^2$Tao Yu is Associate Professor,
Department of Statistics \& Applied Probability,
National University of Singapore, Singapore, 117546\\
(Email: \emph{stayt@nus.edu.sg})\\
$^3$Baojiang Chen is Associate Professor,
Department of Biostatistics, University of Nebraska Medical Center, Omaha, Nebraska 68198, U.S.A.\\
(Email: \emph{baojiang.chen@unmc.edu})\\
$^4$Jing Qin is Mathematical Statistician, National Institute of Allergy and Infectious Diseases, National Institutes of Health, MD 20892, U.S.A.\\
 (Email: \emph{jingqin@niaid.nih.gov})\\
[5mm]
\end{center}

\vspace{4mm}
\begin{center}
{\bf Abstract}
\end{center}
The Box--Cox transformation model has been widely applied for many years. The parametric version of this model assumes that the random error follows a parametric distribution, say the normal distribution, and estimates the model parameters using the maximum likelihood method.  The semiparametric version assumes that the distribution of the random error is completely unknown; {existing methods either need strong assumptions, or are less effective when the distribution of the random error significantly deviates from the normal distribution.} We adopt the semiparametric assumption and 
propose a maximum profile binomial likelihood method. We theoretically establish the joint distribution of the estimators of the model parameters. Through extensive numerical studies, we demonstrate that our method has an advantage over existing methods, especially when the distribution of the random error deviates from the normal distribution. Furthermore, we compare the performance of our method and existing methods on an HIV data set. 
 \vspace*{0.2in}

\noindent \textsc{Keywords}: Binomial likelihood; Box--Cox transformation; empirical processes; M-estimation;  semiparametric inference; U-processes

\section{Introduction\label{introduction}}

Since the seminal work of Box and Cox (1964), the Box--Cox power transformation model has been extensively studied and applied in various disciplines. Let $(Y_i, X_i), i=1,\ldots, n$ be independent and identically distributed (i.i.d.) observations with $Y_i$ the response and $X_i = (X_{i1}, \ldots, X_{ip})^T$ the corresponding covariates. The Box--Cox model assumes that
\begin{equation}\label{Complete_Box--Cox}
Y_i^{(\lambda)}=\gamma + X_i^T\beta+\epsilon_i,
\end{equation}
where $Y^{(\lambda)} =({Y^{\lambda} - 1})/{\lambda}$ if $\lambda \neq 0$ and {$\log Y $} otherwise; $\lambda$, $\gamma$, and $\beta$ are the parameters of interest; and $\epsilon_i, i=1,\ldots, n$, are i.i.d.~mean 0 random errors.

When the distribution of $\epsilon_i$ is assumed to be known only up to an unknown finite-dimensional parameter, we have the parametric Box--Cox power transformation model. This model has been studied extensively under the assumption that the $\epsilon_i$'s are i.i.d. equal-variance normal random variables; see, for example, Box and Cox (1964), Bickel and Doksum (1981), Hinkley and Runger (1984), Carroll and Ruppert (1985), Taylor (1985a, 1985b, 1987), and Sakia (1992).  The maximum likelihood principle has proved to be a powerful tool, but the parametric assumption may be too strong. It could be severely violated in many practical applications, leading to biased inference results; see our numerical studies for details.

{It is not uncommon for the distribution of the random error in the Box--Cox transformation model to deviate from normal. For example, in survival analysis, the well-known proportional hazard model (Cox 1972, 1975) is equivalent to the Box--Cox transformation model with the error following an extreme value distribution if the baseline hazard function is the Weibull distribution. See Lancaster (1990) and Flinn and Heckman (1982) for more discussion of the connection between the Box--Cox transformation model and the proportional hazard model. The proportional odds model (Bennett 1983a, 1983b) is another example. It assumes that $\log[\{1-S_0(Y)\}/S_0(Y)] = X^T \beta + \epsilon$, where $S_0(\cdot)$ is the baseline survival function; the random error $\epsilon$ follows the logistic distribution. Therefore, when $\log[\{1-S_0(Y)\}/S_0(Y)]$ is assumed to be a power function of $Y$, this is the Box--Cox transformation model with the error following the logistic distribution. 
}

%

{In this paper, we assume that the distribution of  $\epsilon_i$ is completely unknown; 
{parametric} models where the error distribution deviates from normal are special cases of our approach.}
Amemiya (1985), Newey (1990), and Robinson (1991) have proposed quasi-likelihood estimating equation methods for this semiparametric Box--Cox power transformation model. However, Foster et~al.~(2001) showed that the root of the expectation of the corresponding estimating equation is generally not unique, and therefore the resulting estimator is not consistent. They instead proposed a ``minimum distance" estimator for $\lambda$ and a least-square estimator for $\beta$, and they established the  joint asymptotic distribution for these estimators. 


{Foster et~al.~(2001) successfully established the asymptotic normality of their $(\lambda, \beta)$ estimator under the assumption that the distribution of $\epsilon_i$ is completely unknown.  However, their approach has two limitations.
First, their estimator for $\beta$ is based on the least-square method. 
This method performs well when the underlying distribution of $\epsilon_i$ is close to normal; 
but if it is not, the estimator may have {less accurate} numerical performance. 
This, in turn, affects the performance of the estimator for $\lambda$. Our simulation study demonstrates this; see Section~\ref{section-simulation} for details.
Second, their method is based on the minimum distance method and does not have a likelihood interpretation.  
We study model (\ref{Complete_Box--Cox}) under the same assumptions used in Foster et~al.~(2001).
We propose a profile binomial likelihood method; we estimate $(\lambda, \beta)$ simultaneously by maximizing the likelihood.  We also establish the joint asymptotic distribution of the estimators for $\lambda$ and $\beta$. Our simulation studies demonstrate that our method achieves more accurate parameter estimates than existing methods, especially when the distribution of $\epsilon_i$ deviates from the normal distribution.}

The paper is organized as follows. Section~\ref{section-literature} gives a brief review of the methods that will be compared with our approach in the numerical studies. Section~\ref{Section_Complete data} proposes the {maximum profile binomial likelihood method for estimating} the parameters under the Box--Cox power transformation model {and presents an algorithm for obtaining our  estimates numerically.} 
Section~\ref{section-asym} studies the joint asymptotic properties of our estimates. Section~\ref{section-simulation} discusses the simulation studies, Section~\ref{Appl} presents the HIV application, and Section~\ref{section-discussion} concludes the paper with a discussion. 
For convenience of presentation, 
the technical details are provided in two Appendices and the supplementary material. 

\section{Existing Methods} \label{section-literature}

With a parametric assumption on the distribution of $\epsilon$, the Box--Cox model \eqref{Complete_Box--Cox} can be analyzed by the classical maximum likelihood principle; see, for example, Box~and~Cox~(1964), Bickel and Doksum (1981), Hinkley and Runger (1984), Carroll and Ruppert (1985), Taylor (1985a, 1985b, 1987), and Sakia (1992). The most popular parametric assumption is that $\epsilon_i, i = 1,\ldots, n$ are i.i.d. $N(0,\sigma^2)$ random variables.
{Under this assumption, the classical maximum likelihood estimators of   $(\lambda,\gamma,\beta,\sigma)$ maximize the log-likelihood function given by
$$
-\frac{1}{2}\sum_{i=1}^n(Y_i^{(\lambda)}-\gamma-X_i^T\beta)^2/\sigma^2-\frac{n}{2}\log(2\pi\sigma^2)+(\lambda-1)\sum_{i=1}^n\log Y_i. 
$$
We can use existing R functions, such as the ``powerTransform" function in the package \textit{car}, to compute these estimates numerically. 
}
In the numerical studies, we will compare this parametric method with our method.  

Foster et~al.~(2001) proposed a semiparametric estimation approach that proceeds as follows. For a given $\lambda$, the model parameters $(\gamma, \beta^T)^T$ in Model \eqref{Complete_Box--Cox} can be estimated by the classical least-square principle, namely, 
\begin{equation}
\label{lse1}
\left(\widehat \gamma(\lambda), \widehat \beta^T(\lambda)\right)^T = \left(\sum_{i=1}^n X_i^* X_i^{*T}\right)^{-1}\sum_{i=1}^n X_i^*Y_i^{(\lambda)},
\end{equation} where 
$X_i^* = (1, X_i^T)^T$. 
Then, since $P(Y \leq t)  = F_\epsilon(t^{(\lambda)} - \gamma -  X_i^T \beta)$ with $F_\epsilon(\cdot)$ being the cumulative distribution function (c.d.f.) of $\epsilon_i$, $\lambda$ can be estimated by a ``minimum distance" estimator that minimizes $S_n(\lambda, \widehat \gamma(\lambda), \widehat \beta(\lambda))$, where 
\begin{eqnarray*}
S_n(\lambda,\gamma, \beta) &=& n^{-1}\sum_{i=1}^n \int_0^\infty \left\{ I(Y_i \leq t) - \widetilde G_{\lambda, \beta}(t^{(\lambda)} - \gamma - X_i^T \beta)\right\}^2 dW(t), \\
\widetilde G_{\lambda, \beta}(t) &=& \frac{1}{n} \sum_{j=1}^n I\left\{Y_j^{(\lambda)} - \gamma -  X_j^T \beta \leq t\right\}, 
\end{eqnarray*}
and $W(\cdot)$ is a positive, differentiable, strictly increasing, deterministic, and bounded weight function. 
{In their numerical study, Foster et~al.~(2001) set $W(\cdot)$ to a normal density with the mean and standard derivation being the sample mean and sample standard error of the $Y_i$'s. 
Since $S_n(\lambda, \widehat \gamma(\lambda), \widehat \beta(\lambda))$ is a function of the one-dimensional parameter $\lambda$, 
a grid search can be used to find this $\lambda$ estimate. 
In the numerical studies, we will also compare this semiparametric method with our approach.  
}

%

\section{{Maximum Profile Binomial Likelihood }Estimation}\label{Section_Complete data}

With the observed data $(Y_i, X_i), i = 1,\ldots, n$, we consider the Box--Cox transformation model (\ref{Complete_Box--Cox}). We assume that the errors $\epsilon_i$ are i.i.d. and independent of $X_i$. 
{Let $F(\cdot)$ be the c.d.f. of $\epsilon^*=\epsilon+\gamma$. 
For any $t>0$, we have 
$$
P(Y_i\leq t|X_i)=P\Big(\epsilon_i^*\leq t^{(\lambda)}- X_i^T\beta\Big| X_i\Big)
=F(t^{(\lambda)}- X_i^T\beta). 
$$
Conditioning on $X_i$, 
$
I(Y_i\leq t)
$
follows a Bernoulli distribution for which the probability of success is $F(t^{(\lambda)}- X_i^T\beta)$; 
here $I(\cdot)$ is the indicator function. 
Therefore, conditioning on $X_i,i=1,\ldots,n$, the log-likelihood of  $\{I(Y_i\leq t)\}_{i=1}^n$ is given by 
\begin{equation*}
\label{bl1}
\tilde l(\lambda,\beta,F;t)=\sum_{i=1}^n \left[ I(Y_i\leq t) \log \left\{F\Big(t^{(\lambda)}-X_i^T\beta\Big)\right\}+I(Y_i>t) \log \left\{1-F\Big(t^{(\lambda)}-X_i^T\beta\Big)\right\} \right].
\end{equation*}
We suggest choosing the values of $t$ as the observed responses $\{Y_j\}_{j=1}^n$ and taking the summation of  $\tilde l(\lambda,\beta,F;Y_j)$ over $j$; this leads to the  binomial likelihood
\begin{equation}
\label{bl.fun}
\tilde l_B(\lambda,\beta,F)=\sum_{j=1}^n\sum_{i=1}^n \left[ I_{i,j} \log \left\{F\Big(Y_j^{(\lambda)}-X_i^T\beta\Big)\right\}+ (1-I_{i,j}) \log \left\{1-F\Big(Y_j^{(\lambda)}-X_i^T\beta\Big)\right\} \right],
\end{equation}
where $I_{i,j} = I(Y_i\leq Y_j)$. 
} 

Note  that $F(\cdot)$ is an infinite-dimensional parameter. Estimating $(F, \lambda, \beta)$ simultaneously by maximizing  $\tilde l_B(\lambda,\beta,F)$ is possible but computationally demanding; this also leads to theoretical difficulties in the subsequent development of the asymptotic distributions of the estimates (Chen et al., 2016). Since $F(\cdot)$ is the distribution function of $\epsilon_i^*$, we can instead use the following profile approach to estimate it by the empirical distribution function. For given $\lambda$ and $\beta$, based on (\ref{Complete_Box--Cox}), we have $\epsilon_i^* = Y_i^{(\lambda)} - X_i^T \beta$; therefore, we consider
\begin{eqnarray}
\widehat G_{\lambda, \beta}(t) &=& \frac{1}{n} \sum_{i=1}^n I\left\{Y_i^{(\lambda)} - X_i^T \beta \leq t\right\}, \label{Complete_CDF0} \\
\widehat F_{\lambda, \beta}(t) &=& \left\{\widehat G_{\lambda, \beta}(t) \vee n^{-2}\right\}\wedge \left(1-n^{-2}\right), \label{Complete_CDF}
\end{eqnarray}
where $n^{-2}$ is added to ensure that $\widehat F_{\lambda, \beta}(\cdot)$ stays away from 0 and 1 to avoid complications in both the numerical analyses and the technical development.
Substituting (\ref{Complete_CDF}) into (\ref{bl.fun}), we obtain the profile  binomial likelihood:
\begin{equation}\label{main.Complete_profile_loglik}
\ell (\lambda,{\beta})=\sum_{j=1}^n\sum_{i=1}^n \left[ I_{i,j}\log\left\{\widehat{F}_{\lambda, \beta}\Big(Y_j^{(\lambda)}-X_i^T{\beta}\Big)\right\}+(1-I_{i,j})\log\left\{1-\widehat{F}_{\lambda, \beta}\Big(Y_j^{(\lambda)}-X_i^T{\beta}\Big)\right\}\right] .
\end{equation}
Consequently, we define
\begin{eqnarray}
\left(\widehat \lambda, \widehat \beta^T\right)^T = {\arg\max}_{\left(\lambda, \beta^T\right)^T\in \Theta} \ell ( \lambda,{\beta}), \label{def-estimates}
\end{eqnarray}
where $\Theta$ is a compact subset of $\RR^{p+1}$, and $\gamma$ is then estimated by 
\begin{eqnarray*}
\widehat \gamma = \frac 1n \sum_{i=1}^n \left\{ Y_i^{(\widehat \lambda)} - X_i^T \widehat \beta \right\}.
\end{eqnarray*}

{The estimator in (\ref{def-estimates}) does not have an explicit form. We implemented the following algorithm in R to compute it numerically.  

\begin{enumerate}
\item[]{\bf Step 1.} For given $\lambda$, we define
\begin{eqnarray}
\beta_\lambda = \arg\max_{\beta} \ell(\lambda, \beta), \label{profile-beta}
\end{eqnarray}
which leads to the profile likelihood for $\lambda$, given by
\begin{eqnarray*}
p\ell ( \lambda) = \ell(\lambda, \beta_\lambda). 
\end{eqnarray*}

In our numerical studies, we solve the optimization \eqref{profile-beta} using {\tt optim()} with the default Nelder--Mead method. For the initial values of $\beta$, we treated $\lambda$ as a constant in the model $Y^{(\lambda)} = X^T\beta + \epsilon$ and considered two possibilities: the 
least-square estimate implemented by {\tt lm()} and the rank-based estimate from {\tt rfit()} in the package {\tt Rfit}.

\item[] {\bf Step 2.} Since $p\ell(\lambda)$ is a function of a one-dimensional parameter $\lambda$, we compute $\widehat\lambda$ via a grid search maximization. 

\item[] {\bf Step 3.} With $\widehat\lambda$, we obtain $\widehat\beta$ from \eqref{profile-beta}.

\end{enumerate}

\begin{remark}
As far as we are aware, the work in the literature that is most closely related to our work is  Foster~et~al.~(2001). We use the same model assumptions and have included the component $I(Y_i\leq t)$ in the objective functions. We incorporate this component to establish the binomial likelihood, while Foster et al.~(2001) use it to construct the $L_2$-distance. We observe that they estimate $(\gamma, \beta)$ by the least-square method for a given $\lambda$, and in the construction of their objective function $S_n(\lambda, \gamma, \beta)$ for the estimation of $\lambda$, they suggest the normal distribution as the weights. These choices do not affect the convergence rates of their estimators and should increase the estimation accuracy of the model parameters when the responses and errors are approximately normally distributed. However, when normality is violated, the performance of their method may be affected. In contrast, our method estimates the model parameters by maximizing a profile binomial likelihood, which is unrelated to the normal distribution. We therefore expect that the method of Foster~et~al.~(2001) may have better performance when both $Y$ and the random errors are close to the normal distribution, but our method may have the advantage when normality is violated. The observations in our numerical studies reinforce this conjecture; see Section~\ref{section-simulation} for details. 
\end{remark}

}
\section{Joint Asymptotic Distribution of Estimators} \label{section-asym}

In this section, we derive the joint asymptotic distribution of $\left(\widehat \lambda, \widehat \beta^T\right)^T$ defined by (\ref{def-estimates}).  
We need the following notation. Let $\theta=(\lambda,\beta^T)^T$ and $\widehat \theta=\left(\widehat \lambda, \widehat \beta^T\right)^T$; and let $\theta_0=(\lambda_0,\beta_0^T)^T$ be the true values of the corresponding parameters.
Denote $V_{\theta} = Y^{(\lambda)} - X^T\beta$, $V_{\theta, i} = Y_i^{(\lambda)} - X_i^T\beta$, and $V_{\theta, i,j} = Y_i^{(\lambda)} - X_j^T\beta$.
Define
\begin{eqnarray}
F_\theta(t) = P(Y^{(\lambda)} - X^T\beta \leq t) = P(V_\theta \leq t). \label{def-F-theta}
\end{eqnarray}
When $\theta= \theta_0$, we write $F_0 = F_{\theta_0}$, $V_0 = V_{\theta_0}$, $V_{0,i} = V_{\theta_0, i}$, $V_{0, i,j} = V_{\theta_0, i,j}$.
Let $\dot{F}_\theta(t) = \frac{\partial F_{\theta}(t)}{\partial \theta}$ and $F_{\theta}'(t) = \frac{\partial F_\theta(t)}{\partial t}$, if they exist; and denote $\dot{F}_0(t) = \dot{F}_{\theta_0}(t)$, $F_0'(t) = F_{\theta_0}'(t)$. Let
\begin{eqnarray}
\dot V_\theta=\frac{\partial V_\theta}{\partial \theta} = \left\{ \begin{array}{ll} \left( \begin{matrix} \lambda^{-2}\left\{\lambda Y^{\lambda} \log Y- Y^\lambda  + 1\right\} \\-X \end{matrix} \right) & \mbox{if} \quad \lambda \neq 0  \\ \left( \begin{matrix} (\log Y)^2/2 \\-X \end{matrix} \right) & \mbox{if} \quad \lambda = 0 \end{array} \right., \label{main.def-V-dot-theta}
\end{eqnarray}
and define $\dot{V}_0$, $\dot{V}_{0,i}$, and $\dot{V}_{0,i,j}$ similarly.
%

Furthermore, {we denote $Z = (Y, X)$ and $\bz = (y, \bx)$}.  Define
\begin{eqnarray}
\varphi(\bz) &=& E\left[\frac{\dot{F}_0(V_{0,2,1}) + F_0'(V_{0,2,1}) \dot{V}_{0, 2,1}}{F_{0}(V_{0, 2,1})\left\{1-F_{0}(V_{0, 2,1})\right\}} \left\{  I(Y_1 \leq Y_2) - F_{0}(V_{0, 2,1})   \right\} \bigg| Z_1 = \bz\right], \label{olddef-varphi}\\
\psi(\bz) &=& -E\left[\frac{\dot{F}_0(V_{0,2,1}) + F_0'(V_{0,2,1}) \dot{V}_{0,2,1}}{F_{0}(V_{0, 2,1})\left\{1-F_{0}(V_{0, 2,1}) \right\}}I\left(V_{0,3} \leq V_{0, 2,1}\right) \Big| Z_3 = \bz\right],\label{olddef-psi}\\
\Sigma_1 &=&  E \left(\left[\frac{\left\{\dot{F}_0(V_{0,2,1}) + F_0'(V_{0,2,1}) \dot{V}_{0, 2,1} \right\} \left\{\dot{F}_0(V_{0,2,1}) + F_0'(V_{0,2,1}) \dot{V}_{0, 2,1} \right\}^T }{F_0(V_{0, 2,1})\left\{1-F_{0}(V_{0, 2,1})\right\}} \right]  \right), \label{olddef-Sigma-1}\\
\Sigma_2 &=& \mbox{var}\left\{\varphi(Z) + \psi(Z)\right\}. \label{olddef-Sigma-2}
\end{eqnarray}

The following theorem establishes the joint asymptotic distribution of $\left(\widehat \lambda, \widehat \beta^T\right)^T$.

\begin{theorem}
\label{theorem-normality}
Assume Conditions 1--5 in Appendix A; then
\begin{eqnarray*}
\sqrt{n}(\widehat \theta - \theta_0) \rightsquigarrow N(0, \Sigma), \label{norm-main-eq-1}
\end{eqnarray*}
where $\Sigma = \frac{1}{4} \Sigma_1^{-1} \Sigma_2 \Sigma_1^{-1}$ with $\Sigma_1$ and $\Sigma_2$ defined by (\ref{olddef-Sigma-1}) and (\ref{olddef-Sigma-2}) respectively.
\end{theorem}

{Note that deriving the asymptotic properties for $\widehat \theta$ is a challenging task. The main difficulty is the complicated structure of the 
profile binomial likelihood $\ell(\cdot)$ defined by \eqref{main.Complete_profile_loglik}. Clearly, it is a U-process, with a plugged-in nonparametric component $\widehat F_{\lambda, \beta}(\cdot)$. Existing U-process theory is not applicable in our context. We use advanced empirical process theory 
(van der Vaart and Wellner, 1996; Kosorok, 2008) to derive the asymptotic normality of $\widehat \theta$ presented in Theorem~\ref{theorem-normality}. For continuity of presentation, we sketch the lengthy proof of this theorem in Appendix~B and relegate the full details to the supplementary document. 
}

\section{Simulation Study} \label{section-simulation}
\subsection{Data simulation}
We use the following simulation examples to examine the numerical performance of our method. We compare our method (labeled ``Our") with the method of Foster et~al. (2001) (``Foster") and the classical parametric method (``Parametric"). 

We simulate the covariates $X_1, X_2, X_3, X_4$ as follows. 
Let 
$S_1=(S_{11},S_{12})^T$ and $S_2=(S_{21},S_{22})^T$ be independent random vectors  
from 
$$
N\left(\left(\begin{matrix} 0 \\ 0 \end{matrix} \right), \left(
\begin{array}{cc}
1&0.6\\
0.6&1\\
\end{array}
\right)\right). 
$$
Set $X_1=-\log\{ 1-\Phi (S_{11})\}$, $X_2=I(S_{21}>0)$, $X_3=-\log\{ 1-\Phi (S_{12})\}$, and $X_4=I(S_{22}>0)$. 
Then $X_1$ and $X_3$ follow the Exponential$(1)$ distribution, while 
$X_2$ and $X_4$ follow the Bernoulli$(0.5)$ distribution. 
Based on these covariates, we consider six simulation models: 
\begin{enumerate}
\item[] {Model 1:} $\log Y =X_1+X_2+\epsilon$;
\item[]  Model 2:  $\log Y =X_1+X_2+X_3+X_4+\epsilon$; 
\item[]  {Model 3:}  $Y=4+2.5X_1+2.5X_2+ \epsilon$;
\item[]  {Model 4:} $Y=4+1.2X_1+1.2X_2+1.2X_3+1.2X_4+ \epsilon$;
\item[]  {Model 5:}  $5/Y=4+2.5X_1+2.5X_2+ \epsilon$;
\item[]  {Model 6:} $5/Y=4+1.2X_1+1.2X_2+1.2X_3+1.2X_4+ \epsilon$. 

\end{enumerate}
For Models 1 and 2, $\lambda = 0$; for Models 3 and 4, $\lambda = 1$; and for Models 5 and 6, $\lambda = -1$. For each model, we consider two distributions for $\epsilon$, $N(0,0.5^2)$ and $0.5(\chi^2_1-1)$, and two sample sizes, $n=100$ and $n=200$. For each scenario, we use 1000 repetitions.

\subsection{Estimation results}

We examine the performance of the three methods by evaluating their bias and mean squared error (MSE) in the estimation of the model parameters $\lambda$, $\beta_1$, and $\beta_2$; here $\beta_1$ and $\beta_2$ are the coefficients of $X_1$ and $X_2$ in our simulation models. The results for $\beta_3$ and $\beta_4$, i.e., the coefficients for $X_3$ and $X_4$ in Models 2, 4, and 6, are similar to those for $\beta_1$ and $\beta_2$ and are omitted. 

Table~\ref{table-normal} presents the results when $\epsilon$ is simulated as $N(0, 0.5^2)$, and we observe that all the methods have small biases. The parametric method results in the smallest MSEs in every scenario. This is not surprising since the assumption that the random error follows the normal distribution is satisfied; the other methods do not need this assumption. For our method and Foster: (1) when $\lambda = 0$ (Models 1 and 2), our method has slightly smaller MSEs; (2) when $\lambda = 1$ (Models 3 and 4), Foster performs slightly better; (3) when $\lambda = -1$, the MSE values are similar. This supports our remark in Section~\ref{Section_Complete data} that Foster may perform well when the distribution of the random error is close to normal. 

 \renewcommand{\baselinestretch}{1.2}
\begin{table}[!htt] 
\begin{center}
\caption{Bias and MSE  for the estimates of  $\lambda$,  $\beta_1$,  and $\beta_2$:  $\epsilon \sim N(0,0.5^2)$.  The reported MSEs for Models 1--4 are MSE$\times 100$; those for Models 5 and 6 are MSE$\times 1000$.  } 
 {\tabcolsep=1.2mm
\begin{tabular}{lc | cc  cc cc | cc cc cc}
\noalign{\smallskip}\hline \noalign{\smallskip}
&&\multicolumn{2}{c}{Parametric}& \multicolumn{2}{c}{Foster}& \multicolumn{2}{c|}{Our}& \multicolumn{2}{c}{Parametric}& \multicolumn{2}{c}{Foster}& \multicolumn{2}{c}{Our}\\
\cline{3-4}\cline{5-6}\cline{7-8}\cline{9-10}\cline{11-12}\cline{13-14}
$n$ &   & Bias &MSE & Bias & MSE & Bias &MSE& Bias &MSE& Bias &MSE& Bias &MSE\\
\noalign{\smallskip}\hline\noalign{\smallskip}
&&\multicolumn{6}{c}{Model 1} & \multicolumn{6}{c}{Model 2}\\
100&$\lambda$  &0.00&0.21&0.00&0.79&0.00&0.37&0.00&0.04&0.01&0.24&0.00&0.07\\
100&$\beta_1$  &0.01&1.33&0.01&4.17&0.01&1.94&0.01&1.06&0.05&5.80&0.01&1.72\\
100&$\beta_2$  &0.00&1.49&0.00&2.64&0.00&2.01&0.01&1.66&0.03&3.81&0.01&2.16\\
200&$\lambda$  &0.01&0.09&0.00&0.37&0.00&0.17&0.00&0.01&0.00&0.10&0.00&0.03\\
200&$\beta_1$  &0.01&0.62&0.02&2.30&0.01&0.96&0.01&0.42&0.02&2.37&0.00&0.72\\
200&$\beta_2$  &0.01&0.72&0.02&1.31&0.01&0.97&0.01&0.74&0.01&1.61&0.00&0.97\\
&&\multicolumn{6}{c}{Model 3}& \multicolumn{6}{c}{Model 4}\\
100&$\lambda$  &0.00&0.71 &0.00&1.01 &-0.01&1.29&0.01&1.36 &0.01 &2.06 &0.00 &2.36 \\   
100&$\beta_1$  &0.05&23.01&0.04&32.75&0.05 &40.13&0.07&11.61&0.08 &20.47&0.08 &20.56\\  
100&$\beta_2$  &0.04&19.77&0.03&27.20&0.03 &33.89&0.07&10.46&0.06 &16.64&0.07 &18.40\\  
200&$\lambda$  &0.01&0.31 &0.00&0.47 &0.01 &0.60&0.00&0.50 &-0.01&0.85 &-0.01&1.05 \\   
200&$\beta_1$  &0.05&10.46&0.05&15.97&0.07 &19.20&0.02&3.83 &0.01 &6.21 &0.01 &7.32 \\   
200&$\beta_2$  &0.05&8.74 &0.05&13.16&0.07 &16.22&0.02&3.76 &0.01 &5.74 &0.02 &6.86 \\   
&&\multicolumn{6}{c}{Model 5}& \multicolumn{6}{c}{Model 6}\\
100&$\lambda$   &0.00 &0.71&0.00 &1.23&0.01 &1.32  &-0.01&1.36&-0.01&2.51&-0.01&2.37     \\
100&$\beta_1$   &0.00 &0.07&0.00 &0.12&0.00 &0.11  &0.00 &0.04&0.00 &0.06&0.00 &0.06     \\
100&$\beta_2$   &0.00 &0.07&0.00 &0.09&0.00 &0.09  &0.00 &0.06&0.00 &0.07&0.00 &0.08     \\
200&$\lambda$   &-0.01&0.31&-0.01&0.55&-0.01&0.59  &0.00 &0.50&0.01 &1.10&0.01 &1.05     \\
200&$\beta_1$   &0.00 &0.03&0.00 &0.06&0.00 &0.05  &0.00 &0.02&0.00 &0.03&0.00 &0.03     \\
200&$\beta_2$   &0.00 &0.03&0.00 &0.04&0.00 &0.04  &0.00 &0.03&0.00 &0.03&0.00 &0.04     \\
\noalign{\smallskip}\hline
\end{tabular}}\label{table-normal}
\end{center}
\end{table}

Table~\ref{table-chisq} presents the results when $\epsilon$ is simulated as $0.5(\chi_1^2 - 1)$; in this scenario the distribution of the random error deviates from normal. The parametric method has larger biases and MSEs than the other methods in every scenario. Our method and Foster continue to have small and comparable biases, but our method has much smaller MSEs, supporting our remark in Section~\ref{Section_Complete data}.

 \renewcommand{\baselinestretch}{1.2}
\begin{table}[!htt]
\begin{center}
\caption{Bias and MSE  for the estimates of  $\lambda$,  $\beta_1$,  and $\beta_2$:  $\epsilon \sim 0.5(\chi^2_1-1)$. The reported MSEs for Models 1--4 are MSE$\times 100$; those for Models 5 and 6 are MSE$\times 1000$.
 } 
 {\tabcolsep=1.2mm
\begin{tabular}{lc | cc  cc cc | cc cc cc}
\noalign{\smallskip}\hline \noalign{\smallskip}
&&\multicolumn{2}{c}{Parametric}& \multicolumn{2}{c}{Foster}& \multicolumn{2}{c|}{Our}& \multicolumn{2}{c}{Parametric}& \multicolumn{2}{c}{Foster}& \multicolumn{2}{c}{Our}\\
\cline{3-4}\cline{5-6}\cline{7-8}\cline{9-10}\cline{11-12}\cline{13-14}
$n$ &   & Bias &MSE & Bias & MSE & Bias &MSE& Bias &MSE& Bias &MSE& Bias &MSE\\
\noalign{\smallskip}\hline\noalign{\smallskip}
&&\multicolumn{6}{c}{Model 1} & \multicolumn{6}{c}{Model 2}\\
100&$\lambda$  &-0.18&4.15 &-0.01&1.02&0.01&0.12  &-0.04&0.29&-0.01&0.39&0.00&0.02     \\
100&$\beta_1$  &-0.31&11.13&-0.01&4.01&0.01&0.21  &-0.15&3.86&-0.01&8.32&0.01&0.33     \\
100&$\beta_2$  &-0.22&6.14 &-0.01&3.44&0.01&0.17  &-0.10&3.36&-0.01&5.51&0.01&0.44     \\
200&$\lambda$  &-0.19&4.11 &-0.01&0.44&0.00&0.03  &-0.04&0.23&-0.01&0.14&0.00&0.01     \\
200&$\beta_1$  &-0.34&11.96&-0.01&2.09&0.00&0.05  &-0.15&3.11&-0.03&2.75&0.01&0.07     \\
200&$\beta_2$  &-0.22&5.73 &-0.01&1.85&0.00&0.04  &-0.10&2.21&-0.02&2.24&0.01&0.09     \\
&&\multicolumn{6}{c}{Model 3}& \multicolumn{6}{c}{Model 4}\\
100&$\lambda$   &-0.13&3.94 &0.00 &0.59 &0.01&0.33 &-0.21&8.23 &-0.01&1.39 &0.02&0.64\\
100&$\beta_1$   &-0.54&59.66&0.02 &15.49&0.07&10.08&-0.38&23.88&0.01 &8.55 &0.06&4.64\\
100&$\beta_2$   &-0.51&51.00&0.02 &14.97&0.06&7.89 &-0.35&21.37&0.02 &10.17&0.06&4.27\\
200&$\lambda$   &-0.14&2.99 &0.00 &0.24 &0.00&0.07 &-0.21&6.63 &0.00 &0.48 &0.01&0.16\\
200&$\beta_1$   &-0.60&53.05&-0.01&5.70 &0.02&1.83 &-0.41&21.59&0.00 &2.90 &0.03&1.04\\
200&$\beta_2$   &-0.55&44.10&-0.01&6.21 &0.02&1.53 &-0.38&19.10&0.00 &3.52 &0.03&0.93\\
&&\multicolumn{6}{c}{Model 5}& \multicolumn{6}{c}{Model 6}\\
100&$\lambda$   &0.13&39.40&0.00&6.26&-0.01&3.31& 0.21&82.28&0.01&15.22&-0.02&6.46   \\
100&$\beta_1$   &0.04&2.65 &0.00&0.42&0.00 &0.17& 0.03&1.30 &0.00&0.45 &0.00 &0.11   \\
100&$\beta_2$   &0.02&1.54 &0.00&0.93&0.00 &0.10& 0.02&1.20 &0.00&1.08 &0.00 &0.16   \\
200&$\lambda$   &0.14&29.93&0.00&2.57&0.00 &0.69& 0.21&66.28&0.01&5.32 &-0.01&1.68   \\
200&$\beta_1$   &0.04&2.27 &0.00&0.16&0.00 &0.04& 0.03&1.00 &0.00&0.17 &0.00 &0.02   \\
200&$\beta_2$   &0.02&1.08 &0.00&0.51&0.00 &0.03& 0.02&0.73 &0.00&0.49 &0.00 &0.03   \\
\noalign{\smallskip}\hline
\end{tabular}}\label{table-chisq}
\end{center}
\end{table}

In summary, we observe that the performance of the parametric method relies heavily on the distribution of the random error. Foster may be slightly better than our method when the distribution of the random error is close to normal. Otherwise, our method has much better performance.

\section{HIV Application}\label{Appl}

We now apply our method to analyze human immunodeficiency virus (HIV) data
from the AIDS Clinical Trials Group Protocol 175 (ACTG175)
(Hammer et al., 1996; Zhang and Wang, 2020)
in which  $n=2139$ HIV-infected patients were enrolled.
The patients were randomly  divided into four arms
according to their treatment regimen:
(I) zidovudine   monotherapy,
(II) zidovudine + didanosine,
(III) zidovudine + zalcitabine,  and
(IV) didanosine monotherapy.
The data record various measurements from each patient,
including age (in years), weight (in kilograms),
CD4 cell count at baseline (cd40),
CD4 cell count at 20$\pm$5 weeks (cd420),
CD4 cell count at 96$\pm$5 weeks (cd496),
CD8 cell count at baseline (cd80),
CD8 cell count at 20$\pm$5 weeks (cd820),
and arm number (arms).
The data are available in the {\tt R} package {\tt speff2trial}.
The effectiveness of an HIV treatment
can be assessed by monitoring the  CD4 cell counts of
HIV-positive patients: an increased count
indicates an improvement in the patient's condition. It is of particular interest to estimate the average CD4 cell count in each arm after 96 weeks of treatment. We take this variable (cd496) { plus 1} as the response variable in our analysis. We consider six covariates, age/10, weight/10, cd40/10, cd420/10, cd80/100, and cd820/100, and focus on the complete data for the patients in arm IV. 

\begin{figure}[!ht]
\centering{\includegraphics[scale=0.4]{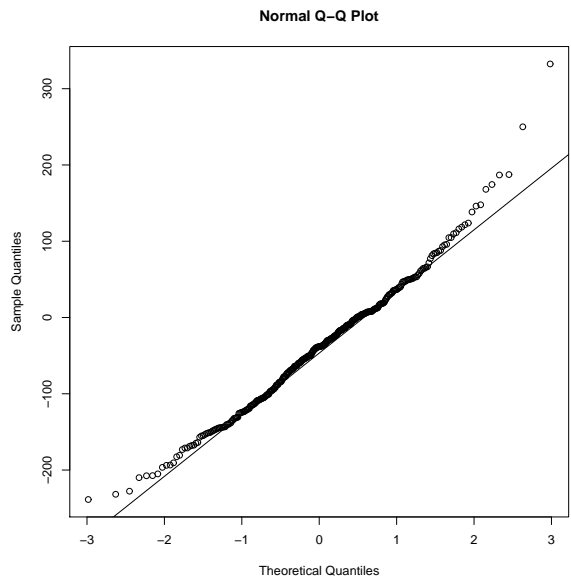}}
  \caption{Q-Q plot of residuals after Box--Cox transformation.}\label{residual1}
\end{figure}

{We apply the three methods from our simulation study to this data set. 
Table~\ref{ACTG1} summarizes the point estimate (Est), the corresponding bootstrap standard deviation (BSD), and the 95\% bootstrap percentile confidence intervals (BCI). Based on the estimates of $\lambda$ and $\beta$ from our method, Figure~\ref{residual1} shows the normal probability plot of the $F$ estimate \eqref{Complete_CDF}. We test the normality of the residuals using the Shapiro--Wilk test, which gives a p-value of 0.0015. Both 
Figure~\ref{residual1} and this test result suggest that the distribution of the random error might deviate from normal. It is therefore not surprising that in Table~\ref{ACTG1}, the estimates of $\lambda$ and $\beta$ based on the parametric method are significantly different from those based on the other methods; the former estimates may not be reliable. Our method and Foster lead to $\lambda$ estimates that are very close to 1 and similar $\beta$ estimates, but our method has much smaller BSD values and shorter BCIs for all the parameter estimates. Since the distribution of the random error might deviate from normal, we expect that our method has produced more accurate results than Foster in this real-data example. 
}

\renewcommand{\baselinestretch}{1}
\begin{table}[!ht]
  \footnotesize
\tabcolsep=1.5mm
\caption{Analysis of  ACTG data}
\label{ACTG1}
\begin{center}
\begin{tabular}{c| ccc| ccc| ccc }
\hline
&\multicolumn{3}{c}{Parametric}&\multicolumn{3}{c}{Foster}&\multicolumn{3}{c}{Our}\\
\hline
          &Est  &BSD  &BCI                  &Est  &BSD  &BCI                &Est  &BSD &BCI                 \\
$\lambda$ &0.76 &0.05 &$(0.68 ,0.89)$       &1.00 &0.13 &$(0.81  ,1.30 )$   &0.95 &0.08&$(0.80  ,1.10 )$    \\
$\beta_1$ &-0.40&2.14 &$(-5.74,3.21)$       &-2.18&15.23&$(-39.14,21.89)$   &-4.17&7.31&$(-22.24,7.60 )$    \\
$\beta_2$ &1.51 &1.51 &$(-0.92,4.88)$       &4.94 &10.89&$(-6.26 ,33.31)$   &3.88 &5.09&$(-3.59 ,14.17)$    \\
$\beta_3$ &0.86 &0.41 &$(0.41 ,2.05)$       &3.36 &5.10 &$(0.85  ,18.49)$   &2.63 &1.55&$(0.85  ,6.62 )$    \\
$\beta_4$ &1.83 &0.65 &$(1.09 ,3.82)$       &7.63 &10.09&$(2.58  ,38.05)$   &5.27 &2.84&$(2.20  ,12.93)$    \\
$\beta_5$ &0.07 &0.81 &$(-1.62,1.38)$       &1.66 &5.55 &$(-5.16 ,12.50)$   &1.19 &2.39&$(-3.52 ,5.87 )$    \\
$\beta_6$ &-0.55&0.74 &$(-2.04,0.67)$       &-3.40&6.76 &$(-27.65,1.85 )$   &-2.65&2.80&$(-8.28 ,1.05 )$    \\
\hline
\end{tabular}

\end{center}
\end{table}

\section{Discussion} \label{section-discussion}

We have focused on the Box--Cox model, which has been extensively studied. Classical methods assume that the distribution of the random error is parametric, say normal, and apply the maximum likelihood method to estimate the model parameters. These methods may give misleading results when the parametric assumption is violated. Semiparametric methods assume that the distribution of the random error is unknown. They may be based on the estimating equation method (Newey, 1990; Robinson, 1991), the validity of which relies on a strong and possibly unrealistic assumption; see 
Foster et~al.~(2001) for a detailed discussion. Alternatively, they may use least-square estimates (Foster et~al.,~2001), with lower efficiency when the distribution of the random error deviates from normal; this has been observed in our numerical studies.

We have adopted the semiparametric assumption and proposed a pseudo-likelihood method for this model. Via extensive numerical analyses, we have compared the performance of our method with the classical parametric method and the method of Foster et~al.~(2001). When the random error is normally distributed, the parametric method performs the best, and Foster is sightly better than our method only when $\lambda = 1$. However, when the distribution of the random error deviates from normal, our method consistently outperforms the other approaches. 

Our proposed pseudo-likelihood \eqref{main.Complete_profile_loglik} is a U-process with a nonparametric plug-in component $\widehat F_{\lambda, \beta}(\cdot)$. The existing theory for U-processes is not applicable, so developing the theoretical properties of the estimators is a challenging task. We have used advanced empirical process techniques. We believe that these developments will benefit research into M-estimators where the objective function is a U-process. Such estimators are not uncommon; they include the objective function from the pairwise likelihood (e.g., Heller and Qin, 2001) and that from the binomial/multinomial likelihood (Tian et~al., 2021).

There are many potential future research topics. For example,  we have assumed that the effect of the covariates on $Y^{(\lambda)}$ is linear. We could explore this assumption by considering models with more complicated structures. We could also consider the Box--Cox model where the  response $Y$ is right censored (Cai et al., 2005; Chen, 2012). Furthermore, we could incorporate smoothing techniques into the estimation of the nonparametric function $F(\cdot)$. 

\begin{center}
{SUPPLEMENTARY MATERIALS}
\end{center}
The supplementary materials contain
the full technical details of the proof of Theorem \ref{theorem-normality}.

\begin{center}
{ACKNOWLEDGEMENTS}
\end{center}

Dr.~Yu was
supported in part by Singapore Ministry Education Academic Research Tier 1 Funds (R-155-000-202-114; R-155-000-157-112). 
Dr.~Li's work is supported in part  by the Natural Sciences and Engineering Research Council of Canada (grant number RGPIN-2020-04964). 
The first two authors contribute equally to this work.

\section*{Appendix A: Regularity Conditions}

We impose the following regularity conditions to establish our asymptotic results. They are not necessarily the weakest possible.

\begin{itemize}

\item[] \underline{Condition 1}: $\theta = (\lambda, \beta)\in \Theta$, which is a compact subset of $\RR^{p+1}$.  $F_X(\bx)$ is supported on $\mathX$ and $F_Y(y)$ is supported on $\mathY$. $\mathZ \equiv \mathX\times \mathY$ is a compact subset of $\RR^{p+1}$. Furthermore, $\inf_{y\in \mathY} |y| >0$.

As a consequence, $t = y^{(\lambda)} - \bx^T \beta$ is supported on $\mathT$, which is a compact subset of $\RR$.

\item[] \underline{Condition 2}:  There exists $\eta_0 >0$ such that $F_\theta(t)$ is second-order continuously differentiable for  $\|\theta-\theta_0\|_2\leq \eta_0$ and $t\in \mathT$. Furthermore,
\begin{eqnarray*}
0< \inf_{\bz \in \mathZ, \|\theta-\theta_0\|_2\leq \eta_0} F_{\theta}(\bv_{\theta}) \leq \sup_{\bz \in \mathZ, \|\theta-\theta_0\|_2\leq \eta_0} F_{\theta}(\bv_{\theta}) < 1
\end{eqnarray*}
and 
\begin{eqnarray*}
\inf_{\|\theta-\theta_0\|_2\leq \eta_0} \left| \frac{\partial F_{\theta}(\bv_{\theta})}{\partial \theta} \right| > 0.
\end{eqnarray*}


\item[] \underline{Condition 3}:  For any $t_1, t_2\in \RR$,
\begin{eqnarray*}
\sup_{\beta\in \mathB}\left| F_{X^T\beta}(t_1) - F_{X^T\beta}(t_2) \right| \lesssim |t_1 - t_2|.
\end{eqnarray*}

\item[] \underline{Condition 4}: If $F_\theta(\bv_\theta) = F_0(\bv_0)$ almost surely in $F_Y(y)F_X(\bx)$, then $\theta = \theta_0$.

\item[] \underline{Condition 5}: Both $\Sigma_1$ and $\Sigma_2$ defined by (\ref{olddef-Sigma-1}) and (\ref{olddef-Sigma-2}) are invertible.

\end{itemize}

\section*{Appendix B: Sketch of the Proof of Theorem 1}

We give a blueprint of the proof of Theorem~\ref{theorem-normality}; the lengthy details are relegated to the supplementary document. 

In addition to the notation of Section~\ref{section-asym}, we need the following.  
Throughout the development, ``$\lesssim$" denotes smaller than, up to a universal constant; $C$ denotes a large universal constant; and $c$ denotes a small positive universal constant.
 
For any positive integer $i,j$, let $Z_{i,j} = (Y_i, X_j)$ and $\bz_{i,j} = (y_i, \bx_j)$.  Therefore, $Z_{i,i} = Z_i = (Y_i, X_i)$ and likewise $\bz_{i,i} = \bz_i = (y_i, \bx_i)$.  Recall that $V_{\theta} = Y^{(\lambda)} - X^T\beta$,  $V_{\theta, i,j} = Y_i^{(\lambda)} - X_j^T\beta$ and define accordingly $\bv_\theta = y^{(\lambda)} - \bx^T\beta$, $\bv_{\theta, i,j} = y_i^{(\lambda)} - \bx_j^T\beta$. Set $\bv_0 = \bv_{\theta_0}$, $\bv_{0, i,j} = \bv_{\theta_0, i,j}$.

Recalling the definition of $\dot{V}_\theta$ given by (\ref{main.def-V-dot-theta}), we define accordingly 
\begin{eqnarray}
\dot \bv_\theta=\frac{\partial \bv_\theta}{\partial \theta} = \left\{ \begin{array}{ll} \left( \begin{matrix} \lambda^{-2}\left\{\lambda y^{\lambda} \log y - y^\lambda  + 1\right\} \\-\bx \end{matrix} \right) & \mbox{if} \quad \lambda \neq 0  \\ \left( \begin{matrix} (\log y)^2/2 \\-\bx \end{matrix} \right) & \mbox{if} \quad \lambda = 0 \end{array} \right., \label{def-dot-v}
\end{eqnarray}
and we define $\dot{\bv}_{\theta, i,j}$, $\dot{\bv}_0$ similarly.

Let $\{ Z_i \}_{i=1,\ldots,n}$ be our observations; recall that we have the following definition in Section~\ref{Section_Complete data}:
\begin{eqnarray}
\widehat G_{\theta}(t) &=& \frac{1}{n} \sum_{i=1}^n I(Y_i^{(\lambda)} - X_i^T \beta \leq t) = \frac{1}{n} \sum_{i=1}^n I(V_{\theta,i} \leq t) \nonumber \\
\widehat F_{\theta}(t) &=& \left\{\widehat G_{\theta}(t) \vee n^{-2}\right\}\wedge (1-n^{-2}). \label{def-F-hat-theta}
\end{eqnarray}
Let $\widehat F_0(t) = \widehat F_{\theta_0}(t)$.

The proof has three main steps. 

\subsection*{Step 1: Consistency of $\widehat \theta$}

In Step 1, we show that
\begin{eqnarray}
\widehat \theta - \theta_0 = o_p(1).  \label{main.consistency-eq-0}
\end{eqnarray}
To this end, we define
\begin{eqnarray*}
M(\theta) = \int\left\{F_0 (y_2^{(\lambda_0)} - \bx_1^T\beta_0) - {F}_\theta\Big(y_2^{(\lambda)}-\bx_1^T{\beta}\Big)  \right\}^2 d F_X(\bx_1) d F_Y(y_2).
\end{eqnarray*}
Then, based on the arguments in Wald (1949), to show (\ref{main.consistency-eq-0}), we need only to show that
\begin{itemize}

\item[(i)] $M(\widehat \theta) = o_p(1)$;
\item[(ii)] $M(\theta) = 0$ implies that $\theta = \theta_0$;
\item[(iii)] $M(\theta)$ is continuous in $\theta \in \Theta$.

\end{itemize}
Note that (ii) holds because of Condition 4 and (iii) holds based on Condition 2. We need to show (i): it follows from Lemmas \ref{lemma-2} and \ref{lemma-3} given below, which are Lemmas 9 and 10 of the supplementary document. Therefore, the proof of Step 1 is complete. 

We need the following notation:
\begin{eqnarray*}
\gamma_1(y, \bx; F, \lambda, \beta) &=&
4\left\{ \sqrt{ \frac{{F}_{\theta}\Big(y^{(\lambda)}-\bx^T{\beta}\Big)}{ {F}_0\Big(y^{(\lambda_0)}-\bx^T{\beta_0}\Big)}} -1 \right\}, \\
\gamma_2(y, \bx; F, \lambda, \beta) &=& 4 \left\{  \sqrt{\frac{1-{F}_{\theta}\Big(y^{(\lambda)}-\bx^T{\beta}\Big)}{ 1- {F}_0\Big(y^{(\lambda_0)}-\bx^T{\beta_0}\Big)}} -1 \right\}.
\end{eqnarray*}

\begin{lemma} \label{lemma-2}
Assume Conditions 1 and 2. We have
\begin{eqnarray*}
&&\int\left\{ F_0(y_2^{(\lambda_0)} - \bx_1^T\beta_0) - {F}_{\widehat \theta}\Big(y_2^{(\widehat\lambda)}-\bx_1^T{\widehat\beta}\Big)  \right\}^2 d F_X(\bx_1) d F_Y(y_2) \\
&\leq& \int \left\{ I(y_1\leq y_2) \gamma_1(y_2, \bx_1; \widehat F, \widehat \lambda, \widehat \beta) + I(y_1>y_2) \gamma_2(y_2, \bx_1; \widehat F, \widehat \lambda, \widehat \beta) \right\} \\ && \hspace{0.7in} \times \Big\{d\FF_{X,Y}(\bx_1, y_1) d\FF_{X,Y}(\bx_2, y_2) - d F_{X,Y}(\bx_1, y_1) d F_{X,Y}(\bx_2, y_2)\Big\} + o_p(1).
\end{eqnarray*}

\end{lemma}

\begin{lemma} \label{lemma-3}

Assume Conditions 1 and 2. We have
\begin{eqnarray*}
&&\int \left\{ I(y_1\leq y_2) \gamma_1(y_2, \bx_1; \widehat F, \widehat \lambda, \widehat \beta) + I(y_1>y_2) \gamma_2(y_2, \bx_1; \widehat F, \widehat \lambda, \widehat \beta) \right\} \\ && \hspace{0.7in} \times \Big\{d\FF_{X,Y}(\bx_1, y_1) d\FF_{X,Y}(\bx_2, y_2) - d F_{X,Y}(\bx_1, y_1) d F_{X,Y}(\bx_2, y_2)\Big\} = o_p(1).
\end{eqnarray*}

\end{lemma}

\subsection*{Step 2: Root $n$ consistency of $\widehat \theta$}

In Step 2, we apply Lemma \ref{lemma-M-rate} below to show that
\begin{eqnarray}
\sqrt{n}\left(\widehat \theta - \theta_0\right) = O_p(1). \label{main.root-n-eq-1}
\end{eqnarray}
This  lemma is adapted from Theorem 3.4.1 of van der Vaart and Wellner (1996).

\begin{lemma} \label{lemma-M-rate}
For each $n$, let $\MM_n$ and $M_n$ be stochastic processes indexed by $\Theta$. Let $0\leq \delta_n < \eta$ be arbitrary. Suppose that for every $n$ and $\delta_n<\delta \leq \eta$
\begin{eqnarray}
\sup_{\delta/2<\|\theta - \theta_0\|_2\leq \delta, \theta \in \Theta} M_n(\theta) - M_n(\theta_0) \lesssim -\delta^2; \label{main.eq-M-rate-cond1}\\
E^*\left[\sup_{\delta/2<\|\theta - \theta_0\|_2\leq \delta, \theta \in \Theta} \sqrt{n} \Big\{ (\MM_n - M_n)(\theta) - (\MM_n - M_n)(\theta_0) \Big\}^{+}\right]\lesssim \phi_n(\delta), \label{main.eq-M-rate-cond2}
\end{eqnarray}
for functions $\phi_n$ such that $\delta\to \phi_n(\delta)/\delta^\tau$ is decreasing on $(\delta_n, \eta)$, for some $\tau<2$. Let $r_n \lesssim \delta_n^{-1}$ satisfy
\begin{eqnarray}
r_n^2 \phi_n\left( \frac{1}{r_n} \right) \leq \sqrt{n}, \qquad \mbox{for every }n. \label{main.eq-M-rate-cond3}
\end{eqnarray}
If $\widehat \theta_n$ takes its values in $\Theta$ and satisfies $\MM_n(\widehat \theta) \geq \MM_n(\theta_0) - O_p(r_n^{-2})$ and $\|\widehat \theta - \theta\|_2$ converges to zero in probability, then $r_n \|\widehat \theta - \theta\|_2 = O_p^*(1)$.
\end{lemma}
Recalling that
\begin{eqnarray*}
\ell(\lambda, \beta) &=& \sum_{j=1}^n\sum_{i=1}^n \left[ I_{i,j}\log\widehat {F}_{\theta}(V_{\theta, j,i})+(1-I_{i,j})\log\left\{1-\widehat {F}_{\theta}(V_{\theta, j,i})\right\}\right],
\end{eqnarray*}
we define
\begin{eqnarray*}
\widetilde \ell(\lambda, \beta) &=& \sum_{j=1}^n\sum_{i=1}^n \left[ I_{i,j}\log F_{\theta}(V_{\theta, j,i})+(1-I_{i,j})\log\left\{1-F_{\theta}(V_{\theta, j,i})\right\}\right].
\end{eqnarray*}
Accordingly,
\begin{eqnarray*}
\ell(\lambda_0, \beta_0) &=& \sum_{j=1}^n\sum_{i=1}^n \left[ I_{i,j}\log\widehat {F}_{0}(V_{0, j,i})+(1-I_{i,j})\log\left\{1-\widehat {F}_{0}(V_{0, j,i})\right\}\right],\\
\widetilde \ell(\lambda_0, \beta_0) &=& \sum_{j=1}^n\sum_{i=1}^n \left[ I_{i,j}\log F_{0}(V_{0, j,i})+(1-I_{i,j})\log\left\{1-F_{0}(V_{0, j,i})\right\}\right].
\end{eqnarray*}
We will apply Lemma \ref{lemma-M-rate} to show (\ref{main.root-n-eq-1}). According to Lemma~\ref{lemma-M-rate}, $\MM_n(\theta)$ and $M_n(\theta)$ are defined to be
\begin{eqnarray*}
\MM_n(\theta) &=& \frac{1}{n^2} \ell(\lambda, \beta) \\
M_n(\theta) &=& \frac{1}{n^2} E\left\{ \widetilde \ell(\theta) \right\}\\
 &=& E \left[ I_{i,j}\log\left\{F_{\theta}(V_{\theta, j,i})\right\}
+(1-I_{i,j})\log\left\{1-F_{\theta}(V_{\theta, j,i})\right\} \right].
\end{eqnarray*}
Then, based on the definition of $\widehat \theta$,
\begin{eqnarray*}
\MM_n(\widehat \theta) \geq \MM_n(\theta_0),
\end{eqnarray*}
and we have shown the consistency of $\widehat \theta$ in Step 1. To apply 
Lemma~\ref{lemma-M-rate} to show the root $n$ consistency of $\widehat \beta$, we need to specify ``$\delta_n$, $\eta$, $\tau$", and verify (\ref{main.eq-M-rate-cond1}) and
(\ref{main.eq-M-rate-cond2}). Furthermore, for $\phi_n(\delta)$ from (\ref{main.eq-M-rate-cond2}), we need to verify that it satisfies (\ref{main.eq-M-rate-cond3}) for $r_n = \sqrt{n}$ and that $\phi_n(\delta)/\delta^\tau$ is decreasing on $(\delta_n,\eta)$.

Note that (\ref{main.eq-M-rate-cond1}) is verified
by Lemma~\ref{lemma-rate-1}, which is Lemma 12 of the supplementary document. To verify (\ref{main.eq-M-rate-cond2}), we decompose
\begin{eqnarray}
&& (\MM_n - M_n)(\theta) - (\MM_n - M_n)(\theta_0)  \nonumber \\
&=& \frac{1}{n^2} \left( \widetilde \ell(\lambda, \beta) - E\left\{\widetilde \ell(\lambda, \beta)\right\} - \left[ \widetilde \ell(\lambda_0, \beta_0) - E\left\{ \widetilde \ell(\lambda_0, \beta_0)\right\} \right] \right) \nonumber \\
 && + \frac{1}{n^2}\left[\ell(\lambda, \beta) -  \widetilde \ell(\lambda, \beta) - \left\{ \ell(\lambda_0, \beta_0) -  \widetilde \ell(\lambda_0, \beta_0) \right\}\right]. \label{main.root-n-eq-2}
\end{eqnarray}
In Lemma~\ref{lemma-rate-2}, which is Lemma 13 of the supplementary document, we verify that for any $\delta < \eta_0$,
\begin{equation}
E \left( \sup_{\theta \in \Theta, \|\theta - \theta_0\|_2 \leq \delta} \left| \widetilde \ell(\lambda, \beta) -  E\left\{\widetilde \ell(\lambda, \beta)\right\} - \left[ \widetilde \ell(\lambda_0, \beta_0) - E\left\{ \widetilde \ell(\lambda_0, \beta_0)\right\} \right] \right| \right) \lesssim n + n^{3/2}\delta. \label{root-n-eq-3}
\end{equation}
Moreover, in Lemma~\ref{lemma-rate-3}, which is Lemma 14 of the supplementary document, we show that
\begin{eqnarray}
&&E\left(\sup_{\theta \in \Theta, \|\theta-\theta_0\|_2 \leq \delta} \left[\ell(\lambda, \beta) -  \widetilde \ell(\lambda, \beta) - \left\{ \ell(\lambda_0, \beta_0) -  \widetilde \ell(\lambda_0, \beta_0) \right\}\right]^+ \right) \nonumber \\
&\lesssim& n\left(1 + \sqrt{\log n} \delta^\alpha + \delta^\alpha \sqrt{-\log\delta}\right) +  n^{3/2} \delta. \label{main.root-n-eq-4}
\end{eqnarray}
Combining (\ref{main.root-n-eq-2})--(\ref{main.root-n-eq-4}), we verify (\ref{main.eq-M-rate-cond2}) with
\begin{eqnarray*}
\phi_n(\delta) = \frac{1 + \sqrt{\log n} \delta^\alpha + \delta^\alpha \sqrt{-\log\delta} }{\sqrt{n}} + \delta,
\end{eqnarray*}
for $\alpha \in (0, 0.25)$. We then have that $\delta \to \phi_n(\delta)/\delta^{1.5}$ is decreasing for $\delta \in (\delta_n, \eta_2)$ for some small $\eta_2 >0$, where $\delta_n$ is defined in the proof of Lemma 14 in the supplementary document. In particular,  $\delta_n= n^{-1/\{2(1-\alpha)\}}$ satisfies $\delta_n^{-1}>\sqrt{n}$. Now set $\eta = \min\{ \eta_0, \eta_1, \eta_2\}$ so that it plays the role of ``$\eta$" in Lemma~\ref{lemma-M-rate}, where $\eta_0$ is given by Condition 2 and  $\eta_1$ is defined by (74) in the proof of Lemma 14 in the supplementary document.  Clearly, $r_n = \sqrt{n}$ satisfies (\ref{main.eq-M-rate-cond3}).
We have finished checking the conditions for Lemma \ref{lemma-M-rate}, and
this completes the proof of Step 2.

\begin{lemma} \label{lemma-rate-1}

Assume Condition 2.  For any $\delta\in (0, \eta_0)$, we have
\begin{eqnarray*}
\sup_{\delta/2 <\|\theta- \theta_0\|_2 \leq \delta, \theta \in \Theta}M_n(\theta) - M_n(\theta_0)\lesssim -\delta^2.
\end{eqnarray*}

\end{lemma}

\begin{lemma} \label{lemma-rate-2}
Assume Conditions 1 and 2. For any $\delta \in (0, \eta_0)$, we have
\begin{eqnarray*}
E \left( \sup_{\|\theta - \theta_0\|_2 \leq \delta} \left| \widetilde \ell(\lambda, \beta) -  E\left\{\widetilde \ell(\lambda, \beta)\right\} - \left[ \widetilde \ell(\lambda_0, \beta_0) - E\left\{ \widetilde \ell(\lambda_0, \beta_0)\right\} \right] \right| \right) \lesssim n + n^{3/2}\delta. \label{root-n-eq-5}
\end{eqnarray*}
\end{lemma}

\begin{lemma} \label{lemma-rate-3}
Assume Conditions 1--3. We have
\begin{eqnarray}
&&E\left(\sup_{\theta \in \Theta, \|\theta-\theta_0\|_2 \leq \delta} \left[\ell(\lambda, \beta) -  \widetilde \ell(\lambda, \beta) - \left\{ \ell(\lambda_0, \beta_0) -  \widetilde \ell(\lambda_0, \beta_0) \right\}\right]^+ \right) \nonumber \\
&\lesssim& n\left(1 + \sqrt{\log n} \delta^\alpha + \delta^\alpha \sqrt{-\log\delta}\right) +  n^{3/2} \delta, \label{root-n-eq-15}
\end{eqnarray}
for some $\alpha \in (0, 0.25)$ and $\delta_n<\delta<\min(\eta_0, \eta_1)$ with $\delta_n= n^{-1/\{2(1-\alpha)\}}$,  $\eta_0$ given by Condition 2, and  $\eta_1$ defined by (74) in the proof of this lemma (i.e., Lemma 14 in the supplementary document).
\end{lemma}

\subsection*{Step 3: Asymptotic normality of $\widehat \theta$}

In Step 3, we establish the asymptotic normality of $\widehat \theta$. In particular, we aim to show that
\begin{eqnarray}
\sqrt{n}(\widehat \theta - \theta_0) \rightsquigarrow N(0, \Sigma), \label{main.norm-main-eq-1}
\end{eqnarray}
where $\Sigma = \frac{1}{4} \Sigma_1^{-1} \Sigma_2 \Sigma_1^{-1}$ with $\Sigma_1$ and $\Sigma_2$ defined by (\ref{olddef-Sigma-1}) and (\ref{olddef-Sigma-2}) respectively.

We need Lemma~\ref{argmax} below, which is adapted from Theorem 14.1 in Kosorok (2008); see also Theorem 3.2.2 in van der Vaart and Wellner (1996).

\begin{lemma} \label{argmax}

Let $\WW_n$, $\WW$ be stochastic processes indexed by a metric space $\mathH$, such that $\WW_n \rightsquigarrow \WW$ in $L^{\infty}(H)$ for every compact $H \subset \mathH$. Suppose also that almost all sample paths $h \mapsto M(h)$ are upper semicontinuous and possess a unique maximum at a (random) point $\widehat h$, which as a random map in $\mathH$ is tight. If the sequence $\widehat h_n$ is uniformly tight and satisfies $\WW_n(\widehat h_n) \geq \sup_{h\in H} \WW_n(h) - o_p(1)$, then $\widehat h_n \rightsquigarrow \widehat h$ in $\mathH$.

\end{lemma}

We apply the argmax theorem above to show (\ref{main.norm-main-eq-1}). Denote $\widehat h_n = \sqrt{n}(\widehat \theta - \theta_0)$ and
let $h = (h_1, h_2^T)^T$, $\theta_{n,h} = \theta_0 + h/\sqrt{n}$, $\lambda_{n,h} = \lambda_0 + h_1/\sqrt{n}$, $\beta_{n,h} = \beta_0 + h_2/\sqrt{n} $. Define
\begin{eqnarray*}
\WW_n(h) = \frac{1}{n}\left\{\ell(\theta_{n, h}) - \ell(\theta_0)\right\}. \label{norm-main-eq-2}
\end{eqnarray*}
Clearly, $\widehat h_n$ is the maximizer of $\WW_n(h)$, and therefore $\WW_n(\widehat h_n) \geq \sup_{h\in \RR^{p+1}} \WW_n(h)$. In Step 2, we have shown that $\widehat h_n$ is uniformly tight.

For $H$ an arbitrary compact subset of $\RR^{p+1}$, consider the process
\begin{eqnarray}
\WW_n(h) = \frac{1}{n}\left\{\ell(\theta_{h, n}) - \ell(\theta_0)\right\} = \WW_{n,1}(h) + \WW_{n,2}(h), \label{main.norm-main-eq-3}
\end{eqnarray}
with $h\in H$, where
\begin{eqnarray*}
\WW_{n,1}(h) &=& \frac{1}{n}\left[\ell(\theta_{n, h}) - \ell(\theta_0) - \left\{\widetilde \ell(\theta_{n, h}) - \widetilde \ell(\theta_0)\right\}\right],\\
\WW_{n,2}(h) &=& \frac 1n \left\{\widetilde \ell(\theta_{n, h}) - \widetilde \ell(\theta_0)\right\}. \label{norm-main-eq-4}
\end{eqnarray*}
We consider $\WW_{n,1}(h)$ and $\WW_{n,2}(h)$ separately. For $\WW_{n,2}(h)$, we show in Lemma~\ref{lemma-norm-0}, which is Lemma 17 of the supplementary document,  that
\begin{eqnarray}
\left\|\WW_{n,2}(h) - \left(h^T\GG_n \varphi - h^T \Sigma_1 h\right) \right\|_{h\in H} =  o_p(1), \label{main.norm-main-eq-5}
\end{eqnarray}
where $\varphi(\cdot)$ is defined by (\ref{olddef-varphi}) and $\Sigma_1$ by (\ref{olddef-Sigma-1}). For $\WW_{n,1}(h)$, we have
\begin{eqnarray}
\WW_{n,1}(h) &=& \frac{1}{n}\left[\ell(\theta_{h, n}) - \ell(\theta_0) - \left\{\widetilde \ell(\theta_{h, n}) - \widetilde \ell(\theta_0)\right\}\right]\nonumber\\
&=& \frac 1n \sum_{j=1}^n\sum_{i=1}^n  I_{i,j}\log\left\{\frac{\widehat F_{\theta_{n,h}}(V_{\theta_{n,h}, j, i}) F_{0}(V_{0,j,i}) }{ \widehat F_{0}(V_{0,j,i}) F_{\theta_{n,h}}(V_{\theta_{n,h}, j, i})}\right\}
\nonumber\\
&&+\frac 1n \sum_{j=1}^n\sum_{i=1}^n (1-I_{i,j})\log\left\{\frac{\left(1-\widehat F_{\theta_{n,h}}(V_{\theta_{n,h}, j, i})\right)(1-F_{0}(V_{0,j,i}))}{ \left(1-\widehat F_{0}(V_{0,j,i})\right) (1-F_{\theta_{n,h}}(V_{\theta_{n,h}, j, i}))}\right\}\nonumber \\
&=& \mathI_5 + \mathI_6. \label{main.norm-main-eq-6}
\end{eqnarray}
Consider $\mathI_5$. By the Taylor expansion for $\log x$ at $x=1$, we have
\begin{eqnarray*}
\mathI_5 &=& \frac 1n \sum_{j=1}^n\sum_{i=1}^n  I_{i,j} \left\{\frac{\widehat F_{\theta_{n,h}}(V_{\theta_{n,h}, j, i}) F_{0}(V_{0,j,i}) }{ \widehat F_{0}(V_{0,j,i}) F_{\theta_{n,h}}(V_{\theta_{n,h}, j, i})} - 1\right\}\\
& &-  \frac 1n \sum_{j=1}^n\sum_{i=1}^n  I_{i,j} \frac{1}{2\xi_{n,h,i,j}}\left\{\frac{\widehat F_{\theta_{n,h}}(V_{\theta_{n,h}, j, i}) F_{0}(V_{0,j,i}) }{ \widehat F_{0}(V_{0,j,i}) F_{\theta_{n,h}}(V_{\theta_{n,h}, j, i})} - 1\right\}^2, \label{norm-main-eq-7}
\end{eqnarray*}
where $\xi_{n,h,i,j}$ is between $\frac{\widehat F_{\theta_{n,h}}(V_{\theta_{n,h}, j, i}) F_{0}(V_{0,j,i}) }{ \widehat F_{0}(V_{0,j,i}) F_{\theta_{n,h}}(V_{\theta_{n,h}, j, i})}$ and 1. Based on Lemma~\ref{lemma-1}, which is  {Lemma 8} of the supplementary document, and Condition 2, when $n$ is sufficiently large, we have
\begin{eqnarray*}
\sup_{1\leq i,j\leq n; h\in H}|\xi_{n,h,i,j} - 1 | \leq \sup_{1\leq i,j\leq n; h\in H} \left|\frac{\widehat F_{\theta_{n,h}}(V_{\theta_{n,h}, j, i}) F_{0}(V_{0,j,i}) }{ \widehat F_{0}(V_{0,j,i}) F_{\theta_{n,h}}(V_{\theta_{n,h}, j, i})}-1 \right| \to 0 \quad \mbox{in probability}. \label{norm-main-eq-8}
\end{eqnarray*}
This implies that
\begin{eqnarray*}
\sup_{1\leq i,j\leq n; h\in H} \frac{1}{\xi_{n,h,i,j}}  = \frac{1}{1-o_p^*(1)}, \label{norm-main-eq-9}
\end{eqnarray*}
where $o_p^*(1)$ is uniform in $1\leq i,j\leq n$ and $h\in H$.
Therefore,
\begin{eqnarray*}
&&\left|\mathI_5 - \frac 1n \sum_{j=1}^n\sum_{i=1}^n  I_{i,j} \left\{\frac{\widehat F_{\theta_{n,h}}(V_{\theta_{n,h}, j, i}) F_{0}(V_{0,j,i}) }{ \widehat F_{0}(V_{0,j,i}) F_{\theta_{n,h}}(V_{\theta_{n,h}, j, i})} - 1\right\} \right|\\
& \lesssim &\frac{n}{1-o_p^*(1)} \sup_{\bz\in \mathZ,h\in H} \left| \frac{\widehat F_{\theta_{n,h}}(\bv_{\theta_{n,h}}) F_0(\bv_{\theta_0})}{\widehat F_0(\bv_{\theta_0}) F_{\theta_{n,h}}(\bv_{\theta_{n,h}})} -1 \right|^2.  \label{norm-main-eq-10}
\end{eqnarray*}
This together with Lemmas \ref{lemma-norm-1} and \ref{lemma-norm-2}, which are Lemmas 18 and 19 in the supplementary document, leads to
\begin{eqnarray}
\sup_{h \in H}\left|\mathI_5 -\sqrt{n} \GG_n \left\{  f_{1,n,h}(\cdot) \right\}\right| = o_p(1), \label{main.norm-main-eq-11}
\end{eqnarray}
where $f_{1,n,h}(\cdot)$ comes from Lemma \ref{lemma-norm-2} and is given by
\begin{eqnarray}
f_{1,n,h}(\bz) = E \left\{ \frac{F_0(V_{0, 2, 1})}{F_{\theta_{n,h}}(V_{\theta_{n,h}, 2,1})} I\left(\bv_{\theta_{n,h}} \leq V_{\theta_{n,h}, 2,1}\right) - I\left(\bv_{0} \leq V_{0, 2,1}\right)\right\}.  \label{main.lemma-norm-2-eq-6-added}
\end{eqnarray}

Using exactly the same derivation, we can verify that
\begin{eqnarray}
\sup_{h \in H}\left|\mathI_6 -\sqrt{n} \GG_n \left\{  f_{2,n,h}(\cdot) \right\}\right| = o_p(1), \label{main.norm-main-eq-13}
\end{eqnarray}
with
\begin{eqnarray*}
f_{2,n,h}(\bz) = E \left[ \frac{1-F_0(V_{0, 2, 1})}{1-F_{\theta_{n,h}}(V_{\theta_{n,h}, 2,1})} \left\{1-I\left(\bv_{\theta_{n,h}} \leq V_{\theta_{n,h}, 2,1}\right)\right\} - \left\{1-I\left(\bv_{0} \leq V_{0, 2,1}\right)\right\}\right]. \label{norm-main-eq-14}
\end{eqnarray*}

Combining (\ref{main.norm-main-eq-6}), (\ref{main.norm-main-eq-11}), and (\ref{main.norm-main-eq-13}) we have
\begin{eqnarray}
\sup_{h\in H}\left|\WW_{n,1}(h) - \sqrt{n} \GG_n \left\{  f_{1,n,h}(\cdot) +  f_{2,n,h}(\cdot) \right\}\right| = o_p(1). \label{main.norm-main-eq-15}
\end{eqnarray}
Furthermore, noting that for any constant $C$, $\GG_n C = 0$, we have
\begin{eqnarray}
\GG_n \left\{  f_{1,n,h}(\cdot) +  f_{2,n,h}(\cdot) \right\} = \GG_n \psi_{n,h}(\cdot), \label{main.norm-main-eq-16}
\end{eqnarray}
where
\begin{eqnarray*}
\psi_{n,h}(\bz) &=& E \left[ \left\{\frac{F_0(V_{0, 2, 1})}{F_{\theta_{n,h}}(V_{\theta_{n,h}, 2,1})} -\frac{1-F_0(V_{0, 2, 1})}{1-F_{\theta_{n,h}}(V_{\theta_{n,h}, 2,1})} \right\}I\left(\bv_{\theta_{n,h}} \leq V_{\theta_{n,h}, 2,1}\right) \right]\\
&=& E\left[\frac{F_0(V_{0, 2, 1}) - F_{\theta_{n,h}}(V_{\theta_{n,h}, 2,1})}{F_{\theta_{n,h}}(V_{\theta_{n,h}, 2,1})\left\{1-F_{\theta_{n,h}}(V_{\theta_{n,h}, 2,1}) \right\}}I\left(\bv_{\theta_{n,h}} \leq V_{\theta_{n,h}, 2,1}\right)\right].
\end{eqnarray*}
Then, based on Lemma~\ref{lemma-norm-3}, which is Lemma 20 in the supplementary document,  we have
\begin{eqnarray}
E\left\|\sqrt{n}\GG_n \psi_{n,h}(\bz) - h^T \GG_n \psi(\bz)\right\|_{h\in H} =  o(1), \label{main.norm-main-eq-17}
\end{eqnarray}
where
\begin{eqnarray*}
\psi(\bz) &=& -E\left[\frac{\dot{F}_0(V_{0,2,1}) + F_0'(V_{0,2,1}) \dot{V}_{0,2,1}}{F_{0}(V_{0, 2,1})\left\{1-F_{0}(V_{0, 2,1}) \right\}}I\left(\bv_{0} \leq V_{0, 2,1}\right)\right],
\end{eqnarray*}
as defined by (\ref{olddef-psi}).
Combining (\ref{main.norm-main-eq-15}), (\ref{main.norm-main-eq-16}), and (\ref{main.norm-main-eq-17}) we have
\begin{eqnarray}
\sup_{h\in H}\left|\WW_{n,1}(h) - h^T \GG_n \psi(\bz)\right| = o_p(1). \label{main.norm-main-eq-18}
\end{eqnarray}
This combined with (\ref{main.norm-main-eq-3}) and (\ref{main.norm-main-eq-5}) gives
\begin{eqnarray*}
\sup_{h\in H} \left| \WW_n(h)  - h^T\GG_n (\varphi+\psi) + h^T \Sigma_1 h \right| =o_p(1).
\end{eqnarray*}
Furthermore, by the central limit theorem and the fact that $\Sigma_2$ is invertible (Condition 5), we have
\begin{eqnarray}
\GG_n (\varphi+\psi) \rightsquigarrow N(0, \Sigma_2), \label{main.norm-main-eq-19}
\end{eqnarray}
where $\Sigma_2$ is given by (\ref{olddef-Sigma-2}).
Now define $\WW(h) = h^T \mathcal{N} - h^T \Sigma_1 h$ where $\mathcal{N}$ is a random vector following the $N(0, \Sigma_2)$ distribution; then $\WW(h)$ has a unique maximum at $\widehat h = 0.5\Sigma_1^{-1} \mathcal{N}$ since $\Sigma_1$ is invertible (Condition 5). 
Combining (\ref{main.norm-main-eq-18}) and (\ref{main.norm-main-eq-19}), we have $\WW_n(h) \rightsquigarrow \WW(h)$, which indicates that $\WW(h)$ plays the role of ``$\WW(h)$" in Lemma \ref{argmax}. This immediately leads to (\ref{main.norm-main-eq-1}) by an application of Lemma \ref{argmax}. Our proof is complete.

\begin{lemma} \label{lemma-1}
Assume Conditions 1 and 2.  For any $\delta\in (0, \eta_0)$, we have, for large $n$,
\begin{eqnarray}
\sqrt{n}E\left\{\sup_{\|\theta-\theta_0\|_2\leq \delta; t\in \mathT} |\widehat F_\theta(t) -  F_\theta(t)|\right\} \lesssim 1,\label{lemma-1-eq-1}\\
\sqrt{n}E\left\{\sup_{\|\theta-\theta_0\|_2\leq \delta; t\in \mathT} |\widehat F_\theta(t) -  F_\theta(t)|^2\right\}\lesssim 1/\sqrt{n}.  \label{lemma-1-eq-2}
\end{eqnarray}
\end{lemma}

\begin{lemma} \label{lemma-norm-0}
Assume Conditions 1 and 2. We have
\begin{eqnarray*}
 \left\|\frac 1n \left\{\widetilde \ell(\theta_{n,h}) - \widetilde \ell(\theta_0)\right\} - \left(h^T\GG_n \varphi - h^T \Sigma_1 h\right) \right\|_{h\in H} =  o_p(1), \label{lemma-norm-0-eq-1}
\end{eqnarray*}
where $\varphi(\cdot)$ is defined by (\ref{olddef-varphi}) and $\Sigma_1$ is defined by (\ref{olddef-Sigma-1}).

\end{lemma}

\begin{lemma} \label{lemma-norm-1}
Assume Conditions 1 and 2. We have
\begin{eqnarray}
\sup_{\bz \in \mathZ,h\in H} \left| \frac{\widehat F_{\theta_{n,h}}(\bv_{\theta_{n,h}}) F_0(\bv_{\theta_0})}{\widehat F_0(\bv_{\theta_0}) F_{\theta_{n,h}}(\bv_{\theta_{n,h}})} -1 \right| = o_p(n^{-1/2}). \label{lemma-norm-1-eq-1}
\end{eqnarray}
\end{lemma}

\begin{lemma} \label{lemma-norm-2}
Assume Conditions 1 and 2.  We have
\begin{eqnarray}
\sup_{h \in H}\left|\frac 1n \sum_{j=1}^n\sum_{i=1}^n  I_{i,j} \left\{\frac{\widehat F_{\theta_{n,h}}(V_{\theta_{n,h}, j,i}) F_0(V_{0,j,i}) }{ \widehat F_0(V_{0,j,i}) F_{\theta_{n,h}}(V_{\theta_{n,h}, j,i})} - 1\right\} -  \sqrt{n} \GG_n \left\{  f_{1,n,h}(\cdot) \right\} \right| = o_p(1), \label{lemma-norm-2-eq-1}
\end{eqnarray}
where $f_{1,n,h}(\cdot)$ is defined by (\ref{main.lemma-norm-2-eq-6-added}).

\end{lemma}

\begin{lemma}\label{lemma-norm-3}

Assume Conditions 1--3. We have
\begin{eqnarray}
E\left\|\sqrt{n}\GG_n \psi_{n,h}(\bz) - h^T \GG_n \psi(\bz)\right\|_{h\in H} =  o(1),\label{lemma-norm-3-eq-1}
\end{eqnarray}
where
\begin{eqnarray*}
\psi_{n,h}(\bz)
&=& E\left[\frac{F_0(V_{0, 2, 1}) - F_{\theta_{n,h}}(V_{\theta_{n,h}, 2,1})}{F_{\theta_{n,h}}(V_{\theta_{n,h}, 2,1})\left\{1-F_{\theta_{n,h}}(V_{\theta_{n,h}, 2,1}) \right\}}I\left(\bv_{\theta_{n,h}} \leq V_{\theta_{n,h}, 2,1}\right)\right]\label{lemma-norm-3-eq-2};\\
\psi(\bz) &=& -E\left[\frac{\dot{F}_0(V_{0,2,1}) + F_0'(V_{0,2,1}) \dot{V}_{0,2,1}}{F_{0}(V_{0, 2,1})\left\{1-F_{0}(V_{0, 2,1}) \right\}}I\left(\bv_{0} \leq V_{0, 2,1}\right)\right]. \label{lemma-norm-3-eq-2-added-1}
\end{eqnarray*}
Note that the definition of $\psi(\bz)$ complies with (\ref{olddef-psi}).

\end{lemma}

\vspace{3em}
\centerline{\bf\sc References}
\begin{description}

\item
Amemiya, T. (1985). Instrumental variable estimator for the nonlinear
errors-in-variable models. {\it Journal of Econometrics}, 38, 273-289.

\item Bennett, S. (1983a). Analysis of survival data by the proportional odds model. {\it Statistics
in Medicine}, 2, 273-277.

\item
Bennett, S. (1983b). Log-logistic regression models for survival data. {\it Applied Statistics},
32, 165-171.

\item
Bickel, P. J.  and Doksum, K. A. (1981). An analysis of transformations
revisited. {\it Journal of the American Statistical Association}, 76, 296-311.

\item
Box, G. E. P. and Cox, D. R. (1964). An analysis of transformations.
{\it Journal of the Royal Statistical Society, Series B}, 26, 211-252.

\item Cai, T., Tian, L., and Wei, L. J. (2005). 
Semiparametric Box--Cox power transformation models for censored survival observations. {\em Biometrika}, 92, 619-632.

\item
Carroll, R. J.  and Ruppert, D. (1985). Transformations in regression: A robust analysis. {\it Technometrics}, 27, 1-12.

\item Chen, B., Li, P., Qin, J., and Yu, T. (2016). Using a monotonic density ratio model to find the asymptotically optimal combination of multiple diagnostic tests. {\it Journal of the American Statistical Association}, 111, 861-874. 

\item Chen, S. (2012). Distribution-free estimation of the Box--Cox regression model with censoring. 
{\it Econometric Theory}, 28, 680-695.

\item
Cox, D. R. (1972). Regression models and life tables.  {\it Journal of the Royal Statistical Society,
Series B}, 34, 187-220.

\item  Cox, D. R. (1975). Partial likelihood. {\it Biometrika}, 62, 269-276.


\item
Flinn, C. and Heckman, J. (1982). 
New methods for analyzing structural models of labor force dynamics. {\it Journal of Econometrics}, 18, 115-168.

\item
Foster, A. M., Tian, L., and Wei, L. J. (2001). Estimation for Box--Cox transformation model without assuming parametric error
distribution. {\em Journal of the American Statistical Association}, 96, 1097-1101.

\item Hammer S. M., Katzenstein D. A., Hughes M. D., Gundacker H., Schooley R. T., Haubrich R. H., Henry W. K., Lederman M. M., 
Phair J. P., Niu M., Hirsch M. S., and Merigan T. C. for the AIDS Clinical Trials Group Study 175 Study Team (1996). 
 A trial comparing nucleoside monotherapy with combination therapy in HIV-infected adults with CD4 cell counts from 200 to 500 per cubic millimeter.  
 {\it New England Journal of Medicine},  335, 1081-1090.


\item
Heller, G. and Qin, J. (2001).
Pairwise rank-based likelihood for estimation and inference on the mixture proportion. {Biometrics}, 57, 813-817.

\item
Hinkley, D. V. and Runger, G. (1984). The analysis of transformed data.
{\it Journal of the American Statistical Association}, 79, 302-309.

\item  Kosorok, M. R. (2008). {\it Introduction to Empirical Processes and Semiparametric Inference.}
New York: Springer.

\item  Lancaster, T. (1990). {\it The Econometric Analysis of Transition Data.} Cambridge: Cambridge University Press.

\item
Newey, W. K. (1990). Efficient instrumental variables estimation of nonlinear
models. {\it Econometrica}, 58, 809-837.


\item
Robinson, P. M. (1991). Best nonlinear three-stage least squares estimation
of certain econometric models. {\it Econometrica}, 59, 755-786.

\item
Sakia, R. M. (1992). The Box--Cox transformation technique: A review.
{\it The Statistician}, 41, 169-178.



\item
Taylor, J. M. G. (1985a). Measures of location of skew distributions
obtained through Box--Cox transformations. {\it Journal of the American
Statistical Association}, 80, 427-432.

\item
Taylor, J. M. G. (1985b). Power transformations to symmetry. {\it Biometrika}, 72, 145-152.

\item
Taylor, J. M. G. (1987). Using a generalized mean as a measure of location.
{\it Biometrical Journal}, 29, 731-738.

\item
Tian, Z., Liang, K., and Li, P. (2021). Maximum multinomial likelihood estimation in compound mixture model with application to malaria study. 
{\it Journal of  Nonparametric Statistics}, DOI 10.1080/10485252.2021.1898609. 

\item  van der Vaart, A. W. and Wellner, J. A. (1996). {\it Weak Convergence and Empirical Processes: With Applications to Statistics.} New York: Springer.

\item Wald, A. (1949). Note on the consistency of the maximum likelihood estimate. {\it Annals of Mathematical Statistics}, 20, 595-601.


\item Zhang, T. and  Wang, L. (2020).  Smoothed empirical likelihood inference and variable selection for quantile regression with nonignorable missing response. {\it Computational Statistics \& Data Analysis}, 144, 106888.


\end{description}

\newpage

\begin{center}
\title{\Large Supplementary materials for \\
 ``Maximum profile binomial likelihood estimation for the semiparametric Box--Cox power transformation model"} \vspace{10mm}
\end{center}

\begin{center}
{\bf Abstract}
\end{center}
This is a supplementary document to the corresponding paper. It contains the technical details for the theoretical results in Section 4 of the main article. 
 \vspace*{0.2in}
 
 \setcounter{equation}{0}
\setcounter{section}{0}
\renewcommand{\theequation} {S.\arabic{equation}}
\setcounter{theorem}{0}
\setcounter{lemma}{0}

\setcounter{remark}{0}
 
\section{Notations, Review of Theorem 1 in the Main Article and the Technical Conditions}

\subsection{Notations and review of Theorem 1 in the main article}

Our proposed $(\lambda, \beta^T)^T$estimator is defined by 
\begin{eqnarray}
\left(\widehat \lambda, \widehat \beta^T\right)^T = {\arg\max}_{\left(\lambda, \beta^T\right)^T\in \Theta} \ell ( \lambda,{\beta}), \label{newdef-estimates}
\end{eqnarray}
where $\Theta$ is a compact subset of $\RR^{p+1}$, and 
\begin{equation}\label{Complete_profile_loglik}
\ell (\lambda,{\beta})=\sum_{j=1}^n\sum_{i=1}^n \left[ I_{i,j}\log\left\{\widehat{F}_{\lambda, \beta}\Big(Y_j^{(\lambda)}-X_i^T{\beta}\Big)\right\}+(1-I_{i,j})\log\left\{1-\widehat{F}_{\lambda, \beta}\Big(Y_j^{(\lambda)}-X_i^T{\beta}\Big)\right\}\right] .
\end{equation}
In Section 4 of the main article, we have introduced the following  notations. Let $\theta=(\lambda,\beta^T)^T$ and $\widehat \theta=\left(\widehat \lambda, \widehat \beta^T\right)^T$; and let $\theta_0=(\lambda_0,\beta_0^T)^T$ be the true values of the corresponding parameters.
Denote $V_{\theta} = Y^{(\lambda)} - X^T\beta$, $V_{\theta, i} = Y_i^{(\lambda)} - X_i^T\beta$, and $V_{\theta, i,j} = Y_i^{(\lambda)} - X_j^T\beta$.
Define
\begin{eqnarray}
F_\theta(t) = P(Y^{(\lambda)} - X^T\beta \leq t) = P(V_\theta \leq t), \label{def-F-theta}
\end{eqnarray}
when $\theta= \theta_0$, we write $F_0 = F_{\theta_0}$, $V_0 = V_{\theta_0}$, $V_{0,i} = V_{\theta_0, i}$, $V_{0, i,j} = V_{\theta_0, i,j}$.
Let $\dot{F}_\theta(t) = \frac{\partial F_{\theta}(t)}{\partial \theta}$ and $F_{\theta}'(t) = \frac{\partial F_\theta(t)}{\partial t}$, if they exist; and denote $\dot{F}_0(t) = \dot{F}_{\theta_0}(t)$, $F_0'(t) = F_{\theta_0}'(t)$. Let
\begin{eqnarray}
\dot V_\theta=\frac{\partial V_\theta}{\partial \theta} = \left\{ \begin{array}{ll} \left( \begin{matrix} \lambda^{-2}\left\{\lambda Y^{\lambda} \log Y- Y^\lambda  + 1\right\} \\-X \end{matrix} \right) & \mbox{if} \quad \lambda \neq 0  \\ \left( \begin{matrix} (\log Y)^2/2 \\-X \end{matrix} \right) & \mbox{if} \quad \lambda = 0 \end{array} \right., \label{def-V-dot-theta}
\end{eqnarray}
and accordingly, with the similar strategies, we can define $\dot{V}_0$, $\dot{V}_{0,i}$, and $\dot{V}_{0,i,j}$.
%

Furthermore, we define
\begin{eqnarray}
\varphi(\bz) &=& E\left[\frac{\dot{F}_0(V_{0,2,1}) + F_0'(V_{0,2,1}) \dot{V}_{0, 2,1}}{F_{0}(V_{0, 2,1})\left\{1-F_{0}(V_{0, 2,1})\right\}} \left\{  I(Y_1 \leq Y_2) - F_{0}(V_{0, 2,1})   \right\} \bigg| Z_1 = \bz\right] \label{def-varphi}\\
\psi(\bz) &=& -E\left[\frac{\dot{F}_0(V_{0,2,1}) + F_0'(V_{0,2,1}) \dot{V}_{0,2,1}}{F_{0}(V_{0, 2,1})\left\{1-F_{0}(V_{0, 2,1}) \right\}}I\left(V_{0,3} \leq V_{0, 2,1}\right) \Big| Z_3 = \bz\right]\label{def-psi}\\
\Sigma_1 &=&  E \left(\left[\frac{\left\{\dot{F}_0(V_{0,2,1}) + F_0'(V_{0,2,1}) \dot{V}_{0, 2,1} \right\} \left\{\dot{F}_0(V_{0,2,1}) + F_0'(V_{0,2,1}) \dot{V}_{0, 2,1} \right\}^T }{F_0(V_{0, 2,1})\left\{1-F_{0}(V_{0, 2,1})\right\}} \right]  \right) \label{def-Sigma-1}\\
\Sigma_2 &=& \mbox{var}\left\{\varphi(Z) + \psi(Z)\right\}. \label{def-Sigma-2}
\end{eqnarray}

With these notations, we have presented the following Theorem in Section 4 of the main article; it establishes the
joint asymptotic distribution of $\left(\widehat \lambda, \widehat \beta^T\right)^T$.

\begin{theorem}
\label{theorem-normality}
Assume Conditions 1--5 in Section \ref{section-condition}; we
have
\begin{eqnarray*}
\sqrt{n}(\widehat \theta - \theta_0) \rightsquigarrow N(0, \Sigma), \label{norm-main-eq-1}
\end{eqnarray*}
where $\Sigma = \frac{1}{4} \Sigma_1^{-1} \Sigma_2 \Sigma_1^{-1}$ with $\Sigma_1$ and $\Sigma_2$ defined by (\ref{def-Sigma-1}) and (\ref{def-Sigma-2}) respectively.
\end{theorem}

Furthermore, to facilitate our technical development, we introduce the following additional notations. They will be used frequently in our subsequent developments. 
Throughout our development, let ``$\lesssim$" denote smaller than, up to a universal constant; use $C$ to denote a generic large universal constant, and use $c$ to denote a generic positive small universal constant.
 
Denote $Z = (Y, X)$, $\bz = (y, \bx)$. For any positive integer $i,j$, let $Z_{i,j} = (Y_i, X_j)$ and $\bz_{i,j} = (y_i, \bx_j)$.  Therefore $Z_{i,i} = Z_i = (Y_i, X_i)$ and likewise $\bz_{i,i} = \bz_i = (y_i, \bx_i)$.  Recall that $V_{\theta} = Y^{(\lambda)} - X^T\beta$,  $V_{\theta, i,j} = Y_i^{(\lambda)} - X_j^T\beta$, define accordingly $\bv_\theta = y^{(\lambda)} - \bx^T\beta$, $\bv_{\theta, i,j} = y_i^{(\lambda)} - \bx_j^T\beta$; and set $\bv_0 = \bv_{\theta_0}$, $\bv_{0, i,j} = \bv_{\theta_0, i,j}$.


Recall the definition of $\dot{V}_\theta$ given by (\ref{def-V-dot-theta}), we define accordingly 
\begin{eqnarray}
\dot \bv_\theta=\frac{\partial \bv_\theta}{\partial \theta} = \left\{ \begin{array}{ll} \left( \begin{matrix} \lambda^{-2}\left\{\lambda y^{\lambda} \log y - y^\lambda  + 1\right\} \\-\bx \end{matrix} \right) & \mbox{if} \quad \lambda \neq 0  \\ \left( \begin{matrix} (\log y)^2/2 \\-\bx \end{matrix} \right) & \mbox{if} \quad \lambda = 0 \end{array} \right., \label{def-dot-v}
\end{eqnarray}
and similarly, we can define $\dot{\bv}_{\theta, i,j}$, $\dot{\bv}_0$.

Let $\{ Z_i \}_{i=1,\ldots,n}$ be our observations; recall that we have the following definition in Section 3 of the main article:
\begin{eqnarray}
\widehat G_{\theta}(t) &=& \frac{1}{n} \sum_{i=1}^n I(Y_i^{(\lambda)} - X_i^T \beta \leq t) = \frac{1}{n} \sum_{i=1}^n I(V_{\theta,i} \leq t) \nonumber \\
\widehat F_{\theta}(t) &=& \left\{\widehat G_{\theta}(t) \vee n^{-2}\right\}\wedge (1-n^{-2}), \label{def-F-hat-theta}
\end{eqnarray}
and let $\widehat F_0(t) = \widehat F_{\theta_0}(t)$.


\subsection{Technical Conditions} \label{section-condition}

In the Appendix of the main article, we have imposed the following regularity conditions that are needed to establish our asymptotic results. They are not necessarily the weakest possible.

\begin{itemize}

\item[] \underline{Condition 1}: $\theta = (\lambda, \beta)\in \Theta$, which is a compact subset of $\RR^{p+1}$.  $F_X(\bx)$ is supported on $\mathX$ and $F_Y(y)$ is supported on $\mathY$. $\mathZ \equiv \mathX\times \mathY$ is a compact subset of $\RR^{p+1}$. Furthermore, $\inf_{y\in \mathY} |y| >0$.

As a consequence, $t = y^{(\lambda)} - \bx^T \beta$ is supported on $\mathT$, which is a compact subset of $\RR$.

\item[] \underline{Condition 2}:  There exists a $\eta_0 >0$, such that $F_\theta(t)$ is second order continuously differentiable for  $\|\theta-\theta_0\|_2\leq \eta_0$ and $t\in \mathT$. Furthermore
\begin{eqnarray*}
0< \inf_{\bz \in \mathZ, \|\theta-\theta_0\|_2\leq \eta_0} F_{\theta}(\bv_{\theta}) \leq \sup_{\bz \in \mathZ, \|\theta-\theta_0\|_2\leq \eta_0} F_{\theta}(\bv_{\theta}) < 1\\
\inf_{\|\theta-\theta_0\|_2\leq \eta_0} \left| \frac{\partial F_{\theta}(\bv_{\theta})}{\partial \theta} \right| > 0.
\end{eqnarray*}


\item[] \underline{Condition 3}:  For any $t_1, t_2\in \RR$,
\begin{eqnarray*}
\sup_{\beta\in \mathB}\left| F_{X^T\beta}(t_1) - F_{X^T\beta}(t_2) \right| \lesssim |t_1 - t_2|.
\end{eqnarray*}

\item[] \underline{Condition 4}: If $F_\theta(\bv_\theta) = F_0(\bv_0)$ almost surely in $F_Y(y)F_X(\bx)$, then $\theta = \theta_0$.

\item[] \underline{Condition 5}: Both $\Sigma_1$ and $\Sigma_2$ defined by (\ref{def-Sigma-1}) and (\ref{def-Sigma-2}) are invertible.

\end{itemize}

\section{Proof of Theorem 1} \label{section-tech-detail}

Our proof for Theorem \ref{theorem-normality} is organised as follows. Section \ref{section-prelim} presents some preliminary results in the literature that are helpful to our development. Section \ref{sec-consistency} shows that $\widehat \theta - \theta_0 = o_p(1)$. Section \ref{proof-rate} verifies that $\widehat \theta$ is root $n$ consistent. Section \ref{section-normal} establishes the asymptotic normality of $\widehat \theta$ claimed in Theorem \ref{theorem-normality}.

\subsection{Preliminaries} \label{section-prelim}
Our technical developments rely heavily on the theory of empirical processes;
we use van der Vaart and Wellner (1996) and Kosorok (2008) as the main references; hereafter we abbreviate them as ``VW" and "Kosorok" respectively. We adapt the commonly used notations in VW and Kosorok. In particular,
for a function $m(\bz)$, we denote
\begin{eqnarray*}
\PP_n\left\{ m(\cdot) \right\}  &=& \frac{1}{n}\sum_{i=1}^n m(Z_i)\\
\PP \left\{ m(\cdot) \right\} &=& \int m(\bz) d F_Z(\bz)\\
\GG_n \left\{ m(\cdot) \right\} &=& \sqrt{n} \left[ \PP_n \left\{ m(\cdot) \right\} - \PP \left\{ m(\cdot) \right\} \right],
\end{eqnarray*}
where $F_Z(\cdot)$ is used to denote the cumulant distribution function for random variable (vector) $Z$. We use ``$\rightsquigarrow$" to denote convergence in distribution, or weak convergence.

We adapt the following conventions to denote norms. For any process or class of functions $\{M(t): t\in T\}$, we use $\|M\|_T$ to denote the supremum norm, namely $\|M\|_T = \sup_{t\in T} |M(t)|$. In particular,
for a class $\mathF$ of functions defined on $\bz\in \mathZ$, we denote
\begin{eqnarray*}
\|\GG_n\|_\mathF = \sup_{m\in \mathF} |\GG_n \left\{m(\cdot)\right\}|.
\end{eqnarray*}
For any $q\geq 1$, we use $\|\cdot\|_{q,\PP}$ to denote $L_q(\PP)$ norm, i.e., for any function $m(\cdot)$ defined on $\mathZ$, $\|m\|_{q, \PP} =  \left[\PP \left\{m^q(\cdot)\right\}\right]^{1/q} $. For any vector $\bz$, $\|\bz\|_q$ denotes the $l_q$ norm in the Euclidean space.

Consider a function class $\mathF$, any $\epsilon>0$, and probability measure $\PP$.  We use $N_{[]}(\epsilon, \mathF, L_r(\PP))$ to denote the bracketing number, i.e., the minimum number of $\epsilon$-brackets in $L_r(\PP)$ needed to ensure that every function $m\in \mathF$ lies in at least one bracket. We use $N(\epsilon, \mathF, L_r(\PP))$ to denote the covering number, i.e., the minimum number of $L_r(P)$ $\epsilon$-balls needed to cover $\mathF$.

We use $\PP^*$ and $E^*$ to denote outer probability and outer expectation, when it is not certain about the measurability of the corresponding random components. Let $T$ be an arbitrary random component, denote by $T^*$ the
minimal measurable majorant of $T$. We refer to Chapter 6 of Kosorok and Chapter  1.2 of VW for more details of these terminologies.
\begin{remark}
We often need  the measurability of the suprema over function classes indexed by parameters in a compact subset of the Euclidean space. In fact,  the measurability is not an obstacle in our analysis. Based on the discussion in Example 1.7.5 in VW, and that measurability is sustained under many computations/transformations (see for example Proposition 1.4 in Shao, 2003),  we observe that in our development the measurability of the suprema over a function class $\{f_\theta(\cdot): \theta\in \Theta\}$ holds if $(Z, \theta)$ is jointly measurable from $\Omega\times \Theta$ to $\RR$, where $\Omega$ is the probability space on which $Z$ is defined. As a consequence, we shall admit the measurability of the superma without a detailed proof when it is appropriate to do so; and  the corresponding $^*$ would be dropped from the superscript of ``$E$", ``$\PP$", and the corresponding random component.
\end{remark}
The following lemmas are adapted from VW. Lemma \ref{lemma-bracketing} is Theorem 2.7.11 in VW;
Lemma \ref{lemma-tail-bound-1} is a subset of Theorem 2.14.2 in VW; and Lemma \ref{lemma-tail-bound-2} is Theorem 3.4.2 of VW.

\begin{lemma} \label{lemma-bracketing}

Let $\mathF = \left\{f_t(\bz): t \in T \right\}$ be a function class. Suppose that $d$ is a metric for the parameter set $T$, and satisfies
\begin{eqnarray}
|f_s(\bz) - f_t(\bz)| \leq d(s,t) \widetilde F(\bz), \label{lemma-bracketing-eq-1}
\end{eqnarray}
for any $s,t\in T$, and some $\widetilde F(\cdot)$ defined on $z\in \mathZ$. Then for any norm $\|\cdot\|$,
\begin{eqnarray*}
N_{[]}(2\epsilon \|\widetilde F\|, \mathF, \|\cdot\|) \leq N(\epsilon, T, d).
\end{eqnarray*}

\end{lemma}

\begin{remark} \label{remark-2}
Note that $\widetilde F$ above is not an envelope function for $\mathF$; however,
if $T$ has diameter $D$ under the metric $d$, then $F(\cdot) = D\cdot \widetilde  F(\cdot)$ is an envelope function for $\mathF - f_{t_0}(\cdot)$, for an arbitrary $f_{t_0}(\cdot) \in \mathF$. Applying the lemma above, we immediately conclude that
\begin{eqnarray*}
N_{[]}\left(\epsilon \|F\|, \mathF-f_{t_0}(\bz), \|\cdot\|\right)  =  N_{[]}\left(\epsilon \|F\|, \mathF, \|\cdot\|\right) \leq N\left( \epsilon/(2D), T, d \right).
\end{eqnarray*}
Furthermore, if $T$ is a compact subset of $\RR^p$ and $d$ is taken as the $\|\cdot\|_2$ distance, then $N\left( \epsilon/(2D), T, d \right)$ given above is bounded by
$C/\epsilon^p$ with $C<\infty$ being a universal constant not depending on $D$. This, together with Lemma \ref{lemma-tail-bound-1} below, is helpful to our development, since for such function class $\mathF$ or $\mathF-f_{t_0}(\bz)$, the bracketing integral $J_{[]} (1, \mathF)$ or $J_{[]}(1, \mathF - f_{t_0}(\bz))$ (defined below) is $\lesssim 1$, up to a universal constant not depending on the diameter $D$.
\end{remark}

We need to define the following bracketing integral:
\begin{eqnarray*}
J_{[]}(\rho, \mathF) = \int_0^\rho \sqrt{1+ \log N_{[]}(\epsilon\|F\|_{2,\PP}, \mathF, L_2(\PP))} d\epsilon
\end{eqnarray*}
with $F$ being an envelope function for function class $\mathF$; and its modified version
\begin{eqnarray*}
\widetilde J_{[]}(\rho, \mathF) =  \int_0^\rho \sqrt{1+ \log N_{[]}(\epsilon, \mathF, L_2(\PP))} d\epsilon
\end{eqnarray*}
which is defined without requiring the existence of an envelope function for the function class $\mathF$.

\begin{lemma} \label{lemma-tail-bound-1}

Let $\mathF$ be a class of measurable functions with measurable envelope function $F$. Then
\begin{eqnarray*}
E^* \left\{ \|\GG_n\|_\mathF \right\} \leq C J_{[]}(1, \mathF) \|F\|_{2, \PP},
\end{eqnarray*}
for some universal constant $C<\infty$.

\end{lemma}

\begin{lemma} \label{lemma-tail-bound-2}

Let $\mathF$ be a class of measurable functions such that $\PP f^2 < \rho^2$ and $\|f\|_{\infty} \leq M$ for every $f\in \mathF$. Then
\begin{eqnarray*}
E^* \left\{ \|\GG_n\|_\mathF \right\} \leq C \widetilde J_{[]}(\rho, \mathF)  \left\{ 1+ \frac{\widetilde J_{[]}(\rho, \mathF)}{\rho^2 \sqrt{n}} M\right\},
\end{eqnarray*}
for some universal constant $C<\infty$.

\end{lemma}

Some available results for U-statistics and U-process in the literature can be applied to facilitate our developments; we summarize them as follows. Let $Z_1, \ldots, Z_n$ be i.i.d. random variables (vectors). Let $k$ be a positive integer and let $\mathF$ be a class of real-valued functions defined on $\mathZ^k = \mathZ\times \ldots \times \mathZ$. For every $m\in \mathF$, a U-statistic of order $k$ is defined to be
\begin{eqnarray*}
\UU_n^k m = \frac{1}{(n)_k}\sum_{i_{(k)}} m(Z_{i_1}, \ldots, Z_{i_k}), \label{prelim-U-eq-1}
\end{eqnarray*}
where $(n)_k = n(n-1)\ldots(n-k+1)$, and $i_{(k)} = (i_1, \ldots, i_k)$ ranges over the $(n)_k$ ordered $k$-tuples of distinct integers from the set $\{1,\ldots, n\}$. The collection $\{\UU_n^k m: m \in \mathF\}$ is called a U-process of order $k$ and is said to be indexed by $\mathF$. Clearly, $\PP_n = \UU_n^1$. We use $\VV_n^k$ and $\PP^k$ to denote the corresponding V-statistic and the products of the probability measures respectively, i.e.,
\begin{eqnarray*}
\VV_n^k m = \frac{1}{n^k} \sum_{i_1 = 1}^n \cdots \sum_{i_k =1}^n m(Z_{i_1}, \ldots, Z_{i_k})\\
\PP^k m = \int \cdots \int m(\bz_1, \ldots, \bz_k) dF_{Z_1} (\bz_1) \cdots dF_{Z_k} (\bz_k).
\end{eqnarray*}
%
%
A function $m$ defined on $\mathZ^k$ is called degenerate, if for every $i=1,\ldots, k$, we have
\begin{eqnarray*}
\PP m(\bz_1, \ldots, \bz_{i-1}, \cdot, \bz_{i+1}, \ldots, \bz_k) = 0. \label{prelim-U-eq-6}
\end{eqnarray*}
Accordingly, $\UU_n^k m$ is called a degenerate U-statistics of order $k$; a function class $\mathF$ is called degenerate on $\mathZ^k$, if every $m \in \mathF$ is degenerate of order $k$.

For any symmetric function $m(\bz_1, \ldots, \bz_k)$ (if it is not symmetric, we can easily make it symmetric; see for example Serfling, 1980, page 172), if $\PP m(\cdot,  \ldots, \cdot) < \infty$, then we can conduct the decomposition (see Serfling, 1980, page 177--178) :
\begin{eqnarray}
\UU_n^k m = \PP^k m(\cdot, \ldots, \cdot) + \PP_n m_1 + \sum_{i=2}^k \UU_n^i m_i, \label{prelim-U-eq-7}
\end{eqnarray}
such that $\UU_n^i m_i, i=2,\ldots, k$ are symmetric and degenerate U-statistics of order $i$. We only provide the explicit formula of $m_1$. The explicit formulae for $m_2, \ldots, m_k$ can also be obtain, but are not needed in our development; we omit them. $m_1$ is given by
\begin{eqnarray}
m_1(\bz) &=&  \PP^{k-1} m(\bz, \cdot, \ldots, \cdot) + \ldots + \PP^{k-1} m(\cdot, \ldots, \cdot, \bz) - k \PP^k m. \label{prelim-U-eq-9}
\end{eqnarray}
We acknowledge that this expression for $m_1$ is valid, even when $m$ is not symmetric.
We need to apply some results in Sherman (1994). We summarize them as follows. The following definition is adapted from Definition 3 in Sherman (1994), where the concept ``packing number" is used. Based on the discussion from VW (page 98), we observe that it can be replaced with covering number.

\begin{defi} \label{def-Euclidean}
A class $\mathF$ of real-valued functions is called Euclidean, if there exists an envelope function $F$ for $\mathF$, and positive universal constants $A$ and $V$ with the following property: for any probability measure $\QQ$  such that $\QQ F^2<\infty$, then
\begin{eqnarray*}
N(\epsilon \|F\|_{2, \QQ}, \mathF, L_2(\QQ)) \leq A\epsilon^{-V}, \qquad \mbox{for any } \epsilon\in (0, 1].
\end{eqnarray*}
\end{defi}

\begin{remark} \label{remark-1}
Combining Remark \ref{remark-2} and Lemma 9.18 in Kosorok, if a function class $\{f_t(x): t\in T\}$, with $T$ being a compact subset of an Euclidean space, satisfies (\ref{lemma-bracketing-eq-1}) with $d = \|\cdot\|_2$, then it is Euclidean.
\end{remark}

Lemmas \ref{lemma-U-stat-1} and \ref{lemma-U-stat-2} below are adapted from Lemma 6 and Corollary 4 in Sherman (1994).

\begin{lemma} \label{lemma-U-stat-1}
If the function class $\mathF$ defined on $\mathZ^k$ is Euclidean for an envelope $F$ satisfying $\PP^k F^2 < \infty$. Let $\mathF_i = \{m_i: m_i \mbox{ given by (\ref{prelim-U-eq-7}) with } m\in \mathF\}$. Then for every $i=1,\ldots, k$, there exists $F_i$ being an envelope for $\mathF_i$ satisfying $\PP^i F_i^2 < \infty$, and $\mathF_i$ is Euclidean for the envelope function $F_i$.
\end{lemma}
\begin{lemma} \label{lemma-U-stat-2}
let $\mathF$ be a class of degenerate functions on $\mathZ^k$, $k\geq 1$. If $\mathF$ is Euclidean for an envelope $F$ satisfying $\PP^k F^2 < \infty$, then the following hold:
\begin{itemize}

\item[(i)] $n^{k/2} \PP \sup_{m \in \mathF} \left| \UU_n^k m \right| = O(1)$;
\item[(ii)] $n^{k/2} \sup_{m\in \mathF} \left| \UU_n^k m \right| = O_p(1)$.

\end{itemize}
\end{lemma}

We need the following decoupling inequality, which is a special case of Theorem 1 in de la Pe$\tilde{n}$a (1992); see also Proposition 2.1 in Arcones and Gin$\acute{e}$ (1993).

\begin{lemma} \label{lemma-decoupling}

Let $\{Z_i\}_{i=1,\ldots,n}$ be independent random variables (vectors), and let $\left\{Z_i^{(r)}\right\}_{i=1,\ldots, n}$ for $r=1,2$  be i.i.d. copies of $\{Z_i\}_{i=1,\ldots,n}$. Let $\mathF$ be a class of uniformly bounded functions defined on $\mathZ\times \mathZ$. Then
\begin{eqnarray*}
E^* \left\| \UU_n^2 m  \right\|_{m\in \mathF} \lesssim E^*\left\| \frac{1}{n(n-1)} \sum_{i\neq j} m\left(Z_i^{(1)}, Z_j^{(2)}\right) \right\|_{m\in \mathF}.
\end{eqnarray*}

\end{lemma}

Furthermore, we use the notation  $\GG_n^{(r)}, r = 1$ or $2$ to denote the corresponding random measure defined based on $\left\{Z_i^{(r)}\right\}_{i=1,\ldots, n}$.

\subsection{Consistency} \label{sec-consistency}

In this section, we shall show that
\begin{eqnarray}
\widehat \theta - \theta_0 = o_p(1).  \label{consistency-eq-0}
\end{eqnarray}
To this end, we define
\begin{eqnarray*}
M(\theta) = \int\left\{F_0 (y_2^{(\lambda_0)} - \bx_1^T\beta_0) - {F}_\theta\Big(y_2^{(\lambda)}-\bx_1^T{\beta}\Big)  \right\}^2 d F_X(\bx_1) d F_Y(y_2).
\end{eqnarray*}
Then, based on the arguments in Wald (1949), to show (\ref{consistency-eq-0}), we need to show only that
\begin{itemize}

\item[(i)] $M(\widehat \theta) = o_p(1)$;
\item[(ii)] $M(\theta) = 0$ implies that $\theta = \theta_0$;
\item[(iii)] $M(\theta)$ is continuous in $\theta \in \Theta$.

\end{itemize}
Note that (ii) holds because of Condition 4; (iii) holds based on Condition 2; therefore, we only need to show (i). The proof of (i) is structured as Lemmas \ref{lemma-2} and \ref{lemma-3} below.

We need the following notations:
\begin{eqnarray*}
\gamma_1(y, \bx; F, \lambda, \beta) &=&
4\left\{ \sqrt{ \frac{{F}_{\theta}\Big(y^{(\lambda)}-\bx^T{\beta}\Big)}{ {F}_0\Big(y^{(\lambda_0)}-\bx^T{\beta_0}\Big)}} -1 \right\} \\
\gamma_2(y, \bx; F, \lambda, \beta) &=& 4 \left\{  \sqrt{\frac{1-{F}_{\theta}\Big(y^{(\lambda)}-\bx^T{\beta}\Big)}{ 1- {F}_0\Big(y^{(\lambda_0)}-\bx^T{\beta_0}\Big)}} -1 \right\}.
\end{eqnarray*}

We need to establish the asymptotic convergence rate for $\widehat F_\theta(t)$ first. This proof relies on the bracketing number of a class of indicator functions, established in Lemm \ref{lemma-entropy-1} below.

\begin{lemma} \label{lemma-entropy-1}
Assume Conditions 1 and 2.
For any $0<C<\infty$ and $\delta \in (0, \eta_0)$, consider the function class
\begin{eqnarray*}
\mathC = \Big\{I\{y^{(\lambda)} - \bx^T\beta \leq t\}: \|\theta-\theta_0\|_2 \leq \delta, |t| \leq  C \Big\},
\end{eqnarray*}
defined on $\mathZ$, we have
\begin{eqnarray*}
N_{[]}(\epsilon, \mathC, L_2(\PP)) \lesssim \frac{1}{\epsilon^{2(p+2)}}.
\end{eqnarray*}

\end{lemma}

\proof Consider the function class
\begin{eqnarray*}
\mathF = \Big\{f_{\theta, t}(\bz) = y^{(\lambda)} - \bx^T\beta - t :  \theta\in \Theta, |t| \leq  C \Big\},
\end{eqnarray*}
defined on $\mathZ$. Based on Condition 1, for any $y\in \mathY, \bx \in \mathX$, $f_{\theta, t} \in \mathF$ is continuously differentiable in $(\theta, t) \in \Theta \times [-C, C]$ with uniformly bounded partial derivatives. Therefore, for any $f_{\theta_1, t_1},  f_{\theta_2, t_2} \in \mathF$, we have
\begin{eqnarray*}
|f_{\theta_1, t_1}(\bz) -  f_{\theta_2, t_2}(\bz)| \lesssim \left\|(\theta_1^T, t_1)^T - (\theta_2^T, t_2)^T \right\|_2.
\end{eqnarray*}
Applying Lemma \ref{lemma-bracketing}, we conclude that there exists a universal constant $C_1>0$, for any $\epsilon>0$,
\begin{eqnarray*}
N_{[]}(\epsilon C_1, \mathF, \|\cdot\|_\infty) \lesssim N\left(\epsilon, \Theta \times [-C, C], \|\cdot\|_2\right) \lesssim \frac{1}{\epsilon^{p+2}}.
\end{eqnarray*}
That is
\begin{eqnarray*}
N_{[]}(\epsilon, \mathF, \|\cdot\|_\infty) \lesssim \frac{1}{\epsilon^{p+2}}.
\end{eqnarray*}
Let $\left\{ [l_j(\bz), u_j(\bz)], j=1,\ldots, N \right\}$ be a set of $\epsilon$-brackets that cover $\mathF$, where $N = N_{[]}(\epsilon, \mathF, \|\cdot\|_\infty)$. We assume that for each $j$, there exists an $f_{\theta_j, t_j}\in \mathF$ such that $l_j(\bz)\leq f_{\theta_j, t_j}(\bz)\leq u_j(\bz)$; otherwise the bracket can be removed from this set. Then, 
$$\Big\{[I\left\{u_j(\bz) \leq 0\right\}, I\left\{l_j(\bz) \leq 0\right\}], j=1,\ldots, N\Big\}$$ 
is a set of brackets that cover $\mathC$, with bracket length
\begin{eqnarray*}
&&\|I\left\{l_j(\bz) \leq 0\right\} - I\left\{u_j(\bz) \leq 0\right\}\|_{2,\PP} = \left\{P(l_j(Z)\leq 0, u_j(Z) >0)\right\}^{1/2}\\
&=& \Big\{P(l_j(Z)\leq 0, u_j(Z) >0, |u_j(Z)-l_j(Z)|\geq \epsilon ) \\ && \ \hspace{0.2in}+ P(l_j(Z)\leq 0, u_j(Z) >0, |u_j(Z)-l_j(Z)|< \epsilon )\Big\}^{1/2}\\
&=& \sqrt{P(l_j(Z)\leq 0, u_j(Z) >0, |u_j(Z)-l_j(Z)|< \epsilon )}\\
&\leq& \sqrt{P( |f_{\theta_j, t_j}(Z)|\leq \epsilon )}\\
&=& \sqrt{P\left(t_j-\epsilon \leq Y^{(\lambda_j)} - X^T \beta_j \leq t_j + \epsilon \right)}\\
&=& \sqrt{F_{\theta_j}(t_j + \epsilon) - F_{\theta_j}(t_j- \epsilon)}\\
&\lesssim & \sqrt{\epsilon},
\end{eqnarray*}
where the last ``$\lesssim$" is because of Condition 2. This completes the proof of this lemma. \epf

\begin{lemma} \label{lemma-1}
Assume Conditions 1 and 2.  For any $\delta\in (0, \eta_0)$, we have, for large $n$,
\begin{eqnarray}
\sqrt{n}E\left\{\sup_{\|\theta-\theta_0\|_2\leq \delta; t\in \mathT} |\widehat F_\theta(t) -  F_\theta(t)|\right\} \lesssim 1,\label{lemma-1-eq-1}\\
\sqrt{n}E\left\{\sup_{\|\theta-\theta_0\|_2\leq \delta; t\in \mathT} |\widehat F_\theta(t) -  F_\theta(t)|^2\right\}\lesssim 1/\sqrt{n}.  \label{lemma-1-eq-2}
\end{eqnarray}
\end{lemma}

\proof  We show (\ref{lemma-1-eq-1}) first. Consider the function class $\mathC$ given in Lemma \ref{lemma-entropy-1}. ``$1$" is an envelope function for $\mathC$, therefore with Lemma \ref{lemma-entropy-1}, we immediately have $J_{[]}(1,\mathC) \lesssim 1$.
Applying Lemma \ref{lemma-tail-bound-1} leads to
\begin{eqnarray*}
E\left( \|\GG_n\|_{\mathC}\right) \lesssim 1,
\end{eqnarray*}
which is equivalent to (\ref{lemma-1-eq-1}), since $\sup_{t\in \mathT; \theta \in \Theta} |\widehat F_{\theta}(t) - \widehat G_{\theta}(t)|\leq n^{-2}$.

We proceed to show (\ref{lemma-1-eq-2}). Let $\left\{Z_i^{(r)}\right\}_{i=1,\ldots, n; r = 1,2}$  be i.i.d. copies of $\{Z_i\}_{i=1,\ldots,n}$ and apply Lemma \ref{lemma-decoupling}, we have
\begin{eqnarray*}
&&\sqrt{n}E\left\{\sup_{\|\theta-\theta_0\|_2\leq \delta; t\in \mathT} |\widehat G_\theta(t) -  F_\theta(t)|^2\right\} = \sqrt{n}E\left\{\sup_{\|\theta-\theta_0\|_2\leq \delta; t\in \mathT} \left| \VV_n^2 f_{\theta, t}(\cdot) f_{\theta, t}(\cdot)\right| \right\}\\
&\leq& \frac{n-1}{\sqrt{n}}E\left\{\sup_{\|\theta-\theta_0\|_2\leq \delta; t\in \mathT} \left| \UU_n^2 f_{\theta, t}(\cdot) f_{\theta, t}(\cdot)\right| \right\} +  \sqrt{n}E\left\{\sup_{\|\theta-\theta_0\|_2\leq \delta; t\in \mathT} \left| \frac{1}{n^2} \sum_{i=1}^n f_{\theta, t}^2(Z_i)  \right| \right\}\\
&\lesssim & \sqrt{n}E\left\{\sup_{\|\theta-\theta_0\|_2\leq \delta; t\in \mathT} \left| \frac{1}{n(n-1)} \sum_{i\neq j} f_{\theta, t}(Z_i^{(1)}) f_{\theta, t}(Z_j^{(2)})\right| \right\} + 1/\sqrt{n}\\
&\lesssim& \sqrt{n}E\left\{\sup_{\|\theta-\theta_0\|_2\leq \delta; t\in \mathT} \left| \frac{1}{n^2} \sum_{i=1}^n \sum_{j=1}^n f_{\theta, t}(Z_i^{(1)}) f_{\theta, t}(Z_j^{(2)})\right| \right\} + 1/\sqrt{n} \\
&\leq& \sqrt{n}E\left\{\left\| \frac{1}{n} \sum_{i=1}^n f_{\theta, t}(Z_i^{(1)})  \right\|_{\|\theta-\theta_0\|_2\leq \delta; t\in \mathT} \cdot \left\| \frac{1}{n} \sum_{j=1}^n f_{\theta, t}(Z_j^{(2)})  \right\|_{\|\theta-\theta_0\|_2\leq \delta; t\in \mathT}\right\} + 1/\sqrt{n} \\
&=& \sqrt{n}E\left\{\left\| \frac{1}{n} \sum_{i=1}^n f_{\theta, t}(Z_i^{(1)})  \right\|_{\|\theta-\theta_0\|_2\leq \delta; t\in \mathT}  \right\} \cdot E\left\{\left\| \frac{1}{n} \sum_{j=1}^n f_{\theta, t}(Z_j^{(2)})  \right\|_{\|\theta-\theta_0\|_2\leq \delta; t\in \mathT}  \right\} \\ && + 1/\sqrt{n}\\
&\lesssim& 1/\sqrt{n},
\end{eqnarray*}
where $f_{\theta, t}(\bz)=I\{y^{(\lambda)} - \bx^T\beta \leq t\} - F_\theta(t)$; to derive the last ``$\lesssim$", we have applied (\ref{lemma-1-eq-1}).
This completes the proof of this Lemma by noting $\sup_{t\in \mathT; \theta \in \Theta} |\widehat F_{\theta}(t) - \widehat G_{\theta}(t)|\leq n^{-2}$. \epf

\begin{lemma} \label{lemma-2}
Assume Conditions 1 and 2. We have
\begin{eqnarray*}
&&\int\left\{ F_0(y_2^{(\lambda_0)} - \bx_1^T\beta_0) - {F}_{\widehat \theta}\Big(y_2^{(\widehat\lambda)}-\bx_1^T{\widehat\beta}\Big)  \right\}^2 d F_X(\bx_1) d F_Y(y_2) \\
&\leq& \int \left\{ I(y_1\leq y_2) \gamma_1(y_2, \bx_1; \widehat F, \widehat \lambda, \widehat \beta) + I(y_1>y_2) \gamma_2(y_2, \bx_1; \widehat F, \widehat \lambda, \widehat \beta) \right\} \\ && \hspace{0.5in} \times \Big\{d\FF_{X,Y}(\bx_1, y_1) d\FF_{X,Y}(\bx_2, y_2) - d F_{X,Y}(\bx_1, y_1) d F_{X,Y}(\bx_2, y_2)\Big\} + o_p(1).
\end{eqnarray*}

\end{lemma}

\proof Based on the definition of $\widehat \theta$, we have
\begin{eqnarray}
0&\geq& \ell (\lambda_0,{\beta_0}) - \ell (\widehat \lambda,{\widehat \beta})  \nonumber \\
&=& -\sum_{j=1}^n\sum_{i=1}^n \Bigg[ I_{i,j}\log\left\{\frac{\widehat{F}_{\widehat \theta}\Big(Y_j^{(\widehat\lambda)}-X_i^T{\widehat\beta}\Big)}{ \widehat{F}_0\Big(Y_j^{(\lambda_0)}-X_i^T{\beta_0}\Big)}\right\}
\nonumber \\ && \hspace{0.8in} +(1-I_{i,j})\log\left\{\frac{1-\widehat{F}_{\widehat \theta}\Big(Y_j^{(\widehat\lambda)}-X_i^T{\widehat\beta}\Big)}{ 1-\widehat{F}_0\Big(Y_j^{(\lambda_0)}-X_i^T{\beta_0}\Big)}\right\}\Bigg] \nonumber \\
&=& - n^2 \int I(y_1\leq y_2) \log\left\{\frac{\widehat{F}_{\widehat \theta}\Big(y_2^{(\widehat\lambda)}-\bx_1^T{\widehat\beta}\Big)}{ \widehat{F}_0\Big(y_2^{(\lambda_0)}-\bx_1^T{\beta_0}\Big)}\right\} \nonumber \\ && \hspace{1.4in}\times d\FF_{X,Y}(\bx_1, y_1) d\FF_{X,Y}(\bx_2, y_2) \nonumber \\
&& - n^2 \int I(y_1> y_2) \log\left\{\frac{1-\widehat{F}_{\widehat \theta}\Big(y_2^{(\widehat\lambda)}-\bx_1^T{\widehat\beta}\Big)}{ 1- \widehat{F}_0\Big(y_2^{(\lambda_0)}-\bx_1^T{\beta_0}\Big)}\right\} \nonumber\\ && \hspace{1.4in} d\FF_{X,Y}(\bx_1, y_1) d\FF_{X,Y}(\bx_2, y_2).  \label{consistency-eq-1}
\end{eqnarray}
Using the fact that $\log x\leq 2(\sqrt{x} - 1)$ for any $x>0$, we have
\begin{eqnarray*}
-\log\left\{\frac{\widehat{F}_{\widehat \theta}\Big(y_2^{(\widehat\lambda)}-\bx_1^T{\widehat\beta}\Big)}{ \widehat{F}_0\Big(y_2^{(\lambda_0)}-\bx_1^T{\beta_0}\Big)}\right\}&\geq& 2\left\{ 1- \sqrt{ \frac{\widehat{F}_{\widehat \theta}\Big(y_2^{(\widehat\lambda)}-\bx_1^T{\widehat\beta}\Big)}{ \widehat{F}_0\Big(y_2^{(\lambda_0)}-\bx_1^T{\beta_0}\Big)}} \right\} \\ &=& -0.5\gamma_1(y_2, \bx_1; \widehat F, \widehat \lambda, \widehat \beta)\nonumber
\end{eqnarray*}
\begin{eqnarray}
-\log\left\{\frac{1-\widehat{F}_{\widehat \theta}\Big(y_2^{(\widehat\lambda)}-\bx_1^T{\widehat\beta}\Big)}{ 1- \widehat{F}_0\Big(y_2^{(\lambda_0)}-\bx_1^T{\beta_0}\Big)}\right\}&\geq& 2 \left\{ 1 - \sqrt{\frac{1-\widehat{F}_{\widehat \theta}\Big(y_2^{(\widehat\lambda)}-\bx_1^T{\widehat\beta}\Big)}{ 1- \widehat{F}_0\Big(y_2^{(\lambda_0)}-\bx_1^T{\beta_0}\Big)}} \right\} \nonumber \\ 
&=& -0.5\gamma_2(y_2, \bx_1; \widehat F, \widehat \lambda, \widehat \beta). \label{consistency-eq-2}
\end{eqnarray}
Combining (\ref{consistency-eq-1}) and (\ref{consistency-eq-2}), we have
\begin{eqnarray}
0&\geq& -\int \left\{ I(y_1\leq y_2) \gamma_1(y_2, \bx_1; \widehat F, \widehat \lambda, \widehat \beta) + I(y_1>y_2) \gamma_2(y_2, \bx_1; \widehat F, \widehat \lambda, \widehat \beta) \right\} \nonumber\\ && \hspace{2.8in} \times d\FF_{X,Y}(\bx_1, y_1) d\FF_{X,Y}(\bx_2, y_2) \nonumber \\
&=& -\int \left\{ I(y_1\leq y_2) \gamma_1(y_2, \bx_1; \widehat F, \widehat \lambda, \widehat \beta) + I(y_1>y_2) \gamma_2(y_2, \bx_1; \widehat F, \widehat \lambda, \widehat \beta) \right\} \nonumber \\ && \hspace{1in} \times \Big\{d\FF_{X,Y}(\bx_1, y_1) d\FF_{X,Y}(\bx_2, y_2) - d F_{X,Y}(\bx_1, y_1) d F_{X,Y}(\bx_2, y_2)\Big\} \nonumber \\
&& -\int \left\{ I(y_1\leq y_2) \gamma_1(y_2, \bx_1; \widehat F, \widehat \lambda, \widehat \beta) + I(y_1>y_2) \gamma_2(y_2, \bx_1; \widehat F, \widehat \lambda, \widehat \beta) \right\} \nonumber\\ && \hspace{2.8in} d F_{X,Y}(\bx_1, y_1) d F_{X,Y}(\bx_2, y_2) \nonumber \\
&=& \mathI_1 + \mathI_2.  \label{consistency-eq-3}
\end{eqnarray}
We consider $\mathI_2$. Note that
\begin{eqnarray}
\int I(y_1\leq y_2) dF_{Y|X_1}(y_1) &=& P(Y_1\leq y_2|X_1)
= P(\epsilon_1^* \leq y_2^{(\lambda_0)} - X_1^T\beta_0|X_1)\nonumber \\
&=& F_0(y_2^{(\lambda_0)} - X_1^T\beta_0). \label{consistency-eq-3-1}
\end{eqnarray}
Therefore
\begin{eqnarray}
&&-\int \left\{ I(y_1\leq y_2) \gamma_1(y_2, \bx_1; \widehat F, \widehat \lambda, \widehat \beta) + I(y_1>y_2) \gamma_2(y_2, \bx_1; \widehat F, \widehat \lambda, \widehat \beta) \right\} \nonumber\\ && \hspace{4in} \times d F_{X,Y}(\bx_1, y_1) d F_{X,Y}(\bx_2, y_2) \nonumber \\
&=& -\int\left\{ F_0(y_2^{(\lambda_0)} - \bx_1^T\beta_0) \gamma_1(y_2, \bx_1; \widehat F, \widehat \lambda, \widehat \beta) + \left( 1- F_0(y_2^{(\lambda_0)} - \bx_1^T\beta_0) \right) \gamma_2(y_2, \bx_1; \widehat F, \widehat \lambda, \widehat \beta) \right\} \nonumber \\
&& \hspace{4in} \times d F_X(\bx_1) d F_Y(y_2) \nonumber \\
&=& -\int\left\{ \widehat F_0(y_2^{(\lambda_0)} - \bx_1^T\beta_0) \gamma_1(y_2, \bx_1; \widehat F, \widehat \lambda, \widehat \beta) + \left( 1- \widehat F_0(y_2^{(\lambda_0)} - \bx_1^T\beta_0) \right) \gamma_2(y_2, \bx_1; \widehat F, \widehat \lambda, \widehat \beta) \right\} \nonumber \\
&& \hspace{4in} \times d F_X(\bx_1) d F_Y(y_2) \nonumber \\
&& -\int\left\{F_0(y_2^{(\lambda_0)} - \bx_1^T\beta_0) - \widehat F_0(y_2^{(\lambda_0)} - \bx_1^T\beta_0) \right\} \left\{\gamma_1(y_2, \bx_1; \widehat F, \widehat \lambda, \widehat \beta) - \gamma_2(y_2, \bx_1; \widehat F, \widehat \lambda, \widehat \beta) \right\} \nonumber \\
&& \hspace{4in} \times d F_X(\bx_1) d F_Y(y_2) \nonumber \\
&\equiv& \mathI_{2,1} + \mathI_{2,2}. \label{consistency-eq-4}
\end{eqnarray}
We consider $\mathI_{2,1}$ and $\mathI_{2,2}$ separately. For $\mathI_{2,1}$:
\begin{eqnarray}
\mathI_{2,1}
 &=& 4 \int \Big\{ 1 -  \sqrt{\widehat F_0(y_2^{(\lambda_0)} - \bx_1^T\beta_0)} \sqrt{\widehat{F}_{\widehat \theta}\Big(y_2^{(\widehat\lambda)}-\bx_1^T{\widehat\beta}\Big)} \nonumber \\
&& - \sqrt{1-\widehat F_0(y_2^{(\lambda_0)} - \bx_1^T\beta_0)} \sqrt{1-\widehat{F}_{\widehat \theta}\Big(y_2^{(\widehat\lambda)}-\bx_1^T{\widehat\beta}\Big)}  \Big\} d F_X(\bx_1) d F_Y(y_2)\nonumber\\
&=& 2 \int \left\{ \sqrt{\widehat F_0(y_2^{(\lambda_0)} - \bx_1^T\beta_0)} - \sqrt{\widehat{F}_{\widehat \theta}\Big(y_2^{(\widehat\lambda)}-\bx_1^T{\widehat\beta}\Big)} \right\}^2 d F_X(\bx_1) d F_Y(y_2) \nonumber \\
&& + 2 \int \left\{ \sqrt{1-\widehat F_0(y_2^{(\lambda_0)} - \bx_1^T\beta_0)} - \sqrt{1-\widehat{F}_{\widehat \theta}\Big(y_2^{(\widehat\lambda)}-\bx_1^T{\widehat\beta}\Big)} \right\}^2 d F_X(\bx_1) d F_Y(y_2)\nonumber \\
&\geq & \int\left\{ \widehat F_0(y_2^{(\lambda_0)} - \bx_1^T\beta_0) - \widehat{F}_{\widehat \theta}\Big(y_2^{(\widehat\lambda)}-\bx_1^T{\widehat\beta}\Big)  \right\}^2 d F_X(\bx_1) d F_Y(y_2)\nonumber \\
&\geq & \int\left\{ F_0(y_2^{(\lambda_0)} - \bx_1^T\beta_0) - {F}_{\widehat \theta}\Big(y_2^{(\widehat\lambda)}-\bx_1^T{\widehat\beta}\Big)  \right\}^2 d F_X(\bx_1) d F_Y(y_2) - o_p(1), \label{consistency-eq-5}
\end{eqnarray}
where the last ``$\geq$" is because of Lemma \ref{lemma-1} and the triangle inequality.
For $\mathI_{2,2}$, using the Cauchy-Schiwaz inequality,
\begin{eqnarray}
|\mathI_{2,2}| &\leq& \int\left\{F_0(y_2^{(\lambda_0)} - \bx_1^T\beta_0) - \widehat F_0(y_2^{(\lambda_0)} - \bx_1^T\beta_0) \right\}^2 d F_X(\bx_1) d F_Y(y_2)\nonumber \\
 && \times \int \left\{\gamma_1(y_2, \bx_1; \widehat F, \widehat \lambda, \widehat \beta) - \gamma_2(y_2, \bx_1; \widehat F, \widehat \lambda, \widehat \beta) \right\}^2 d F_X(\bx_1) d F_Y(y_2) = o_p(1),\nonumber \\ \label{consistency-eq-6}
\end{eqnarray}
because of Condition 2 and Lemma \ref{lemma-1}. Combining (\ref{consistency-eq-3})--(\ref{consistency-eq-6}), we complete the proof of this lemma. \epf

\begin{lemma} \label{lemma-3}

Assume Conditions 1 and 2. We have
\begin{eqnarray*}
&&\int \left\{ I(y_1\leq y_2) \gamma_1(y_2, \bx_1; \widehat F, \widehat \lambda, \widehat \beta) + I(y_1>y_2) \gamma_2(y_2, \bx_1; \widehat F, \widehat \lambda, \widehat \beta) \right\} \\ && \hspace{1in} \times \Big\{d\FF_{X,Y}(\bx_1, y_1) d\FF_{X,Y}(\bx_2, y_2) - d F_{X,Y}(\bx_1, y_1) d F_{X,Y}(\bx_2, y_2)\Big\} = o_p(1).
\end{eqnarray*}

\end{lemma}

\proof Note that we only need to show
\begin{eqnarray}
&& \int  I(y_1\leq y_2) \gamma_1(y_2, \bx_1; \widehat F, \widehat \lambda, \widehat \beta)\nonumber \\ &&\times  \Big\{d\FF_{X,Y}(\bx_1, y_1) d\FF_{X,Y}(\bx_2, y_2) - d F_{X,Y}(\bx_1, y_1) d F_{X,Y}(\bx_2, y_2)\Big\} = o_p(1). \label{consistency-eq-7}
\end{eqnarray}
The same arguments can be applied to show
\begin{eqnarray}
&&\int  I(y_1>y_2) \gamma_2(y_2, \bx_1; \widehat F, \widehat \lambda, \widehat \beta) \nonumber \\ && \times \Big\{d\FF_{X,Y}(\bx_1, y_1) d\FF_{X,Y}(\bx_2, y_2) - d F_{X,Y}(\bx_1, y_1) d F_{X,Y}(\bx_2, y_2)\Big\} = o_p(1). \label{consistency-eq-8}
\end{eqnarray}
Based on Condition 2 and Lemma \ref{lemma-1}, we can have
\begin{eqnarray*}
\sup_{y_2 \in \mathY; \bx_1 \in \mathX}\left| \gamma_1(y_2, \bx_1; \widehat F, \widehat \lambda, \widehat \beta) -  \gamma_1(y_2, \bx_1; F, \widehat \lambda, \widehat \beta)\right| = o_p(1).
\end{eqnarray*}
Therefore
\begin{eqnarray}
&&\int  I(y_1\leq y_2) \gamma_1(y_2, \bx_1; \widehat F, \widehat \lambda, \widehat \beta) \nonumber\\ && \times  \Big\{d\FF_{X,Y}(\bx_1, y_1) d\FF_{X,Y}(\bx_2, y_2) - d F_{X,Y}(\bx_1, y_1) d F_{X,Y}(\bx_2, y_2)\Big\} \nonumber \\
&=& \int  I(y_1\leq y_2) \gamma_1(y_2, \bx_1;  F, \widehat \lambda, \widehat \beta)   \Big\{d\FF_{X,Y}(\bx_1, y_1) d\FF_{X,Y}(\bx_2, y_2) - d F_{X,Y}(\bx_1, y_1)\nonumber \\ && \times d F_{X,Y}(\bx_2, y_2)\Big\} + o_p(1). \label{consistency-eq-9}
\end{eqnarray}
We consider the function classes
\begin{eqnarray*}
\mathF_1 &=& \left\{ I(y_1\leq y_2)\gamma_1(y_2, \bx_1; F, \lambda, \beta):   y_2 \in \mathY, \lambda\in \Lambda, \beta\in \mathB \right\}\\
\mathF_2 &=& \left\{ \int I(y_1\leq y_2)\gamma_1(y_2, \bx_1; F, \lambda, \beta) d F_{X,Y}(\bx_1, y_1): \lambda\in \Lambda, \beta\in \mathB \right\},
\end{eqnarray*}
defined on $\mathZ$.
Based on Conditions 1 and 2 it is straightforward to check that for $r=1,2$,
\begin{eqnarray*}
N_{[]}(\epsilon, \mathF_r, L_1(\PP)) \lesssim 1/\epsilon^{A_r}< \infty,
\end{eqnarray*}
for some universal constant $A_r<\infty$. Therefore based on Theorem 2.2 in Kosorok, we conclude that both $\mathF_1$ and $\mathF_2$ are P-Glivenko- Cantelli. As a consequence
\begin{eqnarray}
\sup_{y_2 \in \mathY} \left|\int  I(y_1\leq y_2) \gamma_1(y_2, \bx_1; F, \widehat \lambda, \widehat \beta)   \Big\{d\FF_{X,Y}(\bx_1, y_1) - d F_{X,Y}(\bx_1, y_1) \Big\} \right| = o_p(1)  \label{consistency-eq-10}\\
\int \left\{\int  I(y_1\leq y_2) \gamma_1(y_2, \bx_1;  F, \widehat \lambda, \widehat \beta) \nonumber  d F_{X,Y}(\bx_1, y_1) \right\} \nonumber \\ \times \Big\{d\FF_{X,Y}(\bx_2, y_2) - d F_{X,Y}(\bx_2, y_2)\Big\}= o_p(1) \label{consistency-eq-11}
\end{eqnarray}
Combining (\ref{consistency-eq-10}) with (\ref{consistency-eq-11}) leads to
\begin{eqnarray*}
&&\int  I(y_1\leq y_2) \gamma_1(y_2, \bx_1;  F, \widehat \lambda, \widehat \beta) \\ && \hspace{0.2in} \times \Big\{d\FF_{X,Y}(\bx_1, y_1) d\FF_{X,Y}(\bx_2, y_2) - d F_{X,Y}(\bx_1, y_1) d F_{X,Y}(\bx_2, y_2)\Big\} = o_p(1),
\end{eqnarray*}
which together with (\ref{consistency-eq-9}) leads to (\ref{consistency-eq-7}). We complete the proof of this lemma. \epf


\subsection{Root $n$ consistency} \label{proof-rate}

In this section, we apply Lemma \ref{lemma-M-rate} below to show that
\begin{eqnarray}
\sqrt{n}\left(\widehat \theta - \theta_0\right) = O_p(1). \label{root-n-eq-1}
\end{eqnarray}
This  lemma is adapted from Theorem 3.4.1 in VW.

\begin{lemma} \label{lemma-M-rate}
For each $n$, let $\MM_n$ and $M_n$ be stochastic processes indexed by $\Theta$. Let $0\leq \delta_n < \eta$ be arbitrary. Suppose that for every $n$ and $\delta_n<\delta \leq \eta$
\begin{eqnarray}
\sup_{\delta/2<\|\theta - \theta_0\|_2\leq \delta, \theta \in \Theta} M_n(\theta) - M_n(\theta_0) \lesssim -\delta^2; \label{eq-M-rate-cond1}\\
E^*\left[\sup_{\delta/2<\|\theta - \theta_0\|_2\leq \delta, \theta \in \Theta} \sqrt{n} \Big\{ (\MM_n - M_n)(\theta) - (\MM_n - M_n)(\theta_0) \Big\}^{+}\right]\lesssim \phi_n(\delta), \label{eq-M-rate-cond2}
\end{eqnarray}
for functions $\phi_n$ such that $\delta\to \phi_n(\delta)/\delta^\tau$ is decreasing on $(\delta_n, \eta)$, for some $\tau<2$. Let $r_n \lesssim \delta_n^{-1}$ satisfy
\begin{eqnarray}
r_n^2 \phi_n\left( \frac{1}{r_n} \right) \leq \sqrt{n}, \qquad \mbox{for every }n. \label{eq-M-rate-cond3}
\end{eqnarray}
If $\widehat \theta_n$ takes its values in $\Theta$ and satisfies $\MM_n(\widehat \theta) \geq \MM_n(\theta_0) - O_p(r_n^{-2})$ and $\|\widehat \theta - \theta\|_2$ converges to zero in probability, then $r_n \|\widehat \theta - \theta\|_2 = O_p^*(1)$.
\end{lemma}
Recall that
\begin{eqnarray*}
\ell(\lambda, \beta) &=& \sum_{j=1}^n\sum_{i=1}^n \left[ I_{i,j}\log\widehat {F}_{\theta}(V_{\theta, j,i})+(1-I_{i,j})\log\left\{1-\widehat {F}_{\theta}(V_{\theta, j,i})\right\}\right],
\end{eqnarray*}
and we define
\begin{eqnarray*}
\widetilde \ell(\lambda, \beta) &=& \sum_{j=1}^n\sum_{i=1}^n \left[ I_{i,j}\log F_{\theta}(V_{\theta, j,i})+(1-I_{i,j})\log\left\{1-F_{\theta}(V_{\theta, j,i})\right\}\right].
\end{eqnarray*}
Accordingly
\begin{eqnarray*}
\ell(\lambda_0, \beta_0) &=& \sum_{j=1}^n\sum_{i=1}^n \left[ I_{i,j}\log\widehat {F}_{0}(V_{0, j,i})+(1-I_{i,j})\log\left\{1-\widehat {F}_{0}(V_{0, j,i})\right\}\right]\\
\widetilde \ell(\lambda_0, \beta_0) &=& \sum_{j=1}^n\sum_{i=1}^n \left[ I_{i,j}\log F_{0}(V_{0, j,i})+(1-I_{i,j})\log\left\{1-F_{0}(V_{0, j,i})\right\}\right].
\end{eqnarray*}
We shall apply Lemma \ref{lemma-M-rate} to show (\ref{root-n-eq-1}). $\MM_n(\theta)$ and $M_n(\theta)$ according to Lemma \ref{lemma-M-rate} are defined to be
\begin{eqnarray*}
\MM_n(\theta) &=& \frac{1}{n^2} \ell(\lambda, \beta) \\
M_n(\theta) &=& \frac{1}{n^2} E\left\{ \widetilde \ell(\theta) \right\}\\
 &=& E \left[ I_{i,j}\log\left\{F_{\theta}(V_{\theta, j,i})\right\}
+(1-I_{i,j})\log\left\{1-F_{\theta}(V_{\theta, j,i})\right\} \right].
\end{eqnarray*}
Then, based on the definition of $\widehat \theta$,
\begin{eqnarray*}
\MM_n(\widehat \theta) \geq \MM_n(\theta_0),
\end{eqnarray*}
and we have shown the consistency of $\widehat \theta$ in Section \ref{sec-consistency}. To apply Lemma
\ref{lemma-M-rate} to show the root $n$ consistency of $\widehat \beta$, we need to specify ``$\delta_n$, $\eta$, $\tau$", and verify (\ref{eq-M-rate-cond1}) and
(\ref{eq-M-rate-cond2}). Furthermore, for $\phi_n(\delta)$ from (\ref{eq-M-rate-cond2}), we need to verify that it satisfies (\ref{eq-M-rate-cond3}) for $r_n = \sqrt{n}$, and $\phi_n(\delta)/\delta^\tau$ is decreasing on $(\delta_n,\eta)$.

Note that (\ref{eq-M-rate-cond1}) is verified by
by Lemma \ref{lemma-rate-1}. To verify (\ref{eq-M-rate-cond2}), we decompose
\begin{eqnarray}
 && (\MM_n - M_n)(\theta) - (\MM_n - M_n)(\theta_0) \nonumber \\ &=& \frac{1}{n^2} \left( \widetilde \ell(\lambda, \beta) - E\left\{\widetilde \ell(\lambda, \beta)\right\} - \left[ \widetilde \ell(\lambda_0, \beta_0) - E\left\{ \widetilde \ell(\lambda_0, \beta_0)\right\} \right] \right) \nonumber \\
 && + \frac{1}{n^2}\left[\ell(\lambda, \beta) -  \widetilde \ell(\lambda, \beta) - \left\{ \ell(\lambda_0, \beta_0) -  \widetilde \ell(\lambda_0, \beta_0) \right\}\right]. \label{root-n-eq-2}
\end{eqnarray}
In Lemma \ref{lemma-rate-2}, we verify that for any $\delta < \eta_0$,
\begin{equation}
E \left( \sup_{\theta \in \Theta, \|\theta - \theta_0\|_2 \leq \delta} \left| \widetilde \ell(\lambda, \beta) -  E\left\{\widetilde \ell(\lambda, \beta)\right\} - \left[ \widetilde \ell(\lambda_0, \beta_0) - E\left\{ \widetilde \ell(\lambda_0, \beta_0)\right\} \right] \right| \right) \lesssim n + n^{3/2}\delta, \label{root-n-eq-3}
\end{equation}
and in Lemma \ref{lemma-rate-3}, we show that
\begin{eqnarray}
&&E\left(\sup_{\theta \in \Theta, \|\theta-\theta_0\|_2 \leq \delta} \left[\ell(\lambda, \beta) -  \widetilde \ell(\lambda, \beta) - \left\{ \ell(\lambda_0, \beta_0) -  \widetilde \ell(\lambda_0, \beta_0) \right\}\right]^+ \right) \nonumber \\
&\lesssim& n\left(1 + \sqrt{\log n} \delta^\alpha + \delta^\alpha \sqrt{-\log\delta}\right) +  n^{3/2} \delta. \label{root-n-eq-4}
\end{eqnarray}
Combining (\ref{root-n-eq-2})--(\ref{root-n-eq-4}), we verified (\ref{eq-M-rate-cond2}) with
\begin{eqnarray*}
\phi_n(\delta) = \frac{1 + \sqrt{\log n} \delta^\alpha + \delta^\alpha \sqrt{-\log\delta} }{\sqrt{n}} + \delta,
\end{eqnarray*}
for an $\alpha \in (0, 0.25)$, which satisfies that $\delta \to \phi_n(\delta)/\delta^{1.5}$ is decreasing for $\delta \in (\delta_n, \eta_2)$ for some small $\eta_2 >0$, where $\delta_n$ is defined by (\ref{def-delta-n}) and satisfies $\delta_n^{-1}>\sqrt{n}$. Now set $\eta = \min\{ \eta_0, \eta_1, \eta_2\}$ so that it plays the role of ``$\eta$" in Lemma \ref{lemma-M-rate}, where $\eta_0$ is given by Condition 2 and  $\eta_1$ is defined in (\ref{root-n-eq-43-added}).  Clearly, $r_n = \sqrt{n}$ satisfies (\ref{eq-M-rate-cond3}).
We have finished checking the conditions for Lemma \ref{lemma-M-rate}.
This completes the proof for (\ref{root-n-eq-1}).

\begin{lemma} \label{lemma-rate-1}

Assume Condition 2.  For any $\delta\in (0, \eta_0)$, we have
\begin{eqnarray*}
\sup_{\delta/2 <\|\theta- \theta_0\|_2 \leq \delta, \theta \in \Theta}M_n(\theta) - M_n(\theta_0)\lesssim -\delta^2.
\end{eqnarray*}

\end{lemma}

\proof Note that for any $x>0$, $\log x \leq 2(\sqrt{x} - 1)$; and applying (\ref{consistency-eq-3-1}), we have
\begin{eqnarray*}
&& M_n(\theta) - M_n(\theta_0)\\ &=& E\left[ I_{i,j}\log\left\{\frac{F_{\theta}(V_{\theta, j,i})}{F_{0}(V_{0, j,i})}\right\} + (1-I_{i,j})\log\left\{\frac{1-F_{\theta}(V_{\theta, j,i})}{1-F_{0}(V_{0, j,i})}\right\}   \right]\\
&\leq& 2 E\left\{ I_{i,j}\left( \sqrt{\frac{F_{\theta}(V_{\theta, j,i})}{F_{0}(V_{0, j,i})}} - 1 \right) + (1-I_{i,j}) \left( \sqrt{\frac{1-F_{\theta}(V_{\theta, j,i})}{1-F_{0}(V_{0, j,i})}} - 1 \right)\right\}\\
&=& -2 \int \Big\{ 1 -  \sqrt{F_0(y_2^{(\lambda_0)} - \bx_1^T\beta_0)} \sqrt{{F_\theta}\Big(y_2^{(\lambda)}-\bx_1^T{\beta}\Big)} \nonumber \\
&& - \sqrt{1-F_0(y_2^{(\lambda_0)} - \bx_1^T\beta_0)} \sqrt{1-{F}_\theta\Big(y_2^{(\lambda)}-\bx_1^T{\beta}\Big)}  \Big\} d F_X(\bx_1) d F_Y(y_2)\\
&=& -\int \left\{ \sqrt{F_0(y_2^{(\lambda_0)} - \bx_1^T\beta_0)} - \sqrt{{F}_\theta\Big(y_2^{(\lambda)}-\bx_1^T{\beta}\Big)} \right\}^2 d F_X(\bx_1) d F_Y(y_2) \nonumber \\
&& - \int \left\{ \sqrt{1- F_0(y_2^{(\lambda_0)} - \bx_1^T\beta_0)} - \sqrt{1-{F}_\theta\Big(y_2^{(\lambda)}-\bx_1^T{\beta}\Big)} \right\}^2 d F_X(\bx_1) d F_Y(y_2)\\
&\leq& - \int\left\{ F_0(y_2^{(\lambda_0)} - \bx_1^T\beta_0) - {F}_\theta\Big(y_2^{(\lambda)}-\bx_1^T{\beta}\Big)  \right\}^2 d F_X(\bx_1) d F_Y(y_2),
\end{eqnarray*}
which together with Condition 2 completes the proof of this lemma. \epf

\begin{lemma} \label{lemma-rate-2}
Assume Conditions 1 and 2. For any $\delta \in (0, \eta_0)$, we have
\begin{eqnarray*}
E \left( \sup_{\|\theta - \theta_0\|_2 \leq \delta} \left| \widetilde \ell(\lambda, \beta) -  E\left\{\widetilde \ell(\lambda, \beta)\right\} - \left[ \widetilde \ell(\lambda_0, \beta_0) - E\left\{ \widetilde \ell(\lambda_0, \beta_0)\right\} \right] \right| \right) \lesssim n + n^{3/2}\delta. \label{root-n-eq-5}
\end{eqnarray*}
\end{lemma}

\proof We can write
\begin{eqnarray*}
&&\widetilde \ell(\lambda, \beta) -  E\left\{\widetilde \ell(\lambda, \beta)\right\} - \left[ \widetilde \ell(\lambda_0, \beta_0) - E\left\{ \widetilde \ell(\lambda_0, \beta_0)\right\} \right] \\
 &=& \sum_{j=1}^n \sum_{i=1}^n m_{\theta}(Z_i, Z_j) + \sum_{j=1}^n \sum_{i=1}^n \widetilde m_{\theta}(Z_i, Z_j)\\
&=& n(n-1) \UU_n^2 m_{\theta} + n(n-1)\UU_n^2 \widetilde m_{\theta} + \sum_{i=1}^n \left\{ m_{\theta}(Z_i, Z_i) + \widetilde m_{\theta}(Z_i, Z_i)\right\} \label{root-n-eq-6}
\end{eqnarray*}
where
\begin{eqnarray*}
m_{\theta}(\bz_1, \bz_2) &=& I(y_1\leq y_2) \log \left\{ \frac{F_\theta(\bv_{\theta, 2,1})}{F_0(\bv_{0,2,1})}\right\}  - E\left\{I_{i,j}\log \left( \frac{F_{\theta}(V_{\theta, j,i})}{F_{0}(V_{0, j,i})} \right)\right\} \\
\widetilde m_{\theta}(\bz_1, \bz_2) &=& \left\{1-I(y_1\leq y_2)\right\} \log \left\{ \frac{1-F_\theta(\bv_{\theta, 2,1})}{1-F_0(\bv_{0,2,1})}\right\} \\ && - E\left\{(1-I_{i,j})\log \left( \frac{1-F_{\theta}(V_{\theta, j,i})}{1-F_{0}(V_{0, j,i})} \right)\right\}. \label{root-n-eq-7}
\end{eqnarray*}
Based on Condition 2, we have
\begin{eqnarray*}
\sup_{\|\theta - \theta_0\|_2 \leq \delta} \left| \sum_{i=1}^n \left\{ m_{\theta}(Z_i, Z_i) + \widetilde m_{\theta}(Z_i, Z_i)\right\} \right|\lesssim n. \label{root-n-eq-8}
\end{eqnarray*}
Therefore the proof of this lemma is completed if we can verify that
\begin{eqnarray}
\sup_{\|\theta - \theta_0\|\leq \delta}|\UU_n^2 m_{\theta}| &\lesssim& \frac{1}{n} + \frac{\delta}{n^{1/2}} \label{root-n-eq-9}\\
\sup_{\|\theta - \theta_0\|\leq \delta}|\UU_n^2 \widetilde m_{\theta}| &\lesssim& \frac{1}{n} + \frac{\delta}{n^{1/2}}. \label{root-n-eq-10}
\end{eqnarray}
In fact, we only need to verify (\ref{root-n-eq-9}), as the proof for (\ref{root-n-eq-10}) is exactly the same. Referring to (\ref{prelim-U-eq-7}) and noting that $\PP^2 m_{\theta} = 0$, we have
\begin{eqnarray}
\UU_n^2 m_{\theta} = \PP_n m_{1, \theta} + \UU_n^2 m_{2,\theta}, \label{root-n-eq-11}
\end{eqnarray}
where, $m_{1, \theta}$, by referring to (\ref{prelim-U-eq-9}), is given by
\begin{eqnarray*}
m_{1, \theta}(\bz_1) =   \PP m_{\theta}(\cdot, \bz_1) + \PP m_{\theta}(\bz_1, \cdot). \label{root-n-eq-12}
\end{eqnarray*}
The explicit form for $m_{2,\theta}$ can also be obtained, but it is not essential to our development and is omitted.

Based on Condition 2, the function class $\{m_{\theta}(\cdot, \cdot): \theta \in \Theta, \|\theta - \theta_0\|_2 \leq \delta\}$ defined on $\mathZ^2$ satisfies (\ref{lemma-bracketing-eq-1}), and because of Condition 1, the discussion in Remark \ref{remark-1} is applicable; therefore, it is Euclidean with envelope function $C \delta$, where $C$ is a universal constant. Applying Lemma \ref{lemma-U-stat-1}, we conclude that the function class $\{m_{2, \theta}: \theta \in \Theta, \|\theta - \theta_0\|_2 \leq \delta\}$ is Euclidean for an envelope function $F_2$, which satisfies $\PP F_2^2 < \infty$. This together with Lemma \ref{lemma-U-stat-2} concludes that
\begin{eqnarray}
E \left\{ \sup_{\|\theta - \theta_0\|_2 \leq \delta }\left|\UU_n^2 m_{2,\theta}\right| \right\} = O(1/n). \label{root-n-eq-13}
\end{eqnarray}

Furthermore, based on Condition 2, the function class
\begin{eqnarray*}
\mathF_1 = \{m_{1, \theta}(\cdot): \theta \in \Theta, \|\theta - \theta_0\|_2 \leq \delta\},
\end{eqnarray*}
defined on $\mathZ$ satisfies (\ref{lemma-bracketing-eq-1}), and has envelope function $C\delta$ for some universal constant $C<\infty$; referring to Remark \ref{remark-2}, it satisfies $J_{[]}(1, \mathF_1)\lesssim 1$. Applying Lemma \ref{lemma-tail-bound-1}, we immediately conclude that
\begin{eqnarray}
E\left\{\sup_{ \|\theta - \theta_0\|_2 \leq \delta } \left|\PP_n m_{1,\theta}\right| \right\} \lesssim \delta/\sqrt{n}. \label{root-n-eq-14}
\end{eqnarray}
Combining (\ref{root-n-eq-11}), (\ref{root-n-eq-13}), and (\ref{root-n-eq-14}) leads to (\ref{root-n-eq-9}); and therefore, we complete the proof of this lemma. \epf

\begin{lemma} \label{lemma-rate-3}
Assume Conditions 1--3. We have
\begin{eqnarray}
&&E\left(\sup_{\theta \in \Theta, \|\theta-\theta_0\|_2 \leq \delta} \left[\ell(\lambda, \beta) -  \widetilde \ell(\lambda, \beta) - \left\{ \ell(\lambda_0, \beta_0) -  \widetilde \ell(\lambda_0, \beta_0) \right\}\right]^+ \right) \nonumber \\
&\lesssim& n\left(1 + \sqrt{\log n} \delta^\alpha + \delta^\alpha \sqrt{-\log\delta}\right) +  n^{3/2} \delta, \label{root-n-eq-15}
\end{eqnarray}
for some $\alpha \in (0, 0.25)$, and $\delta_n<\delta<\min(\eta_0, \eta_1)$ with $\delta_n$ defined by (\ref{def-delta-n}),  $\eta_0$ given by Condition 2 and  $\eta_1$ is defined in (\ref{root-n-eq-43-added}).
\end{lemma}

\proof
Consider
\begin{eqnarray*}
&&\ell(\lambda, \beta) -  \widetilde \ell(\lambda, \beta) - \left\{ \ell(\lambda_0, \beta_0) -  \widetilde \ell(\lambda_0, \beta_0) \right\}\\
&=& \sum_{j=1}^n\sum_{i=1}^n \left[ I_{i,j}\log\left\{\frac{\widehat{F}_{\theta}(V_{\theta, j,i}) {F}_{0}(V_{0,j,i})}{ \widehat{F}_0(V_{0,j,i}){F}_{\theta}(V_{\theta, j,i})}\right\} \right] \\ &&
+\sum_{j=1}^n\sum_{i=1}^n \left[ (1-I_{i,j})\log\left\{\frac{\left\{1-\widehat{F}_{\theta}(V_{\theta, j,i})\right\} \left\{1- {F}_{0}(V_{0,j,i})\right\}}{ \left\{1-\widehat{F}_0(V_{0,j,i})\right\}\left\{ 1 - {F}_{\theta}(V_{\theta, j,i})\right\}}\right\} \right]\\
&\equiv& \mathI_3 + \mathI_4. \label{root-n-eq-16}
\end{eqnarray*}
Therefore, to show (\ref{root-n-eq-15}), we only need to show
\begin{eqnarray}
E\left(\sup_{\|\theta-\theta_0\|_2 \leq \delta} \mathI_3^+\right) &\lesssim& n\left(1 + \sqrt{\log n} \delta^\alpha + \delta^\alpha \sqrt{-\log\delta}\right) +  n^{3/2} \delta \label{root-n-eq-17}\\
E\left(\sup_{ \|\theta-\theta_0\|_2 \leq \delta} \mathI_4^+\right) &\lesssim& n\left(1 + \sqrt{\log n} \delta^\alpha + \delta^\alpha \sqrt{-\log\delta}\right) +  n^{3/2} \delta.  \label{root-n-eq-18}
\end{eqnarray}
We show only (\ref{root-n-eq-17}), as the proof for (\ref{root-n-eq-18}) takes exactly the same procedure.

Using the inequality $\log x \leq 2(\sqrt{x} - 1)$ for any $x>0$, we have
\begin{eqnarray}
\mathI_3 &\leq& 2 \sum_{j=1}^n \sum_{i=1}^n \left[ I_{i,j} \left\{ \sqrt{\frac{\widehat{F}_{\theta}(V_{\theta, j,i}) {F}_{0}(V_{0, j,i})}{ \widehat{F}_{0}(V_{0, j,i}){F}_{\theta}(V_{\theta, j,i})}} -1 \right\} \right] \nonumber \\
&\equiv& \mathI_{3,1} + \mathI_{3,2}, \label{root-n-eq-19}
\end{eqnarray}
where
\begin{eqnarray*}
\mathI_{3,1} &=&   \sum_{j=1}^n \sum_{i=1}^n I_{i,j}\left\{\frac{\widehat{F}_{\theta}(V_{\theta, j,i}) }{{F}_{\theta}(V_{\theta, j,i})} -  \frac{ \widehat{F}_{0}(V_{0, j,i})}{{F}_{0}(V_{0, j,i})}\right\} \label{root-n-eq-20}\\
\mathI_{3,2} &=& 2 \sum_{j=1}^n \sum_{i=1}^n I_{i,j} \Big\{ \widehat{F}_{\theta}(V_{\theta, j,i}) {F}_{0}(V_{0, j,i})- \widehat{F}_{0}(V_{0, j,i}){F}_{\theta}(V_{\theta, j,i}) \Big\} \\
 &&  \times \Bigg\{\frac{1}{ \sqrt{ \widehat{F}_{0}(V_{0, j,i}){F}_{\theta}(V_{\theta, j,i})}\left(\sqrt{\widehat{F}_{\theta}(V_{\theta, j,i}) {F}_{0}(V_{0, j,i})} + \sqrt{ \widehat{F}_{0}(V_{0, j,i}){F}_{\theta}(V_{\theta, j,i})}\right)}  \\ && \hspace{0.3in} - \frac{1}{2{F}_{0}(V_{0, j,i}){F}_{\theta}(V_{\theta, j,i})} \Bigg\}. \label{root-n-eq-21}
\end{eqnarray*}
We consider $\mathI_{3,2}$ first. Based on Condition 2, set $0<c = 0.5 \inf_{\bz \in \mathZ, \|\theta-\theta_0\|_2\leq \eta_0 } F_{\theta}(\bv_\theta)$, we have
\begin{eqnarray}
&& E\left(\sup_{\|\theta-\theta_0\|_2\leq \delta}|\mathI_{3,2}|\right) \nonumber \\ &\lesssim& n^2E\left\{\sup_{\|\theta-\theta_0\|_2\leq \delta; \bz\in \mathZ}\frac{(\widehat F_{\theta}(\bv_\theta) - F_{\theta}(\bv_\theta))^2 + (\widehat F_{0}(\bv_0) - F_{0}(\bv_0))^2}{\widehat F_0(\bv_0) } \right\} \nonumber\\
&\lesssim & n^2 E\left\{\sup_{\|\theta-\theta_0\|_2\leq \delta; \bz\in \mathZ}\frac{(\widehat F_{\theta}(\bv_\theta) - F_{\theta}(\bv_\theta))^2 }{\widehat F_0(\bv_0) } \right\} \nonumber \\
&\lesssim &  n^2 E\left\{\sup_{\|\theta-\theta_0\|_2\leq \delta; \bz\in \mathZ}\frac{(\widehat F_{\theta}(\bv_\theta) - F_{\theta}(\bv_\theta))^2 }{c}I(\widehat F_0(\bv_0)> c) \right\} \nonumber \\
&& + n^2 E\left\{\sup_{\|\theta-\theta_0\|_2\leq \delta; \bz\in \mathZ}\frac{(\widehat F_{\theta}(\bv_\theta) - F_{\theta}(\bv_\theta))^2 }{\widehat F_0(\bv_0) } I(\widehat F_0(\bv_0)\leq c)\right\} \nonumber \\
&\lesssim& n + n^4 P\left(\sup_{\bz \in \mathZ} \widehat F_0(\bv_0)\leq c\right) \nonumber \\
&\leq &  n + n^4 P\left(\sup_{\bz \in \mathZ} \left|\widehat F_0(\bv_0) - F_0(\bv_0) \right|\geq c\right) \nonumber \\
&\leq & n + 2 n^4 \exp\left\{ -2 n c^2 \right\},  \label{root-n-eq-22-added}
\end{eqnarray}
where the first ``$\lesssim$" is based on Condition 2 and straightforward computations;  the fourth ``$\lesssim$" is based on Lemma \ref{lemma-1} and  $\inf_{\theta\in \Theta, \bz\in \mathZ}\widehat F_{\theta}(\bv_\theta)\geq n^{-2}$ because of the definition given by (\ref{def-F-hat-theta}); and the last ``$\leq$" is an application of Theorem 11.6 in Kosorok. With (\ref{root-n-eq-22-added}), we have verified that when $n$ is sufficiently large,
\begin{eqnarray}
E\left(\sup_{\|\theta-\theta_0\|_2\leq \delta}|\mathI_{3,2}|\right) \lesssim n. \label{root-n-eq-22}
\end{eqnarray}
We proceed to consider $\mathI_{3,1}$. Recalling the definition of $\widehat F_\theta(\cdot)$ given by (\ref{def-F-hat-theta}), we have
\begin{eqnarray*}
\mathI_{3,1} &=& \frac{1}{n}\sum_{k=1}^n  \sum_{j=1}^n \sum_{i=1}^n I(Y_i\leq Y_j) \\
&&\times\left\{ \frac{I(Y_k^{(\lambda)} - X_k^T \beta \leq Y_j^{(\lambda)}-X_i^T{\beta}) }{ {F}_\theta \Big(Y_j^{(\lambda)}-X_i^T{\beta}\Big)} -  \frac{ I(Y_k^{(\lambda_0)} - X_k^T \beta_0 \leq Y_j^{(\lambda_0)}-X_i^T{\beta_0})}{{F}_0\Big(Y_j^{(\lambda_0)}-X_i^T{\beta_0}\Big)} \right\} \\ && \hspace{4.8 in} + O(1)\\
&=& \frac{1}{n} \sum_{k=1}^n  \sum_{j=1}^n \sum_{i=1}^n f_{\theta}(Z_i,Z_j, Z_k) +O(1), \label{root-n-eq-23}
\end{eqnarray*}
where the $O(1)$ above is uniform in $\theta \in \Theta$, and
\begin{eqnarray}
&&f_{\theta}(\bz_1, \bz_2, \bz_3) \label{root-n-eq-24} \\&& = I(y_1\leq y_2) \left\{ \frac{I\left(y_3^{(\lambda)} - \bx_3^T \beta \leq y_2^{(\lambda)}-\bx_1^T{\beta}\right) }{ {F}_\theta\Big(y_2^{(\lambda)}-\bx_1^T{\beta}\Big)} -  \frac{ I\left(y_3^{(\lambda_0)} - \bx_3^T \beta_0 \leq y_2^{(\lambda_0)}-\bx_1^T{\beta_0}\right)}{{F}_0\Big(y_2^{(\lambda_0)}-\bx_1^T{\beta_0}\Big)} \right\}. \nonumber 
\end{eqnarray}
Based on the definition of $F_{\theta}(\cdot)$ given by (\ref{def-F-theta}), we have $\PP f_\theta(\bz_1, \bz_2,\cdot) = 0$; therefore
\begin{eqnarray}
\mathI_{3,1} - O(1) &=& n^2 \VV_n^3 f_{\theta} = n^{3/2} \VV_n^2 \widetilde f_\theta = \sum_{j=1}^n \GG_n \widetilde f_\theta(\cdot, Z_j) + \sqrt{n}\sum_{j=1}^n \PP \widetilde f_\theta(\cdot, Z_j)\nonumber \\
&=& \sum_{j=1}^n \GG_n \widetilde f_\theta(\cdot, Z_j) + n \int \GG_n \widetilde f_\theta(\bz_1, \cdot) dF_{Z_1}(\bz_1) + n^{3/2} \PP^2 \widetilde f_\theta(\cdot, \cdot)\nonumber \\
&\equiv& \mathI_{3,1,1} + \mathI_{3,1,2} + \mathI_{3,1,3}, \label{root-n-eq-25}
\end{eqnarray}
where
\begin{eqnarray}
\widetilde f_\theta(\bz_1, \bz_2) = \GG_n f_\theta(\bz_1, \bz_2, \cdot). \label{root-n-eq-26}
\end{eqnarray}
The rest of the proof for this lemma proceeds as follows. If we can show
\begin{eqnarray}
E(\|\mathI_{3,1,1}\|_{\|\theta-\theta_0\|_2\leq \delta}) &\lesssim& n \left(1 + \sqrt{\log n} \delta^\alpha + \delta^\alpha \sqrt{-\log\delta}\right) \label{root-n-eq-26-added-1}\\
E(\|\mathI_{3,1,2}\|_{\|\theta-\theta_0\|_2\leq \delta}) &\lesssim& n \left(1 + \sqrt{\log n} \delta^\alpha + \delta^\alpha \sqrt{-\log\delta}\right)\label{root-n-eq-26-added-2}\\
E(\|\mathI_{3,1,3}\|_{\|\theta-\theta_0\|_2\leq \delta}) &\lesssim& n^{3/2} \delta, \label{root-n-eq-26-added-3}
\end{eqnarray}
then combining (\ref{root-n-eq-25}) with (\ref{root-n-eq-26-added-1})--(\ref{root-n-eq-26-added-3}) leads to
\begin{eqnarray*}
E(\|\mathI_{3,1}\|_{\|\theta-\theta_0\|_2\leq \delta}) \lesssim n \left(1 + \sqrt{\log n} \delta^\alpha + \delta^\alpha \sqrt{-\log\delta}\right) +  n^{3/2} \delta,
\end{eqnarray*}
which combined with (\ref{root-n-eq-22}) and (\ref{root-n-eq-19}) concludes (\ref{root-n-eq-17}); and therefore the proof of this lemma is completed.

We consider the proof of (\ref{root-n-eq-26-added-1}) and (\ref{root-n-eq-26-added-2}) first. In fact, based on the definition of $\mathI_{3,1,1}$ and $\mathI_{3,1,2}$ given in (\ref{root-n-eq-25}), it suffices to show that
\begin{eqnarray}
&&E\left\{\left\|\GG_n \widetilde f_\theta(\cdot, \bz_2)\right\|_{\|\theta-\theta_0\|_2\leq \delta, \bz_2 \in \mathZ} \right\} \nonumber \\&\lesssim& \left(1 + \sqrt{\log n} \delta^\alpha + \delta^\alpha \sqrt{-\log\delta}\right) \label{root-n-eq-26-added-4}\\
&& E\left\{\left\|\GG_n \int \widetilde f_\theta(\bz_1, \cdot) dF_{Z_1}(\bz_1)\right\|_{\|\theta-\theta_0\|_2\leq \delta, \bz_1 \in \mathZ} \right\} \nonumber \\ &\lesssim& \left(1 + \sqrt{\log n} \delta^\alpha + \delta^\alpha \sqrt{-\log\delta}\right). \label{root-n-eq-26-added-5}
\end{eqnarray}
We show (\ref{root-n-eq-26-added-4}) only, since the proof for (\ref{root-n-eq-26-added-5}) follows a similar procedure. Referring to (\ref{root-n-eq-26}), we can write
\begin{eqnarray}
\GG_n \widetilde f_\theta(\cdot, \bz_2) &=& \frac{1}{n} \sum_{k=1}^n \sum_{i=1}^n \left[ f_\theta(Z_i, \bz_2, Z_k) - \PP f_\theta( \cdot, \bz_2, Z_k) \right] \nonumber \\
&=& (n-1) \UU_n^2 f_{\theta, \bz_2} + \frac{1}{n} \sum_{k=1}^n f_{\theta, \bz_2} (Z_k, Z_k), \label{root-n-eq-27}
\end{eqnarray}
where
\begin{eqnarray*}
f_{\theta,\bz_2}(\bz_1, \bz_3) = f_\theta(\bz_1, \bz_2, \bz_3) - \PP f_\theta(\cdot, \bz_2, \bz_3). \label{root-n-eq-28}
\end{eqnarray*}
Based on Condition 2, $f_{\theta, \bz_2}(\bz_1, \bz_3)$ is uniformly bounded over $\|\theta-\theta_0\|_2\leq \delta, \bz_1, \bz_2, \bz_3 \in \mathZ$, therefore
\begin{eqnarray}
\left\|\frac{1}{n} \sum_{k=1}^n f_{\theta, \bz_2} (Z_k, Z_k) \right\|_{\|\theta-\theta_0\|_2\leq \delta, \bz_2 \in \mathZ} \lesssim 1. \label{root-n-eq-29}
\end{eqnarray}
Consider $\UU_n^2 f_{\theta, \bz_2}$.  We apply Lemma \ref{lemma-decoupling}: for $\left\{Z_i^{(r)}\right\}_{i=1,\ldots, n; r = 1,2}$ being  i.i.d. copies of $\{Z_i\}_{i=1,\ldots, n}$, we have
\begin{eqnarray}
&&E\left\|\UU_n^2 f_{\theta, \bz_2} \right\|_{\|\theta-\theta_0\|_2\leq \delta; \bz_2 \in \mathZ} \nonumber \\ &\lesssim& E\left\| \frac{1}{n(n-1)}\sum_{k\neq i} f_{\theta, \bz_2} \left(Z_i^{(1)}, Z_k^{(2)} \right)  \right\|_{\|\theta-\theta_0\|_2\leq \delta; \bz_2 \in \mathZ} \nonumber\\
&\lesssim & E\left\| \frac{1}{n^2}\sum_{k=1}^n \sum_{i=1}^n f_{\theta, \bz_2} \left( Z_i^{(1)}, Z_k^{(2)} \right)  \right\|_{\|\theta-\theta_0\|_2\leq \delta; \bz_2 \in \mathZ} + 1, \label{root-n-eq-30}
\end{eqnarray}
since
\begin{eqnarray*}
\left\|\frac{1}{n} \sum_{k=1}^n f_{\theta, \bz_2} \left(Z_k^{(1)}, Z_k^{(2)}\right)\right\|_{\theta \in \Theta; \bz_2 \in \mathZ} \lesssim 1. \label{root-n-eq-31}
\end{eqnarray*}
Furthermore,
\begin{eqnarray}
E\left\| \frac{1}{n}\sum_{k=1}^n \sum_{i=1}^n f_{\theta, \bz_2} \left( Z_i^{(1)}, Z_k^{(2)} \right)  \right\|_{\|\theta-\theta_0\|_2\leq \delta; \bz_2 \in \mathZ} = E\left\| \GG_n^{(1)} \widetilde f_{\theta}^{(2)}(\cdot, \bz_2) \right\|_{\|\theta-\theta_0\|_2\leq \delta; \bz_2 \in \mathZ}, \label{root-n-eq-32}
\end{eqnarray}
where
\begin{eqnarray}
\widetilde f_\theta^{(2)}(\bz_1, \bz_2) = \GG_n^{(2)} f_\theta(\bz_1, \bz_2, \cdot). \label{root-n-eq-33}
\end{eqnarray}
Combining (\ref{root-n-eq-27})--(\ref{root-n-eq-32}) leads to
\begin{eqnarray}
E\left\| \GG_n \widetilde f_\theta(\cdot, \bz_2) \right\|_{\|\theta-\theta_0\|_2\leq \delta; \bz_2 \in \mathZ}\lesssim E\left\| \GG_n^{(1)} \widetilde f_{\theta}^{(2)}(\cdot, \bz_2) \right\|_{\|\theta-\theta_0\|_2\leq \delta; \bz_2 \in \mathZ} + 1, \label{root-n-eq-33-added}
\end{eqnarray}
with $\widetilde f_{\theta}^{(2)}(\cdot, \bz_2)$ defined by (\ref{root-n-eq-33}). We need to derive the bound for the right hand side of (\ref{root-n-eq-33-added}).

Consider the function class
\begin{eqnarray*}
\mathF = \{f_\theta(\bz_1, \bz_2, \bz_3): \bz_1\in \mathZ, \bz_2 \in \mathZ, \|\theta-\theta_0\|_2\leq \delta\}, \label{root-n-eq-34}
\end{eqnarray*}
defined on $\mathZ$. With the similar strategy as the proof for Lemma \ref{lemma-entropy-1}, we can check that there exists a universal constant $A>0$, such that
\begin{eqnarray*}
N_{[]}(\epsilon, \mathF, L_2(\PP)) \lesssim 1/\epsilon^A. \label{root-n-eq-35}
\end{eqnarray*}
Furthermore, based on Condition 2, we can check that every function $f_\theta(\cdot) \in \mathF$ satisfies $\PP f_\theta^2 \lesssim  \delta$ and $\|f_\theta\|_\infty \leq M$ for some universal constant $M<\infty$. Applying Lemma \ref{lemma-tail-bound-2}, when $n$ is sufficiently large, we have
\begin{eqnarray}
E\left\| \widetilde f_\theta^{(2)} (\bz_1, \bz_2) \right \|_{\|\theta-\theta_0\|_2\leq \delta; \bz_1 \in \mathZ; \bz_2\in \mathZ} \lesssim \delta^{\alpha}\left( 1+ \frac{\delta^\alpha}{\delta \sqrt{n}}M \right), \label{root-n-eq-36}
\end{eqnarray}
for an arbitrarily given $0<\alpha<1/4$. Set
\begin{eqnarray}
\delta_n = n^{-1/\{2(1-\alpha)\}}, \label{def-delta-n}
\end{eqnarray}
which plays the role of ``$\delta_n$" given in Lemma \ref{lemma-M-rate}. Clearly $\delta_n^{-1} > n^{1/2}$. When $\delta> \delta_n$, $\frac{\delta^\alpha}{\delta \sqrt{n}} \leq 1$, plugging in which to (\ref{root-n-eq-36}) leads to
\begin{eqnarray}
E\left\| \widetilde f_\theta^{(2)} (\bz_1, \bz_2) \right \|_{\|\theta-\theta_0\|_2\leq \delta; \bz_1 \in \mathZ; \bz_2\in \mathZ} \lesssim \delta^\alpha. \label{root-n-eq-37}
\end{eqnarray}

For any given values of $\left\{Z_i^{(2)}\right\}_{i=1,\ldots,n}$, and $\widetilde f_\theta^{(2)} (\bz_1, \bz_2) $ defined by (\ref{root-n-eq-33}), consider the function class
\begin{eqnarray}
\mathF_{n, \delta} \left(Z_1^{(2)}, \ldots, Z_n^{(2)}\right) = \left \{\widetilde f_{\theta, \bz_2}^{(2)} (\bz_1)\equiv \widetilde f_{\theta}^{(2)} (\bz_1, \bz_2):  \bz_2\in \mathZ, \|\theta-\theta_0\|_2\leq \delta \right\}, \label{root-n-eq-38}
\end{eqnarray}
which is a subset of the function class $\mathF_n\left(Z_1^{(2)}, \ldots, Z_n^{(2)}\right)$ defined in Lemma \ref{lemma-bracketing-F-n-1}.
Note that for every function in this function class, $\widetilde f_{\theta, \bz_2}^{(2)} (Z_i^{(1)})$ for $i=1,\ldots, n$ are i.i.d., conditioning on $\left\{Z_k^{(2)}\right\}_{k=1,\ldots,n}$.  Let
\begin{eqnarray}
\bar F(\bz_1) = \left\|\widetilde f_{\theta, \bz_2}^{(2)} (\bz_1) \right\|_{\bz_2 \in \mathZ; \|\theta-\theta_0\|_2\leq \delta } \label{root-n-eq-38-added}
\end{eqnarray}
be an envelope function for $\mathF_{n, \delta} \left(Z_1^{(2)}, \ldots, Z_n^{(2)}\right)$.  Applying Lemma \ref{lemma-tail-bound-1}, for sufficiently large $n$, we have
\begin{eqnarray}
&&E\left( \left\|\GG_n^{(1)}\right\|_{\mathF_{n, \delta}\left(Z_1^{(2)}, \ldots, Z_n^{(2)}\right)} \Big| Z_1^{(2)}, \ldots, Z_n^{(2)}  \right) \nonumber\\ &\lesssim& J_{[]}\left(1, \mathF_{n, \delta} \left(Z_1^{(2)}, \ldots, Z_n^{(2)}\right)\right) \left\| \bar F \right\|_{2, \PP}, \label{root-n-eq-39}
\end{eqnarray}
Based on Lemma \ref{lemma-bracketing-F-n-1}, for large $n$,
\begin{eqnarray}
&& J_{[]}\left(1, \mathF_{n, \delta}\left(Z_1^{(2)}, \ldots, Z_n^{(2)}\right)\right) \nonumber \\&=& \int_0^1 \sqrt{1+ \log N_{[]}\left(\epsilon\|\bar F\|_{2,\PP}, \mathF_{n, \delta}\left(Z_1^{(2)}, \ldots, Z_n^{(2)}\right), L_2(\PP)\right) } d\epsilon \nonumber\\
&\lesssim & \int_0^1 \sqrt{ 1 + (p+2)\log n  - 2(p+2) \log \|\bar F\|_{2, \PP} - 2(p+2) \log \epsilon } d \epsilon \nonumber \\
&\lesssim & \int_0^1 \sqrt{\log n} d\epsilon + \int_0^1 \sqrt{\left|\log \epsilon\right|} d\epsilon + \int_0^1 \sqrt{\left( - \log \| \bar F \|_{2,\PP} \right)^+} d \epsilon \nonumber  \\
&\lesssim & \sqrt{\log n} +  \sqrt{\left( -\log \| \bar F \|_{2,\PP} \right)^+}. \label{root-n-eq-40}
\end{eqnarray}
Note that there exists a constant $0<c<1$, such that the function $x \sqrt{\log(1/x)}$ is concave and strictly increasing when $x\in (0, c)$, and it is bounded when $x\in [c, 1)$. We have,
\begin{eqnarray}
&& E\left\| \GG_n^{(1)} \widetilde f_{\theta}^{(2)}(\cdot, \bz_2) \right\|_{\|\theta-\theta_0\|_2\leq \delta; \bz_2 \in \mathZ} \nonumber \\ &=& E \left[E\left\{ \left\|\GG_n^{(1)}\right\|_{\mathF_{n, \delta}\left(Z_1^{(2)}, \ldots, Z_n^{(2)}\right)} \Big| Z_1^{(2)}, \ldots, Z_n^{(2)} \right\}   \right] \label{root-n-eq-41-1} \\
&\lesssim& E\left[\left\{\sqrt{\log n} +  \sqrt{\left( -\log \| \bar F \|_{2,\PP} \right)^+} \right\} \| \bar F \|_{2,\PP} \right] \label{root-n-eq-41-2} \\
& = & \sqrt{\log n} E\left(\|\bar F\|_{2, \PP}\right) + E \left\{\| \bar F \|_{2,\PP} \sqrt{\left( -\log \| \bar F \|_{2,\PP} \right)^+}\right\} \nonumber\\
&=& \sqrt{\log n} E\left(\|\bar F\|_{2, \PP}\right) + E \left\{I (\| \bar F \|_{2,\PP}<1) \| \bar F \|_{2,\PP} \sqrt{ \log \frac{1}{\| \bar F \|_{2,\PP}} }\right\} \nonumber \\
&=& \sqrt{\log n} E(\|\bar F\|_{2,\PP}) + E \left\{I (c\leq \| \bar F \|_{2,\PP}<1) \| \bar F \|_{2,\PP} \sqrt{ \log \frac{1}{\| \bar F \|_{2,\PP}} }\right\} \nonumber \\
&& + E \left\{I (\| \bar F \|_{2,\PP}<c) \| \bar F \|_{2,\PP} \sqrt{ \log \frac{1}{\| \bar F \|_{2,\PP}} }\right\} \label{root-n-eq-41-3}\\
&\lesssim& \sqrt{\log n} E(\|\bar F\|_{2,\PP}) + P\left(c\leq \|\bar  F \|_{2,\PP}<1\right) \nonumber \\
&&+ \frac{E\left\{ I (\| \bar F \|_{2,\PP}<c) \| \bar F \|_{2,\PP} \right\} }{P(\| \bar F \|_{2,\PP}<c)}\sqrt{ \log \frac{P(\| \bar F \|_{2,\PP}<c)}{E\left\{ I (\| \bar F \|_{2,\PP}<c) \| \bar F \|_{2,\PP} \right\}} }, \label{root-n-eq-41}
\end{eqnarray}
where we have combined (\ref{root-n-eq-39}) and (\ref{root-n-eq-40}) to derive from (\ref{root-n-eq-41-1}) to (\ref{root-n-eq-41-2}); we have applied Jensen's inequality to get from (\ref{root-n-eq-41-3}) to (\ref{root-n-eq-41}).
We need to derive some properties of $\| \bar F \|_{2,\PP}$ so that we can find an upper bound for (\ref{root-n-eq-41}). Recalling the definition of $\| \bar F \|_{2,\PP}$ given by (\ref{root-n-eq-38-added}), and the definition of $\widetilde f_{\theta, \bz_2}^{(2)} (\bz_1)$ given in (\ref{root-n-eq-38}), we have
\begin{eqnarray}
\| \bar F \|_{2,\PP} &=& \left\|\left\|\widetilde f_{\theta, \bz_2}^{(2)} (\cdot) \right\|_{\bz_2 \in \mathZ; \|\theta-\theta_0\|_2\leq \delta} \right\|_{2,\PP} \nonumber \\
&\leq & \left\|\widetilde f_{\theta, \bz_2}^{(2)} (\bz_1) \right\|_{\bz_1 \in \mathZ, \bz_2 \in \mathZ; \|\theta-\theta_0\|_2\leq \delta} \nonumber \\
&=& \left\|\widetilde f_{\theta}^{(2)} (\bz_1, \bz_2) \right\|_{\bz_1 \in \mathZ, \bz_2 \in \mathZ; \|\theta-\theta_0\|_2\leq \delta}. \label{root-n-eq-42}
\end{eqnarray}
Based on Chebyshev's inequality and (\ref{root-n-eq-37}), for any $0<\alpha<1/4$,
\begin{eqnarray}
P(c\leq \| \bar F \|_{2,\PP}<1) &\leq & P(\| \bar F \|_{2,\PP} \geq c) \nonumber\\
&\leq & P\left(\left\|\widetilde f_{\theta}^{(2)} (\bz_2, \bz_3) \right\|_{\bz_1 \in \mathZ, \bz_2 \in \mathZ; \|\theta-\theta_0\|_2\leq \delta} \geq c\right) \nonumber\\
&\leq& \frac{E\left\| \widetilde f_\theta^{(2)} (\bz_1, \bz_2) \right \|_{\|\theta-\theta_0\|_2\leq \delta; \bz_1 \in \mathZ; \bz_2\in \mathZ}}{c} \nonumber \\
&\lesssim& \delta^{\alpha}/c, \label{root-n-eq-43}
\end{eqnarray}
 based on which, we can also conclude that
\begin{eqnarray}
\mbox{there exists a small universal constant $\eta_1>0$, such that for any $\delta< \eta_1$} \nonumber \\
P(\| \bar F \|_{2,\PP} < c) = 1- P(\| \bar F \|_{2,\PP} \geq c) \geq 1/2. \label{root-n-eq-43-added}
\end{eqnarray}
Furthermore, based on  (\ref{root-n-eq-37}) and (\ref{root-n-eq-42}), we have
\begin{eqnarray}
E\left\{ I (\| \bar F \|_{2,\PP}<c) \| \bar F \|_{2,\PP} \right\} &\leq& E\left\{ \| \bar F \|_{2,\PP} \right\}
\nonumber\\
&\leq& E \left\{ \left\|\widetilde f_{\theta}^{(2)} (\bz_2, \bz_3) \right\|_{\bz_2 \in \mathZ, \bz_3 \in \mathZ; \|\theta-\theta_0\|_2\leq \delta} \right\} \nonumber \\
&\lesssim & \delta^\alpha. \label{root-n-eq-44}
\end{eqnarray}
Combining (\ref{root-n-eq-41}) with (\ref{root-n-eq-43})--(\ref{root-n-eq-44}) and noting that $x \sqrt{\log(1/x)}$ is strictly increasing when $0<x<c$, we conclude that there exists an $\eta_1>0$ satisfying (\ref{root-n-eq-43-added}), such that for any $0<\delta<\eta_1$
\begin{eqnarray}
E\left\| \GG_n^{(1)} \widetilde f_{\theta}^{(2)}(\cdot, \bz_2) \right\|_{\|\theta-\theta_0\|_2\leq \delta; \bz_2 \in \mathZ} \lesssim \sqrt{\log n} \delta^\alpha + \delta^\alpha \sqrt{-\log\delta}.  \label{root-n-eq-45}
\end{eqnarray}
Now combining (\ref{root-n-eq-33-added}) with (\ref{root-n-eq-45}), we have completed the proof for (\ref{root-n-eq-26-added-4}), and therefore (\ref{root-n-eq-26-added-1}) is valid.

It is left to show (\ref{root-n-eq-26-added-3}).
Recalling the definition of $\mathI_{3,1,3}$ in (\ref{root-n-eq-25}), we have
\begin{eqnarray}
\mathI_{3,1,3} &=& n^{3/2} \PP^2 \widetilde f_\theta(\cdot, \cdot)= n \sum_{k=1}^n \PP^2 f_\theta(\cdot, \cdot, Z_k) = n \sum_{k=1}^n f_{1, \theta}(Z_k) = n^{3/2} \GG_n f_{1, \theta}(\cdot), \label{root-n-eq-46}
\end{eqnarray}
where
\begin{eqnarray*}
f_{1, \theta}(\bz_3)&=& \PP^2 f_\theta(\cdot, \cdot, \bz_3)\\
&=& E\left[E\left\{f_\theta(Z_1, Z_2, Z_3)\Big|Z_3, X_1, Y_2\right\}\Big| Z_3 = \bz_3\right]\nonumber \\
&=& E\Bigg[ \Bigg\{  \frac{ {F}_0\Big(Y_2^{(\lambda_0)}-X_1^T{\beta_0}\Big) }{ {F}_\theta\Big(Y_2^{(\lambda)}-X_1^T{\beta}; \lambda,{\beta}\Big)} I\left(Y_3^{(\lambda)} - X_3^T \beta \leq Y_2^{(\lambda)}-X_1^T{\beta}\right) \nonumber \\
&&\hspace{0.4in} -  I\left(Y_3^{(\lambda_0)} - X_3^T \beta_0 \leq Y_2^{(\lambda_0)}-X_1^T{\beta_0}\right) \Bigg\} \bigg| Z_3 = \bz_3   \Bigg], \label{root-n-eq-47}
\end{eqnarray*}
with $Z_1, Z_2, Z_3$ being independent copies of $Z$;
here, we have used the fact that
\begin{eqnarray}
E(I(Y_1\leq Y_2)|Z_3, X_1, Y_2) &=& E(I(Y_1 \leq Y_2)|X_1, Y_2) \nonumber \\
&=& P(Y_1\leq Y_2|X_1, Y_2) = P(\epsilon_2^* \leq Y_2^{(\lambda_0)} - X_1 \beta_0|X_1, Y_2) \nonumber\\
&=& F_0(Y_2^{(\lambda_0)} - X_1 \beta_0). \label{root-n-eq-48}
\end{eqnarray}
We can further decompose
\begin{eqnarray}
f_{1, \theta}(\bz_3) = f_{1,1,\theta}(\bz_3) + f_{1,2,\theta}(\bz_3), \label{root-n-eq-49}
\end{eqnarray}
with
\begin{eqnarray*}
&&f_{1,1,\theta}(\bz_3)\\ &=&  E\left[ \left\{  \frac{ {F}_0\Big(Y_2^{(\lambda_0)}-X_1^T{\beta_0}\Big) }{ {F}_\theta\Big(Y_2^{(\lambda)}-X_1^T{\beta}\Big)} - 1 \right\} I\left(Y_3^{(\lambda)} - X_3^T \beta \leq Y_2^{(\lambda)}-X_1^T{\beta}\right) \bigg| Z_3 = \bz_3\right] \\
&&f_{1,2,\theta}(\bz_3) \\&=& E\left\{ I\left(Y_3^{(\lambda)} - X_3^T \beta \leq Y_2^{(\lambda)}-X_1^T{\beta}\right)  -  I\left(Y_3^{(\lambda_0)} - X_3^T \beta_0 \leq Y_2^{(\lambda_0)}-X_1^T{\beta_0}\right) \bigg| Z_3 = \bz_3 \right\}.  \label{root-n-eq-50}
\end{eqnarray*}
Based on Conditions 2 and 3, we can verify that for $r=1,2$,
\begin{eqnarray*}
\sup_{\|\theta-\theta_0\|_2\leq \delta} |f_{1,r,\theta}(\bz_1)| \lesssim \delta.
\end{eqnarray*}
By considering the function classes
\begin{eqnarray*}
\left\{f_{1,r,\theta}(\bz_1): \|\theta-\theta_0\|_2\leq \delta \right\},
\end{eqnarray*}
defined on $\mathZ$ with envelope function $C\delta$ for $C<\infty$ being a universal constant, we can conclude
\begin{eqnarray}
E\left\{\|\GG_n f_{1,r,\theta}\|_{\|\theta-\theta_0\|_2\leq \delta} \right\} \lesssim \delta, \label{root-n-eq-51}
\end{eqnarray}
by Lemma \ref{lemma-tail-bound-1} and the discussion given in Remark \ref{remark-2}.  Combining (\ref{root-n-eq-46}), (\ref{root-n-eq-49}), and (\ref{root-n-eq-51}), we immediately conclude (\ref{root-n-eq-26-added-3}). We have completed the proof for this lemma. \epf

\begin{lemma} \label{lemma-bracketing-F-n-1}
Let $\widetilde \bz_1, \ldots, \widetilde \bz_n \in \mathZ$ be arbitrary, where $\widetilde \bz_k = (\widetilde y_k, \widetilde \bx_k)$, for $k=1,\ldots, n$. Consider $\widetilde f_\theta(\bz_1, \bz_2)$ defined by (\ref{root-n-eq-26}) but evaluated according to $Z_1 = \widetilde \bz_1, \ldots, Z_n = \widetilde \bz_n$. That is
\begin{eqnarray*}
\widetilde f_\theta(\bz_1, \bz_2) = \frac{1}{\sqrt{n}}\sum_{k=1}^n f_\theta(\bz_1, \bz_2, \widetilde \bz_k),
\end{eqnarray*}
where $f_\theta$ is defined by (\ref{root-n-eq-24}).
Consider the function class
\begin{eqnarray*}
\mathF_n(\widetilde \bz_1, \ldots, \widetilde \bz_n) = \left\{ \widetilde f_{\theta, \bz_2} (\bz_1)\equiv \widetilde f_{\theta} (\bz_1, \bz_2): \bz_2 \in \mathZ, \|\theta-\theta_0\|_2\leq \eta_0 \right\},
\end{eqnarray*}
defined on $\mathZ$, where $\eta_0$ is given in Condition 2. Assume Conditions 1--3, we have
\begin{eqnarray*}
N_{[]}\left(\epsilon, \mathF_n\left(\widetilde \bz_1, \ldots, \widetilde \bz_n \right), L_2(\PP)\right) \lesssim \frac{n^{p+2}}{\epsilon^{2(p+2)}},
\end{eqnarray*}
up to a constant not depending on the values of $\widetilde \bz_k, k=1,\ldots,n$.
\end{lemma}

\proof Let $\Theta_0 = \{\theta: \|\theta-\theta_0\|_2\leq \eta_0\}$.  Based on Condition 1 that  $\mathY$ is compact, for every $\epsilon>0$, there exist $(\theta_1, y_{2,1}), \ldots, (\theta_{N_1}, y_{2, N_1}) \in \Theta_0 \times \mathY$ with $N_1 \lesssim \frac{1}{\epsilon^{p + 2}}$; for any $(\theta, y_2) \in \Theta_0 \times \mathY$, there exists $s\in \{1,\ldots, N_1\}$, such that $\|(\theta, y_2) - (\theta_s, y_{2, s})\|_2\leq \epsilon$. Based on Condition 2, there exists a universal constant $C$, such that for any $(\theta, y_2), (\bar \theta, \bar y_2) \in \Theta_0\times \mathY$, $\bx_1\in \mathX$, and $k=1,\ldots,n$,
\begin{eqnarray*}
&&\left|\frac{1}{F_{\theta}\left(y_2^{(\lambda)}- \bx_1^T\beta\right)}- \frac{1}{F_{\bar \theta}\left(\bar y_2^{(\bar \lambda)}- \bx_1^T\bar \beta\right)}\right| \\
&\leq& C \left\|(\theta, y_2) - (\bar \theta, \bar y_2)\right\|_2\\
&&\left|\widetilde y_k^{(\lambda)} - \widetilde \bx_k^T\beta - \left\{ y_2^{(\lambda)} - \bx_1^T\beta\right\} - \left[\widetilde y_k^{(\bar\lambda)} - \widetilde \bx_k^T\beta - \left\{ \bar y_2^{(\bar \lambda)} - \bx_1^T \bar\beta\right\}\right] \right| \\
&\leq& C \left\|(\theta, y_2) - (\bar \theta, \bar y_2)\right\|_2.
\end{eqnarray*}
Therefore, it is straightforward to check that for any $k=1,\ldots, n$, the set of brackets
\begin{eqnarray*}
\left\{\left[l_{k,s}(\bz_1), u_{k,s}(\bz_1)\right]: s=1,\ldots, N_1 \right\}
\end{eqnarray*}
covers the function class
\begin{eqnarray*}
\left\{f_{\theta}(\bz_1, \bz_2, \widetilde \bz_k): \bz_2 \in \mathZ, \theta \in \Theta_0 \right\}
\end{eqnarray*}
defined on $\bz_1 \in \mathZ$, where
\begin{eqnarray*}
u_{k,s}(\bz_1) &=& I(y_1\leq  y_{2,s} + \epsilon)\left[I_{\theta_s, -\epsilon}(\bz_1, y_{2,s}, \widetilde \bz_k)\left\{\frac{1}{F_{\theta_s}\left(y_{2,s}^{(\lambda_s)} - \bx_1^T \beta_s \right)} + C\epsilon\right\} \right.\\
&&\hspace{1in} \left.- I_{0, +\epsilon}(\bz_1, y_{2,s}, \widetilde \bz_k)\left\{\frac{1}{F_0\left(y_{2,s}^{(\lambda_0)} - \bx_1^T \beta_0 \right)} - C\epsilon \right\} \right]\\
l_{k,s}(\bz_1) &=& I(y_1\leq  y_{2,s} - \epsilon)\left[I_{\theta_s, +\epsilon}(\bz_1, y_{2,s}, \widetilde \bz_k)\left\{\frac{1}{F_{\theta_s}\left(y_{2,s}^{(\lambda_s)} - \bx_1^T \beta_s \right)} - C\epsilon\right\} \right.\\
&&\hspace{1in} \left.- I_{0, -\epsilon}(\bz_1, y_{2,s}, \widetilde \bz_k)\left\{\frac{1}{F_0\left(y_{2,s}^{(\lambda_0)} - \bx_1^T \beta_0 \right)} + C\epsilon \right\} \right]\\
I_{\theta_s, \widetilde \epsilon}(\bz_1, y_{2,s}, \widetilde \bz_k) &=& I\left\{ \widetilde y_k^{(\lambda_s)} - \widetilde \bx_k^T\beta_s - \left\{y_{2,s}^{(\lambda_s)} - \bx_1^T\beta_{s}\right\} + \widetilde \epsilon C \leq 0 \right\}\\
I_{0, \widetilde \epsilon} (\bz_1, y_{2,s}, \widetilde \bz_k) &=& I_{\theta_0, \widetilde \epsilon}(\bz_1, y_{2,s}, \widetilde \bz_k),
\end{eqnarray*}
with $\widetilde \epsilon = \epsilon$ or $-\epsilon$. Based on Conditions 2 and 3, we can check that the bracket length is given by
\begin{eqnarray*}
\|u_{k,s}(\bz_1) - l_{k,s}(\bz_1)\|_{2,\PP}\lesssim \sqrt{\epsilon},
\end{eqnarray*}
up to a constant not depending on the values of $\widetilde \bz_k, k=1,\ldots,n$.
Furthermore, the set of brackets
\begin{eqnarray*}
\left\{ \left[\frac{1}{\sqrt{n}}\sum_{k=1}^n l_{k,s}(\bz_1), \frac{1}{\sqrt{n}}\sum_{k=1}^n u_{k,s}(\bz_1)\right]: s=1,\ldots, N_1  \right\}
\end{eqnarray*}
covers $\mathF_n\left(\widetilde \bz_1, \ldots, \widetilde \bz_n\right)$, with bracket length
\begin{eqnarray*}
&&\left\|\frac{1}{\sqrt{n}}\sum_{k=1}^n u_{k,s}(\bz_1) -  \frac{1}{\sqrt{n}}\sum_{k=1}^n l_{k,s}(\bz_1) \right\|_{2,\PP}\\
&\lesssim& \frac{1}{\sqrt{n}} \sum_{k=1}^n \| u_{k,s}(\bz_1) - l_{k,s}(\bz_1)\| \lesssim \sqrt{n} \sqrt{\epsilon}.
\end{eqnarray*}
This indicates
\begin{eqnarray*}
N_{[]}(\sqrt{n\epsilon}, \mathF_n\left(\widetilde \bz_1, \ldots, \widetilde \bz_n\right), L_2(\PP)) \lesssim 1/\epsilon^{p+2},
\end{eqnarray*}
up to a constant not depending on the values of $\widetilde \bz_k, k=1,\ldots,n$. This completes the proof of the lemma. \epf

\subsection{Asymptotic Normality} \label{section-normal}

In this section, we establish the asymptotic normality of $\widehat \theta$. In particular, we aim to show that
\begin{eqnarray}
\sqrt{n}(\widehat \theta - \theta_0) \rightsquigarrow N(0, \Sigma), \label{norm-main-eq-1}
\end{eqnarray}
where $\Sigma = \frac{1}{4} \Sigma_1^{-1} \Sigma_2 \Sigma_1^{-1}$ with $\Sigma_1$ and $\Sigma_2$ defined by (\ref{def-Sigma-1}) and (\ref{def-Sigma-2}) respectively.

We need the following Lemma \ref{argmax}, which is the argmax theorem adapted from Theorem 14.1 in Kosorok (2008); see also Theorem 3.2.2 in VW.

\begin{lemma} \label{argmax}

Let $\WW_n$, $\WW$ be stochastic processes indexed by a metric space $\mathH$, such that $\WW_n \rightsquigarrow \WW$ in $L^{\infty}(H)$ for every compact $H \subset \mathH$. Suppose also that almost all sample paths $h \mapsto M(h)$ are upper semicontinuous and possess a unique maximum at a (random) point $\widehat h$, which as a random map in $\mathH$ is tight. If the sequence $\widehat h_n$ is uniformly tight and satisfies $\WW_n(\widehat h_n) \geq \sup_{h\in H} \WW_n(h) - o_p(1)$, then $\widehat h_n \rightsquigarrow \widehat h$ in $\mathH$.

\end{lemma}

We shall apply the argmax theorem above to show (\ref{norm-main-eq-1}). Denote $\widehat h_n = \sqrt{n}(\widehat \theta - \theta_0)$, and
let $h = (h_1, h_2^T)^T$, $\theta_{n,h} = \theta_0 + h/\sqrt{n}$, $\lambda_{n,h} = \lambda_0 + h_1/\sqrt{n}$, $\beta_{n,h} = \beta_0 + h_2/\sqrt{n} $. Define
\begin{eqnarray*}
\WW_n(h) = \frac{1}{n}\left\{\ell(\theta_{n, h}) - \ell(\theta_0)\right\}. \label{norm-main-eq-2}
\end{eqnarray*}
Clearly, $\widehat h_n$ is the maximizer of $\WW_n(h)$, and therefore $\WW_n(\widehat h_n) \geq \sup_{h\in \RR^{p+1}} \WW_n(h)$. In Section \ref{proof-rate}, we have shown that $\widehat h_n$ is uniformly tight.


For $H$ being an arbitrary compact subset of $\RR^{p+1}$, consider the process
\begin{eqnarray}
\WW_n(h) = \frac{1}{n}\left\{\ell(\theta_{h, n}) - \ell(\theta_0)\right\} = \WW_{n,1}(h) + \WW_{n,2}(h), \label{norm-main-eq-3}
\end{eqnarray}
with $h\in H$, where
\begin{eqnarray*}
\WW_{n,1}(h) &=& \frac{1}{n}\left[\ell(\theta_{n, h}) - \ell(\theta_0) - \left\{\widetilde \ell(\theta_{n, h}) - \widetilde \ell(\theta_0)\right\}\right]\\
\WW_{n,2}(h) &=& \frac 1n \left\{\widetilde \ell(\theta_{n, h}) - \widetilde \ell(\theta_0)\right\}. \label{norm-main-eq-4}
\end{eqnarray*}
We consider $\WW_{n,1}(h)$ and $\WW_{n,2}(h)$ separately. For $\WW_{n,2}(h)$, we have derived in Lemma \ref{lemma-norm-0} that
\begin{eqnarray}
\left\|\WW_{n,2}(h) - \left(h^T\GG_n \varphi - h^T \Sigma_1 h\right) \right\|_{h\in H} =  o_p(1), \label{norm-main-eq-5}
\end{eqnarray}
where $\varphi(\cdot)$ is defined by (\ref{def-varphi}) and derived by (\ref{lemma-norm-0-eq-6-added}), and $\Sigma_1$ is defined by (\ref{def-Sigma-1}) and derived by  (\ref{lemma-norm-0-eq-14}). For $\WW_{n,1}(h)$, we have
\begin{eqnarray}
\WW_{n,1}(h) &=& \frac{1}{n}\left[\ell(\theta_{h, n}) - \ell(\theta_0) - \left\{\widetilde \ell(\theta_{h, n}) - \widetilde \ell(\theta_0)\right\}\right]\nonumber\\
&=& \frac 1n \sum_{j=1}^n\sum_{i=1}^n  I_{i,j}\log\left\{\frac{\widehat F_{\theta_{n,h}}(V_{\theta_{n,h}, j, i}) F_{0}(V_{0,j,i}) }{ \widehat F_{0}(V_{0,j,i}) F_{\theta_{n,h}}(V_{\theta_{n,h}, j, i})}\right\}
\nonumber\\
&&+\frac 1n \sum_{j=1}^n\sum_{i=1}^n (1-I_{i,j})\log\left\{\frac{\left(1-\widehat F_{\theta_{n,h}}(V_{\theta_{n,h}, j, i})\right)(1-F_{0}(V_{0,j,i}))}{ \left(1-\widehat F_{0}(V_{0,j,i})\right) (1-F_{\theta_{n,h}}(V_{\theta_{n,h}, j, i}))}\right\}\nonumber \\
&=& \mathI_5 + \mathI_6. \label{norm-main-eq-6}
\end{eqnarray}
Consider $\mathI_5$. By Taylor expansion for $\log x$ at $x=1$, we have
\begin{eqnarray*}
\mathI_5 &=& \frac 1n \sum_{j=1}^n\sum_{i=1}^n  I_{i,j} \left\{\frac{\widehat F_{\theta_{n,h}}(V_{\theta_{n,h}, j, i}) F_{0}(V_{0,j,i}) }{ \widehat F_{0}(V_{0,j,i}) F_{\theta_{n,h}}(V_{\theta_{n,h}, j, i})} - 1\right\}  \\&& - \frac 1n \sum_{j=1}^n\sum_{i=1}^n  I_{i,j} \frac{1}{2\xi_{n,h,i,j}}\left\{\frac{\widehat F_{\theta_{n,h}}(V_{\theta_{n,h}, j, i}) F_{0}(V_{0,j,i}) }{ \widehat F_{0}(V_{0,j,i}) F_{\theta_{n,h}}(V_{\theta_{n,h}, j, i})} - 1\right\}^2, \label{norm-main-eq-7}
\end{eqnarray*}
where $\xi_{n,h,i,j}$ is in between $\frac{\widehat F_{\theta_{n,h}}(V_{\theta_{n,h}, j, i}) F_{0}(V_{0,j,i}) }{ \widehat F_{0}(V_{0,j,i}) F_{\theta_{n,h}}(V_{\theta_{n,h}, j, i})}$ and 1. Based on Lemma \ref{lemma-1} and Condition 2, when $n$ is sufficiently large, we have
\begin{eqnarray*}
\sup_{1\leq i,j\leq n; h\in H}|\xi_{n,h,i,j} - 1 | \leq \sup_{1\leq i,j\leq n; h\in H} \left|\frac{\widehat F_{\theta_{n,h}}(V_{\theta_{n,h}, j, i}) F_{0}(V_{0,j,i}) }{ \widehat F_{0}(V_{0,j,i}) F_{\theta_{n,h}}(V_{\theta_{n,h}, j, i})}-1 \right| \to 0 \quad \mbox{in probability}, \label{norm-main-eq-8}
\end{eqnarray*}
which implies that
\begin{eqnarray*}
\sup_{1\leq i,j\leq n; h\in H} \frac{1}{\xi_{n,h,i,j}}  = \frac{1}{1-o_p^*(1)}, \label{norm-main-eq-9}
\end{eqnarray*}
where $o_p^*(1)$ is uniform in $1\leq i,j\leq n$ and $h\in H$.
Therefore
\begin{eqnarray*}
&&\left|\mathI_5 - \frac 1n \sum_{j=1}^n\sum_{i=1}^n  I_{i,j} \left\{\frac{\widehat F_{\theta_{n,h}}(V_{\theta_{n,h}, j, i}) F_{0}(V_{0,j,i}) }{ \widehat F_{0}(V_{0,j,i}) F_{\theta_{n,h}}(V_{\theta_{n,h}, j, i})} - 1\right\} \right| \\ &\lesssim& \frac{n}{1-o_p^*(1)} \sup_{\bz\in \mathZ,h\in H} \left| \frac{\widehat F_{\theta_{n,h}}(\bv_{\theta_{n,h}}) F_0(\bv_{\theta_0})}{\widehat F_0(\bv_{\theta_0}) F_{\theta_{n,h}}(\bv_{\theta_{n,h}})} -1 \right|^2,  \label{norm-main-eq-10}
\end{eqnarray*}
which together with Lemmas \ref{lemma-norm-1} and \ref{lemma-norm-2} concludes
\begin{eqnarray}
\sup_{h \in H}\left|\mathI_5 -\sqrt{n} \GG_n \left\{  f_{1,n,h}(\cdot) \right\}\right| = o_p(1), \label{norm-main-eq-11}
\end{eqnarray}
where $f_{1,n,h}(\cdot)$ defined in (\ref{lemma-norm-2-eq-6-added}) is given by
\begin{eqnarray*}
f_{1,n,h}(\bz) = E \left\{ \frac{F_0(V_{0, 2, 1})}{F_{\theta_{n,h}}(V_{\theta_{n,h}, 2,1})} I\left(\bv_{\theta_{n,h}} \leq V_{\theta_{n,h}, 2,1}\right) - I\left(\bv_{0} \leq V_{0, 2,1}\right)\right\}.  \label{norm-main-eq-12}
\end{eqnarray*}

Using exactly the same derivation, we can verify
\begin{eqnarray}
\sup_{h \in H}\left|\mathI_6 -\sqrt{n} \GG_n \left\{  f_{2,n,h}(\cdot) \right\}\right| = o_p(1), \label{norm-main-eq-13}
\end{eqnarray}
with
\begin{eqnarray*}
f_{2,n,h}(\bz) = E \left[ \frac{1-F_0(V_{0, 2, 1})}{1-F_{\theta_{n,h}}(V_{\theta_{n,h}, 2,1})} \left\{1-I\left(\bv_{\theta_{n,h}} \leq V_{\theta_{n,h}, 2,1}\right)\right\} - \left\{1-I\left(\bv_{0} \leq V_{0, 2,1}\right)\right\}\right] \label{norm-main-eq-14}
\end{eqnarray*}

Combining (\ref{norm-main-eq-6}), (\ref{norm-main-eq-11}), and (\ref{norm-main-eq-13}) we have
\begin{eqnarray}
\sup_{h\in H}\left|\WW_{n,1}(h) - \sqrt{n} \GG_n \left\{  f_{1,n,h}(\cdot) +  f_{2,n,h}(\cdot) \right\}\right| = o_p(1). \label{norm-main-eq-15}
\end{eqnarray}
Furthermore, noting that for any constant $C$, $\GG_n C = 0$, we have
\begin{eqnarray}
\GG_n \left\{  f_{1,n,h}(\cdot) +  f_{2,n,h}(\cdot) \right\} = \GG_n \psi_{n,h}(\cdot), \label{norm-main-eq-16}
\end{eqnarray}
where
\begin{eqnarray*}
\psi_{n,h}(\bz) &=& E \left[ \left\{\frac{F_0(V_{0, 2, 1})}{F_{\theta_{n,h}}(V_{\theta_{n,h}, 2,1})} -\frac{1-F_0(V_{0, 2, 1})}{1-F_{\theta_{n,h}}(V_{\theta_{n,h}, 2,1})} \right\}I\left(\bv_{\theta_{n,h}} \leq V_{\theta_{n,h}, 2,1}\right) \right]\\
&=& E\left[\frac{F_0(V_{0, 2, 1}) - F_{\theta_{n,h}}(V_{\theta_{n,h}, 2,1})}{F_{\theta_{n,h}}(V_{\theta_{n,h}, 2,1})\left\{1-F_{\theta_{n,h}}(V_{\theta_{n,h}, 2,1}) \right\}}I\left(\bv_{\theta_{n,h}} \leq V_{\theta_{n,h}, 2,1}\right)\right].
\end{eqnarray*}
Then based on Lemma \ref{lemma-norm-3}, we have
\begin{eqnarray}
E\left\|\sqrt{n}\GG_n \psi_{n,h}(\bz) - h^T \GG_n \psi(\bz)\right\|_{h\in H} =  o(1), \label{norm-main-eq-17}
\end{eqnarray}
where
\begin{eqnarray*}
\psi(\bz) &=& -E\left[\frac{\dot{F}_0(V_{0,2,1}) + F_0'(V_{0,2,1}) \dot{V}_{0,2,1}}{F_{0}(V_{0, 2,1})\left\{1-F_{0}(V_{0, 2,1}) \right\}}I\left(\bv_{0} \leq V_{0, 2,1}\right)\right],
\end{eqnarray*}
as defined by (\ref{def-psi}).
Combining (\ref{norm-main-eq-15}), (\ref{norm-main-eq-16}), and (\ref{norm-main-eq-17}) leads to
\begin{eqnarray*}
\sup_{h\in H}\left|\WW_{n,1}(h) - h^T \GG_n \psi(\bz)\right| = o_p(1), \label{norm-main-eq-18}
\end{eqnarray*}
which combined with (\ref{norm-main-eq-3}) and (\ref{norm-main-eq-5}) concludes
\begin{eqnarray*}
\sup_{h\in H} \left| \WW_n(h)  - h^T\GG_n (\varphi+\psi) + h^T \Sigma_1 h \right| =o_p(1).
\end{eqnarray*}
Furthermore, based on Central Limit Theorem, and Condition 5 that $\Sigma_2$ is invertible, we have
\begin{eqnarray*}
\GG_n (\varphi+\psi) \rightsquigarrow N(0, \Sigma_2), \label{norm-main-eq-19}
\end{eqnarray*}
where $\Sigma_2$ is given by (\ref{def-Sigma-2}).
Now define $\WW(h) = h^T \mathcal{N} - h^T \Sigma_1 h$ where $\mathcal{N}$ is a random vector following the $N(0, \Sigma_2)$ distribution; then $\WW(h)$ has a unique maximum at $\widehat h = 0.5\Sigma_1^{-1} \mathcal{N}$ based on Condition 5 that $\Sigma_1$ is invertible. Combining (\ref{norm-main-eq-18}) and (\ref{norm-main-eq-19}), we have $\WW_n(h) \rightsquigarrow \WW(h)$, which indicates that $\WW(h)$ plays the role of ``$\WW(h)$" in Lemma \ref{argmax}. This immediately leads to (\ref{norm-main-eq-1}) by an application of Lemma \ref{argmax}. Our proof is completed.

\begin{lemma} \label{lemma-norm-0}
Assume Conditions 1 and 2. We have
\begin{eqnarray*}
 \left\|\frac 1n \left\{\widetilde \ell(\theta_{n,h}) - \widetilde \ell(\theta_0)\right\} - \left(h^T\GG_n \varphi - h^T \Sigma_1 h\right) \right\|_{h\in H} =  o_p(1), \label{lemma-norm-0-eq-1}
\end{eqnarray*}
where $\varphi(\cdot)$ is defined by (\ref{def-varphi}) and $\Sigma_1$ is defined by (\ref{def-Sigma-1}).

\end{lemma}

\proof Based on (\ref{prelim-U-eq-7}), we have
\begin{eqnarray}
&&\frac 1n \left\{\widetilde \ell(\theta_{n,h}) - \widetilde \ell(\theta_0)\right\} \nonumber \\ &=& \frac{1}{n}\sum_{j=1}^n\sum_{i=1}^n \left[ I_{i,j}\log \left\{ \frac{F_{\theta_{n,h}}(V_{\theta_{n,h}, j,i})}{F_{0}(V_{0, j,i})}\right\}+(1-I_{i,j})\log\left\{\frac{1-F_{\theta_{n,h}}(V_{\theta_{n,h}, j,i})}{1-F_{0}(V_{0, j,i})}\right\}\right] \nonumber \\
&=& \frac{1}{n\sqrt{n}} \sum_{j=1}^n \sum_{i=1}^n m_{n,h}(Z_i, Z_j) \nonumber \\
&=& \frac{n-1}{\sqrt{n}} \UU_n^2 m_{n,h} + \frac{1}{n\sqrt{n}} \sum_{i=1}^n m_{n,h}(Z_i, Z_i) \nonumber\\
&=& \frac{n-1}{\sqrt{n}} \left( \PP^2 m_{n,h} + \PP_n m_{n,h,1} + \UU_n^2 m_{n,h,2}\right) + \frac{1}{n\sqrt{n}} \sum_{i=1}^n m_{n,h}(Z_i, Z_i), \label{lemma-norm-0-eq-2}
\end{eqnarray}
where $m_{n,h,1}$ and $m_{n,h,2}$ are  the decomposed functions in (\ref{prelim-U-eq-7}) based on $m_{n,h}$; $m_{n,h}$ is given by
\begin{eqnarray*}
&&m_{n,h}(\bz_1, \bz_2) \\ &=& \sqrt{n}I(y_1 \leq y_2) \log \left\{ \frac{F_{\theta_{n,h}}(\bv_{\theta_{n,h}, 2,1})}{F_{0}(\bv_{0, 2,1})}\right\}+\sqrt{n}I(y_1 > y_2) \log\left\{\frac{1-F_{\theta_{n,h}}(\bv_{\theta_{n,h}, 2,1})}{1-F_{0}(\bv_{0, 2,1})}\right\}. \label{lemma-norm-0-eq-3}
\end{eqnarray*}

We first derive $\frac{\partial m_{n,h}(\bz_1, \bz_2)}{\partial h}\Big|_{h = 0}$, $E\left\{\frac{\partial m_{n,h}(Z_1, Z_2)}{\partial h}\Big|_{h = 0}\bigg| Z_2\right\}$ and $E\left\{\frac{\partial^2 m_{n,h}(Z_1, Z_2)}{\partial h \partial h^T}\Big|_{h=0} \right\}$; and then derive the asymptotic properties for each term on the far right of (\ref{lemma-norm-0-eq-2}) separately. Consider
\begin{eqnarray}
&&\frac{\partial m_{n,h}(\bz_1, \bz_2)}{\partial h}\nonumber\\ &=& \sqrt{n}I(y_1 \leq y_2) \frac{ \frac{\partial F_{\theta_{n,h}}(\bv_{\theta_{n,h}, 2,1})}{\partial h}}{F_{\theta_{n,h}}(\bv_{\theta_{n,h}, 2,1})}+\sqrt{n}I(y_1 > y_2) \frac{- \frac{\partial F_{\theta_{n,h}}(\bv_{\theta_{n,h}, 2,1})}{\partial h}}{1-F_{\theta_{n,h}}(\bv_{\theta_{n,h}, 2,1})}\nonumber\\
&=&   \left\{\frac{I(y_1 \leq y_2)}{F_{\theta_{n,h}}(\bv_{\theta_{n,h}, 2,1})}-\frac{I(y_1 > y_2)}{1-F_{\theta_{n,h}}(\bv_{\theta_{n,h}, 2,1})} \right\} \left\{\dot F_{\theta_{n,h}}(\bv_{\theta_{n,h}, 2,1}) + F_{\theta_{n,h}}'(\bv_{\theta_{n,h}, 2,1}) \dot{\bv}_{\theta_{n,h}, 2,1} \right\}, \nonumber \\ \label{lemma-norm-0-eq-11}
\end{eqnarray}
where $\dot{\bv}_{\theta_{n,h}, 2,1}$ is given by  (\ref{def-dot-v}). Setting $h=0$ in (\ref{lemma-norm-0-eq-11}) leads to
\begin{eqnarray}
\frac{\partial m_{n,h}(\bz_1, \bz_2)}{\partial h}\Big|_{h = 0} = \left\{\frac{I(y_1 \leq y_2)}{F_0(\bv_{0, 2,1})}-\frac{I(y_1 > y_2)}{1-F_{0}(\bv_{0, 2,1})} \right\} \left\{\dot{F}_0(\bv_{0,2,1}) + F_0'(\bv_{0,2,1}) \dot{\bv}_{0, 2,1} \right\}. \label{lemma-norm-0-eq-12}
\end{eqnarray}
We observe that $\dot{F}_0(\bv_{0,2,1}) + F_0'(\bv_{0,2,1}) \dot{\bv}_{0, 2,1} $, appeared on the right hand side of (\ref{lemma-norm-0-eq-12}), depends only on $y_2$ and $\bx_1$; furthermore, by noting (\ref{root-n-eq-48}), we observe that conditioning on $Y_2, X_1$, the expectation of the expression $\frac{I(y_1 \leq y_2)}{F_0(\bv_{0, 2,1})}-\frac{I(y_1 > y_2)}{1-F_{0}(\bv_{0, 2,1})}$  on the right hand side of (\ref{lemma-norm-0-eq-12}) by replacing $\bz_1, \bz_2$ with $Z_1, Z_2$, is zero. As a consequence,
\begin{eqnarray}
E\left\{\frac{\partial m_{n,h}(Z_1, Z_2)}{\partial h}\Big|_{h = 0}\bigg| Z_2\right\} = 0. \label{lemma-norm-0-eq-13}
\end{eqnarray}
With straightforward computations and similar arguments for deriving (\ref{lemma-norm-0-eq-13}), we can also establish
\begin{eqnarray}
&&E\left\{\frac{\partial^2 m_{n,h}(Z_1, Z_2)}{\partial h \partial h^T}\Big|_{h=0} \right\} \nonumber \\ &=& - E \left(\left[\frac{\left\{\dot{F}_0(V_{0,2,1}) + F_0'(V_{0,2,1}) \dot{V}_{0, 2,1} \right\} \left\{\dot{F}_0(V_{0,2,1}) + F_0'(V_{0,2,1}) \dot{V}_{0, 2,1} \right\}^T }{\sqrt{n} F_0(V_{0, 2,1})\left\{1-F_{0}(V_{0, 2,1})\right\}} \right]  \right) \nonumber \\
&=& - \frac{1}{\sqrt{n}} \Sigma_1,  \label{lemma-norm-0-eq-14}
\end{eqnarray}
by referring to the definition of $\Sigma_1$ given in (\ref{def-Sigma-1}).

We now derive the asymptotic properties for each term on the far right of (\ref{lemma-norm-0-eq-2}) separately.
Based on Condition 2 and referring to the discussion in Remark \ref{remark-1}, it is straightforward to check that there exists a universal constant $C< \infty$, such that the function class
$\{ m_{n,h}: h \in H \}$
defined on $\mathZ^2$ is Euclidean with envelope function equal to $C$, where the universal constants ``$A$" and ``$V$" in Definition \ref{def-Euclidean} do not rely on $n$. Applying Lemmas \ref{lemma-U-stat-1} and \ref{lemma-U-stat-2}, we have
\begin{eqnarray}
\frac{n-1}{\sqrt{n}}\left|\UU_n^2 m_{n,h,2}\right\|_{h \in H} = O_p(n^{-1/2}). \label{lemma-norm-0-eq-4}
\end{eqnarray}
Referring to Remark \ref{remark-2} and applying Lemma \ref{lemma-tail-bound-1}, we can also conclude
\begin{eqnarray}
\frac{1}{n\sqrt{n}} \left\|\sum_{i=1}^n m_{n,h}(Z_i, Z_i)\right\|_{h\in H} = O_p(n^{-1/2}). \label{lemma-norm-0-eq-5}
\end{eqnarray}
Based on (\ref{prelim-U-eq-9}),
\begin{eqnarray*}
m_{n,h,1}(\bz) = \PP m_{n,h}(\cdot, \bz) + \PP m_{n,h}(\bz, \cdot) - 2\PP^2 m_{n, h}. \label{lemma-norm-0-eq-6}
\end{eqnarray*}
Set $\dot{m}_{n, 0, 1}(\bz) = \frac{\partial m_{n,h,1}(\bz)}{\partial h}\Big|_{h=0}$; then based on Condition 2 and referring to (\ref{lemma-norm-0-eq-12}) and (\ref{lemma-norm-0-eq-13}), we have
\begin{eqnarray}
&&\dot{m}_{n, 0, 1}(\bz)\nonumber \\&=& \frac{\partial m_{n,h,1}(\bz)}{\partial h}\Big|_{h=0} \nonumber\\
&=& E\left\{\frac{\partial m_{n,h}(Z_1, Z_2)}{\partial h}\Big|_{h = 0}\bigg| Z_2 = \bz\right\} + E\left\{\frac{\partial m_{n,h}(Z_1, Z_2)}{\partial h}\Big|_{h = 0}\bigg| Z_1 = \bz\right\} \nonumber \\
&& -  2 E\left\{\frac{\partial m_{n,h}(Z_1, Z_2)}{\partial h}\Big|_{h = 0}\right\} \nonumber  \\
&=& E\left\{\frac{\partial m_{n,h}(Z_1, Z_2)}{\partial h}\Big|_{h = 0}\bigg| Z_1 = \bz\right\} \nonumber \\
&=& E\left[\left\{\frac{I(Y_1 \leq Y_2)}{F_0(V_{0, 2,1})}-\frac{I(Y_1 > Y_2)}{1-F_{0}(V_{0, 2,1})} \right\} \left\{\dot{F}_0(V_{0,2,1}) + F_0'(V_{0,2,1}) \dot{V}_{0, 2,1} \right\} \bigg| Z_1 = \bz\right]. \label{lemma-norm-0-eq-6-added}
\end{eqnarray}
By comparing (\ref{lemma-norm-0-eq-6-added}) with (\ref{def-varphi}), we observe $\varphi(\bz) = \dot{m}_{n, 0, 1}(\bz)$.
Since $\PP m_{n,h,1} = 0$, we have
\begin{eqnarray}
\sqrt{n}\PP_n m_{n,h,1} = \GG_n \left(m_{n,h,1} -h^T \dot{m}_{n, 0, 1}\right) + h^T\GG_n \dot{m}_{n, 0, 1}, \label{lemma-norm-0-eq-7}
\end{eqnarray}
where $\dot{m}_{n, 0, 1}(\bz)$ is given by (\ref{lemma-norm-0-eq-6-added}). Based on Condition 2, it is straightforward to verify that when $n$ is sufficiently large, every function in the function class
\begin{eqnarray*}
\mathM_{n,1} = \left\{m_{n,h,1} - h^T\dot{m}_{n, 0, 1}: h \in H \right\}, \label{lemma-norm-0-eq-8}
\end{eqnarray*}
satisfies (\ref{lemma-bracketing-eq-1}) with ``$\widetilde F(\bz) = C$", where $C$ is a universal constant; therefore applying Lemma \ref{lemma-bracketing} and noting that $H$ is compact subset of $\RR^{p+1}$, we have
\begin{eqnarray*}
N_{[]}(\epsilon, \mathM_{n,1}, L_2(\PP)) \lesssim 1/\epsilon^{p+1}.
\end{eqnarray*}
Furthermore, based on Condition 2 and
the fact $m_{n,0,1} = 0$, we have as  $n\to\infty$,
\begin{eqnarray*}
\alpha_n &\equiv& \sup_{h\in H, \bz\in \mathZ}|m_{n,h,1}(\bz) - h^T\dot{m}_{n, 0, 1}(\bz)| \\ &=& \sup_{h\in H, \bz\in \mathZ}|m_{n,h,1}(\bz) - m_{n,0,1}(\bz) - h^T\dot{m}_{n, 0, 1}(\bz)| \to 0.
\end{eqnarray*}
Clearly $\alpha_n$ can serve as an envelope function for $\mathM_{n,1}$.  We have
\begin{eqnarray*}
J_{[]}(1, \mathM_{n,1}) &=& \int_0^1 \sqrt{1+\log N_{[]}(\epsilon\cdot\|\alpha_n\|_{2,\PP}, \mathM_{n,1}, L_2(\PP))}\\
&\lesssim & \int_0^1 \sqrt{1+ \log(1/(\epsilon\alpha_n)^{p+1}} d\epsilon \lesssim \sqrt{-\log \alpha_n}.
\end{eqnarray*}
Applying Lemma \ref{lemma-tail-bound-1}, we have,
\begin{eqnarray}
E\left\{\sup_{h\in H} |\GG_n \left(m_{n,h,1} - h^T\dot{m}_{n, 0, 1}\right)|\right\} \lesssim \alpha_n \sqrt{-\log\alpha_n}\to 0, \label{lemma-norm-0-eq-10}
\end{eqnarray}
as $n\to \infty$.
Finally, by Condition 2, when $n$ is sufficiently large, $m_{n,h}$ is second order continuously differentiable in $h\in H$, and noting that $m_{n,0} = 0$, (\ref{lemma-norm-0-eq-13}), (\ref{lemma-norm-0-eq-14}), we have
\begin{eqnarray}
\PP^2 m_{n,h} = - \frac{1}{\sqrt{n}} h^T \Sigma_1 h  + o(n^{-1/2}), \label{lemma-norm-0-eq-15}
\end{eqnarray}
by Taylor's expansion,
where $o(\cdot)$ is uniform in $h\in H$.

Now combining (\ref{lemma-norm-0-eq-2}), (\ref{lemma-norm-0-eq-4}), (\ref{lemma-norm-0-eq-5}), (\ref{lemma-norm-0-eq-7}), (\ref{lemma-norm-0-eq-10}), and (\ref{lemma-norm-0-eq-15}) leads to
\begin{eqnarray*}
\frac 1n \left\{\widetilde \ell(\theta_{n,h}) - \widetilde \ell(\theta_0)\right\} &=& \frac{n-1}{n} \left(h^T\GG_n \dot{m}_{n, 0, 1} - h^T \Sigma_1 h\right)  + o_p(1)\\
&=& h^T\GG_n \dot{m}_{n, 0, 1} - h^T \Sigma_1 h + o_p(1),
\end{eqnarray*}
where the $o_p(1)$ above is uniform in $h\in H$. This together with the definition of $\varphi(\cdot)$ in (\ref{lemma-norm-0-eq-6-added}) completes the proof of this lemma. \epf

\begin{lemma} \label{lemma-norm-1}
Assume Conditions 1 and 2. We have
\begin{eqnarray}
\sup_{\bz \in \mathZ,h\in H} \left| \frac{\widehat F_{\theta_{n,h}}(\bv_{\theta_{n,h}}) F_0(\bv_{\theta_0})}{\widehat F_0(\bv_{\theta_0}) F_{\theta_{n,h}}(\bv_{\theta_{n,h}})} -1 \right| = o_p(n^{-1/2}). \label{lemma-norm-1-eq-1}
\end{eqnarray}
\end{lemma}
\proof Based on Condition 2 and Lemma \ref{lemma-1},
up to a universal constant not depending on $\bz$ and $ h$, for sufficiently large $n$, by noting the definition of $\widehat F_{\theta}(\cdot)$ given by (\ref{def-F-hat-theta}), we have
\begin{eqnarray}
&&\left| \frac{\widehat F_{\theta_{n,h}}(\bv_{\theta_{n,h}}) F_0(\bv_{\theta_0})}{\widehat F_0(\bv_{\theta_0}) F_{\theta_{n,h}}(\bv_{\theta_{n,h}})} -1 \right| \nonumber \\
&\lesssim& \left|\widehat F_{\theta_{n,h}}(\bv_{\theta_{n,h}}) F_0(\bv_{\theta_0}) - \widehat F_0(\bv_{\theta_0}) F_{\theta_{n,h}}(\bv_{\theta_{n,h}}) \right| \nonumber \\
&\leq& \left| \PP_n \bar f_{\bz, n,h} \right| + n^{-2}, \label{lemma-norm-1-eq-1-added}
\end{eqnarray}
where
\begin{eqnarray}
\bar f_{\bz, n,h}(\bz_1) &=& I\left(y_1^{(\lambda_{n,h})} - \bx_1 \beta_{n,h} \leq y^{(\lambda_{n,h})}- \bx \beta_{n,h}\right) F_0(\bv_{\theta_0}) \nonumber \\
&& - I\left(y_1^{(\lambda_0)} - \bx_1 \beta_0 \leq y^{(\lambda_0)}- \bx \beta_0\right)F_{\theta_{n,h}}(\bv_{\theta_{n,h}}) \nonumber
\end{eqnarray}
Note that $\PP \bar f_{\bz, n,h} = 0$, therefore
\begin{equation}
\left| \PP_n \bar f_{\bz, n,h} \right| = \frac{1}{\sqrt{n}}\GG_n \bar f_{\bz, n,h}
\leq \frac{1}{\sqrt{n}}\Big|\GG_n \bar f_{1, \bz, n,h} \Big|F_0(\bv_{\theta_0}) + \frac{1}{\sqrt{n}}\Big|\GG_n \bar f_{2, \bz} \Big|\cdot |F_{\theta_{n,h}}(\bv_{\theta_{n,h}}) - F_0(\bv_{\theta_0})|, \label{lemma-norm-1-eq-2}
\end{equation}
where
\begin{eqnarray}
\bar f_{1, \bz, n,h} (\bz_1) &=& I\left(y_1^{(\lambda_{n,h})} - \bx_1 \beta_{n,h} \leq y^{(\lambda_{n,h})}- \bx \beta_{n,h}\right) - I\left(y_1^{(\lambda_0)} - \bx_1 \beta_0 \leq y^{(\lambda_0)}- \bx \beta_0\right) \nonumber \\
\bar f_{2, \bz} (\bz_1) &=& I\left(y_1^{(\lambda_0)} - \bx_1 \beta_0 \leq y^{(\lambda_0)}- \bx \beta_0\right) - F_0(\bv_{\theta_0}). \label{lemma-norm-1-eq-3} \nonumber
\end{eqnarray}

 Based on compactness of $H$, Condition 2, and Lemma \ref{lemma-1}, we immediately have
\begin{eqnarray}
\sup_{\bz \in \mathZ} \Big|\GG_n \bar f_{2, \bz} \Big| &=& O_p(1)\nonumber \\
\sup_{\bz \in \mathZ,h\in H}|F_{\theta_{n,h}}(\bv_{\theta_{n,h}}) - F_0(\bv_{\theta_0})| &=& O(n^{-1/2}). \label{lemma-norm-1-eq-4}
\end{eqnarray}
Combining (\ref{lemma-norm-1-eq-1-added}), (\ref{lemma-norm-1-eq-2}), and (\ref{lemma-norm-1-eq-4}), to show (\ref{lemma-norm-1-eq-1}), we only need to show
\begin{eqnarray}
\sup_{\bz \in \mathZ,h\in H} \left| \GG_n \bar f_{1, \bz, n,h} \right| = o_p(1). \label{lemma-norm-1-eq-5}
\end{eqnarray}
Consider the function class
\begin{eqnarray*}
\bar \mathF_n = \{\bar f_{1, \bz, n,h} : \bz \in \mathZ, h \in H\}, \label{lemma-norm-1-eq-6}
\end{eqnarray*}
which is a subset of the function class $\mathC-\mathC$ with $\mathC$ defined in Lemma \ref{lemma-entropy-1}. Therefore, by Lemma \ref{lemma-entropy-1} and applying Lemma 9.25 in Kosorok, we have
\begin{eqnarray*}
N_{[]}(\epsilon, \bar \mathF_n, L_2(\PP)) \lesssim \frac{1}{\epsilon^{4(p+2)}}. \label{lemma-norm-1-eq-7}
\end{eqnarray*}
Furthermore, 
$
y^{(\lambda)} - \bx^T \beta
$
as a function of $(\lambda, \beta)$ satisfies
\begin{eqnarray*}
\sup_{\bz \in \mathZ}|y^{(\lambda_1)} - \bx^T \beta_1 - (y^{(\lambda_2)} - \bx^T \beta_2)| \leq C \|\theta_1 - \theta_2\|_2, \label{lemma-norm-1-eq-8}
\end{eqnarray*}
for a univeral constant $C>0$ not depending on $(y, \bx)$, and any $\theta_1, \theta_2 \in \Theta$. Therefore
for every $\bar f_{1, \bz, n,h}  \in \bar \mathF_n$, we have $\|\bar f_{1, \bz, n,h}\|_\infty \leq 1$ and based on Condition 2,
\begin{eqnarray}
&&\PP \bar{f}_{1, \bz, n,h} ^2 \nonumber \\ &=& E\left\{ I\left(Y_1^{(\lambda_{n,h})} - X_1 \beta_{n,h} \leq y^{(\lambda_{n,h})}- \bx \beta_{n,h}\right) - I\left(Y_1^{(\lambda_0)} - X_1 \beta_0 \leq y^{(\lambda_0)}- \bx \beta_0\right) \right\}^2 \nonumber\\
&=& P\left(Y_1^{(\lambda_{n,h})} - X_1 \beta_{n,h} \leq y^{(\lambda_{n,h})}- \bx \beta_{n,h}; \ \  Y_1^{(\lambda_0)} - X_1 \beta_0 > y^{(\lambda_0)}- \bx \beta_0 \right) \nonumber \\
&& + P\left(Y_1^{(\lambda_{n,h})} - X_1 \beta_{n,h} > y^{(\lambda_{n,h})}- \bx \beta_{n,h}; \ \  Y_1^{(\lambda_0)} - X_1 \beta_0 \leq y^{(\lambda_0)}- \bx \beta_0 \right)\nonumber \\
&=& P\Big(Y_1^{(\lambda_{n,h})} - X_1 \beta_{n,h} \leq y^{(\lambda_{n,h})}- \bx \beta_{n,h}; \ \  Y_1^{(\lambda_0)} - X_1 \beta_0 > y^{(\lambda_0)}- \bx \beta_0; \nonumber \\
&& \hspace{0.2in} |Y_1^{(\lambda_{n,h})} - X_1^T \beta_{n,h} - (Y_1^{(\lambda_0)} - X_1 \beta_0)| \leq Ch/\sqrt{n} \Big)\nonumber \\
&&+ P\Big(Y_1^{(\lambda_{n,h})} - X_1 \beta_{n,h} > y^{(\lambda_{n,h})}- \bx \beta_{n,h}; \ \  Y_1^{(\lambda_0)} - X_1 \beta_0 \leq y^{(\lambda_0)}- \bx \beta_0; \nonumber \\
&& \hspace{0.2in} |Y_1^{(\lambda_{n,h})} - X_1^T \beta_{n,h} - (Y_1^{(\lambda_0)} - X_1 \beta_0)| \leq Ch/\sqrt{n} \Big) \nonumber \\
&\leq& P\left(y^{(\lambda_0)}- \bx \beta_0 - Ch/\sqrt{n} < Y_1^{(\lambda_{n,h})} - X_1 \beta_{n,h} \leq y^{(\lambda_{n,h})}- \bx \beta_{n,h}\right) \nonumber \\
&& + P\left( y^{(\lambda_{n,h})}- \bx \beta_{n,h} <  Y_1^{(\lambda_{n,h})} - X_1 \beta_{n,h} \leq y^{(\lambda_0)}- \bx \beta_0 + Ch/\sqrt{n} \right) \nonumber \\
&=& F_{\theta_{n,h}}(y^{(\lambda_{n,h})}- \bx \beta_{n,h})- F_{\theta_{n,h}}(y^{(\lambda_0)}- \bx \beta_0 - Ch/\sqrt{n}) \nonumber \\
&& + F_{\theta_{n,h}}(y^{(\lambda_0)}- \bx \beta_0 + Ch/\sqrt{n}) - F_{\theta_{n,h}}(y^{(\lambda_{n,h})}- \bx \beta_{n,h}) \nonumber \\
&\lesssim & h/\sqrt{n}. \label{lemma-norm-1-eq-9}
\end{eqnarray}
Applying Lemma \ref{lemma-tail-bound-2}, we have
\begin{eqnarray}
 E \left(\| \GG_n \|_{\bar \mathF_n}\right)   \lesssim  \widetilde J_{[]}(\delta, \bar\mathF_n, L_2(\PP))\left[ 1 + \frac{\widetilde J_{[]}(\delta, \bar\mathF_n, L_2(\PP))}{\delta^2 \sqrt{n}}\cdot 1  \right], \label{lemma-norm-1-eq-10}
\end{eqnarray}
with $\delta = \sqrt{Ch/\sqrt{n}} = C^{1/2} h^{1/2}/n^{0.25}$ for some universal constant $C>0$, and
\begin{eqnarray*}
\widetilde J_{[]}(\delta, \bar \mathF_n, L_2(\PP)) &=& \int_0^\delta \sqrt{1+ \log N_{[]}(\epsilon, \bar\mathF_n, L_2(\PP))} d\epsilon \lesssim  \int_0^\delta \sqrt{1+\log \left(1/\epsilon^{4(p+2)} \right)}\\
&\lesssim & \int_0^\delta \sqrt{-\log\epsilon} d \epsilon = \int_{-\log\delta}^\infty t^{1/2} e^{-t} dt = o(1), \label{lemma-norm-1-eq-11}
\end{eqnarray*}
which together with (\ref{lemma-norm-1-eq-10}) leads to (\ref{lemma-norm-1-eq-5}); we complete the proof of this lemma. \epf


\begin{lemma} \label{lemma-norm-2}
Assume Conditions 1 and 2.  We have
\begin{eqnarray}
\sup_{h \in H}\left|\frac 1n \sum_{j=1}^n\sum_{i=1}^n  I_{i,j} \left\{\frac{\widehat F_{\theta_{n,h}}(V_{\theta_{n,h}, j,i}) F_0(V_{0,j,i}) }{ \widehat F_0(V_{0,j,i}) F_{\theta_{n,h}}(V_{\theta_{n,h}, j,i})} - 1\right\} -  \sqrt{n} \GG_n \left\{  f_{1,n,h}(\cdot) \right\} \right| = o_p(1), \label{lemma-norm-2-eq-1}
\end{eqnarray}
where $f_{1,n,h}(\cdot)$ is defined by (\ref{lemma-norm-2-eq-6-added}).

\end{lemma}

\proof We can write
\begin{eqnarray}
&&\frac 1n \sum_{j=1}^n\sum_{i=1}^n  I_{i,j} \left\{\frac{\widehat F_{\theta_{n,h}}(V_{\theta_{n,h}, j,i}) F_0(V_{0,j,i}) }{ \widehat F_0(V_{0,j,i}) F_{\theta_{n,h}}(V_{\theta_{n,h}, j,i})} - 1\right\} \nonumber \\
&=& \frac 1n \sum_{j=1}^n\sum_{i=1}^n  I_{i,j} \left\{ \frac{\widehat F_{\theta_{n,h}}(V_{\theta_{n,h}, j,i}) F_0(V_{0,j,i}) - \widehat F_0(V_{0,j,i}) F_{\theta_{n,h}}(V_{\theta_{n,h}, j,i})}{F_0(V_{0,j,i})F_{\theta_{n,h}}(V_{\theta_{n,h}, j,i})} \right\} \nonumber \\
&& + \frac 1n \sum_{j=1}^n\sum_{i=1}^n  I_{i,j} \left\{\widehat F_{\theta_{n,h}}(V_{\theta_{n,h}, j,i}) F_0(V_{0,j,i}) - \widehat F_0(V_{0,j,i}) F_{\theta_{n,h}}(V_{\theta_{n,h}, j,i})\right\}\nonumber \\
&& \times \hspace{0.5in}\left\{ \frac{1}{\widehat F_0(V_{0,j,i}) F_{\theta_{n,h}}(V_{\theta_{n,h}, j,i})} - \frac{1}{F_0(V_{0,j,i}) F_{\theta_{n,h}}(V_{\theta_{n,h}, j,i})}\right\} \nonumber \\
&\equiv& \mathI_7 + \mathI_8.    \label{lemma-norm-2-eq-2}
\end{eqnarray}
With Lemmas \ref{lemma-1} and \ref{lemma-norm-1}, and Condition 2, we have
\begin{eqnarray}
\sup_{h\in H} |\mathI_8| = o_p(1). \label{lemma-norm-2-eq-3}
\end{eqnarray}
For $\mathI_7$, recall the definition of $\widehat F_{\theta}(\cdot)$ given by (\ref{def-F-hat-theta}), we can write
\begin{eqnarray*}
\mathI_7 &=& \frac 1n \sum_{j=1}^n\sum_{i=1}^n  I_{i,j} \left\{ \frac{\widehat F_{\theta_{n,h}}(V_{\theta_{n,h}, j,i})}{F_{\theta_{n,h}}(V_{\theta_{n,h}, j,i})} -\frac{ \widehat F_0(V_{0,j,i}) }{F_0(V_{0,j,i})} \right\}\\
&=&\frac{1}{n^2} \sum_{j=1}^n \sum_{i=1}^n \sum_{k=1}^n f_{n,h}(Z_i,Z_j, Z_k) + O(n^{-1}), \label{lemma-norm-2-eq-4}
\end{eqnarray*}
where $O(n^{-1})$ is uniform in $h\in H$; $f_{n,h}(\cdot ,\cdot , \cdot) = f_{\theta_{n,h}}(\cdot ,\cdot , \cdot)$, with ``$f_{\theta}(\cdot ,\cdot , \cdot)$" defined by (\ref{root-n-eq-24}).
Note that $\PP f_{n,h}(\bz_1, \bz_2, \cdot) = 0$, therefore
\begin{eqnarray}
\mathI_7 &=& n \VV_n^3 f_{n,h} = \sqrt{n} \VV_n^2 \widetilde f_{n,h} = \frac{1}{n} \sum_{j=1}^n \GG_n \widetilde f_{n,h}(\cdot, Z_j) + \frac{1}{\sqrt{n}} \sum_{j=1}^n \PP \widetilde f_{n,h}(\cdot, Z_j)\nonumber \\
&=& \frac{1}{n} \sum_{j=1}^n \GG_n \widetilde f_{n,h}(\cdot, Z_j) +  \int \GG_n \widetilde f_{n,h}( \bz_1 ,\cdot) dF_{Z_1}(\bz_1) + \sqrt{n} \PP^2 \widetilde f_{n,h}(\cdot, \cdot) \nonumber \\
&\equiv& \mathI_{7,1} + \mathI_{7,2} + \mathI_{7,3}, \label{lemma-norm-2-eq-5}
\end{eqnarray}
where
\begin{eqnarray}
\widetilde f_{n,h}(\bz_1, \bz_2) = \GG_n f_{n,h}(\bz_1, \bz_2, \cdot).  \label{lemma-norm-2-eq-6}
\end{eqnarray}
We consider $\mathI_{7,3}$ first:
\begin{eqnarray}
\mathI_{7,3} &=& \sqrt{n} \PP^2 \widetilde f_{n,h}(\cdot, \cdot)= \sum_{k=1}^n  \PP^2 f_{n,h}(\cdot, \cdot, Z_k) \nonumber \\ &=& \sum_{k=1}^n f_{1,n,h}(Z_k) = \sqrt{n} \GG_n \left\{  f_{1,n,h}(\cdot) \right\}, \label{lemma-norm-2-eq-6-added-1}
\end{eqnarray}
since $\PP f_{1,n,h} = 0$, where
\begin{eqnarray}
f_{1,n,h}(\bz_3) &=&   E\left\{f_{n,h}(Z_1, Z_2, \bz_3)\right\} \nonumber \\
&=& E \left\{ \frac{F_0(V_{0, 2, 1})}{F_{\theta_{n,h}}(V_{\theta_{n,h}, 2,1})} I\left(\bv_{\theta_{n,h}, 3,3} \leq V_{\theta_{n,h}, 2,1}\right) - I\left(\bv_{0, 3,3} \leq V_{0, 2,1}\right)\right\}. \label{lemma-norm-2-eq-6-added}
\end{eqnarray}
Therefore, the proof of this lemma is completed if we can show
\begin{eqnarray}
\|\mathI_{7,1}\|_{h \in H} = o_p(1)   \label{lemma-norm-2-eq-7}        \\
\|\mathI_{7,2}\|_{h \in H} = o_p(1)   \label{lemma-norm-2-eq-8},
\end{eqnarray}
since  (\ref{lemma-norm-2-eq-5}), (\ref{lemma-norm-2-eq-6-added-1}),  (\ref{lemma-norm-2-eq-7}),  and (\ref{lemma-norm-2-eq-8}) imply
\begin{eqnarray*}
\sup_{h\in H}\left|\mathI_7 - \sqrt{n} \GG_n \left\{  f_{1,n,h}(\cdot) \right\}\right| = o_p(1),
\end{eqnarray*}
which together with (\ref{lemma-norm-2-eq-2}) and (\ref{lemma-norm-2-eq-3}) leads to (\ref{lemma-norm-2-eq-1}).

To show (\ref{lemma-norm-2-eq-7}) and (\ref{lemma-norm-2-eq-8}), it suffices to show
\begin{eqnarray}
\left\| \GG_n \widetilde f_{n,h}(\cdot, \bz_2) \right\|_{h\in H; \bz_2 \in \mathZ} &=& o_p(1) \label{lemma-norm-2-eq-9}\\
\left\| \GG_n \left\{\int \widetilde f_{n,h}(\bz_1, \cdot) dF_{Z_1}(\bz_1)\right\} \right\|_{h\in H; \bz_1 \in \mathZ} &=& o_p(1) \label{lemma-norm-2-eq-10}.
\end{eqnarray}
In fact, we only need to show (\ref{lemma-norm-2-eq-9}), since a very similar procedure can be used to show (\ref{lemma-norm-2-eq-10}).
Referring to the definition of $\widetilde f_{n,h}(\bz_1, \bz_2)$ given by (\ref{lemma-norm-2-eq-6}), we can write
\begin{eqnarray}
\GG_n \widetilde f_{n,h}(\cdot, \bz_2) &=& \frac{1}{n} \sum_{k=1}^n \sum_{i=1}^n \left[ f_{n,h}(Z_i, \bz_2, Z_k) - \PP f_{n,h}( \cdot, \bz_2, Z_k) \right] \nonumber \\
&=& (n-1) \UU_n^2 f_{n,h, \bz_2} + \frac{1}{n} \sum_{k=1}^n f_{n,h, \bz_2} (Z_k, Z_k), \label{lemma-norm-2-eq-11}
\end{eqnarray}
where
\begin{eqnarray*}
f_{n,h,\bz_2}(\bz_1, \bz_3) = f_{n,h}(\bz_1, \bz_2, \bz_3) - \PP f_{n,h}(\cdot, \bz_2, \bz_3). \label{lemma-norm-2-eq-12}
\end{eqnarray*}
With Condition 2 and Lemma \ref{lemma-tail-bound-1}, by working on the function class $\left\{f_{n,h, \bz_2} (\bz, \bz): \bz_2 \in \mathZ, h \in H \right\}$ defined on $\bz \in \mathZ$, we can show
\begin{eqnarray}
\left\|\frac{1}{n} \sum_{k=1}^n f_{n,h, \bz_2} (Z_k, Z_k)\right\|_{h\in H; \bz_2 \in \mathZ} = o_p(1). \label{lemma-norm-2-eq-13}
\end{eqnarray}
Consider $\UU_n^2 f_{n,h, \bz_2}$. We apply Lemma \ref{lemma-decoupling}: let $\left\{Z_i^{(r)}\right\}_{i=1,\ldots,n}$ for $r=1,2$ be i.i.d. copies of $\left\{ Z_i \right\}_{i=1,\ldots, n}$; we have
\begin{eqnarray}
E\left\|\UU_n^2 f_{n,h, \bz_2} \right\|_{h\in H; \bz_2 \in \mathZ} &\lesssim& E\left\| \frac{1}{n(n-1)}\sum_{k\neq i} f_{n,h, \bz_2} ( Z_i^{(1)}, Z_k^{(2)} )  \right\|_{h \in H; \bz_2 \in \mathZ} \nonumber \\
&\lesssim & E\left\| \frac{1}{n^2}\sum_{k=1}^n \sum_{i=1}^n f_{n,h, \bz_2} ( Z_i^{(1)}, Z_k^{(2)} )  \right\|_{h \in H; \bz_2 \in \mathZ} + o(n), \label{lemma-norm-2-eq-14}
\end{eqnarray}
where the second ``$\lesssim$" is because that
with Condition 2 and Lemma \ref{lemma-tail-bound-1}, by working on the function class $\left\{f_{n,h, \bz_2} (\bz^{(1)}, \bz^{(2)}): \bz_2 \in \mathZ, h \in H \right\}$ defined on $(\bz^{(1)}, \bz^{(2)}) \in \mathZ^2$, we can check
\begin{eqnarray*}
E\left\|\frac{1}{n} \sum_{k=1}^n f_{n,h, \bz_2} (Z_k^{(1)}, Z_k^{(2)})\right\|_{h\in H; \bz_2 \in \mathZ} = o(1). \label{lemma-norm-2-eq-15}
\end{eqnarray*}
Furthermore,
\begin{eqnarray}
E\left\| \frac{1}{n}\sum_{k=1}^n \sum_{i=1}^n f_{n,h, \bz_2} ( Z_i^{(1)}, Z_k^{(2)} )  \right\|_{h \in H; \bz_2 \in \mathZ} = E\left\| \GG_n^{(1)} \widetilde f_{n,h}^{(2)}(\cdot, \bz_2) \right\|_{h\in H; \bz_2 \in \mathZ}, \label{lemma-norm-2-eq-16}
\end{eqnarray}
where
\begin{eqnarray}
\widetilde f_{n,h}^{(2)}(\bz_1, \bz_2) = \GG_n^{(2)} f_{n,h}(\bz_1, \bz_2, \cdot). \label{lemma-norm-2-eq-17}
\end{eqnarray}
Combining (\ref{lemma-norm-2-eq-11})--(\ref{lemma-norm-2-eq-16}) leads to
\begin{eqnarray}
E\left\{ \left\|\GG_n \widetilde f_{n,h}(\cdot, \bz_2)\right\|_{h \in H; \bz_2 \in \mathZ}\right\} \lesssim E\left\| \GG_n^{(1)} \widetilde f_{n,h}^{(2)}(\cdot, \bz_2) \right\|_{h\in H; \bz_2 \in \mathZ} + o(1), \label{lemma-norm-2-eq-18}
\end{eqnarray}
with $\widetilde f_{n,h}^{(2)}(\bz_1, \bz_2)$ being defined by (\ref{lemma-norm-2-eq-17}). We need to show the term on the right hand side of (\ref{lemma-norm-2-eq-18}) is $o(1)$.

Consider the function class
\begin{eqnarray*}
\mathF = \{f_{n,h}(\bz_1, \bz_2, \bz_3): \bz_1\in \mathZ, \bz_2 \in \mathZ, h \in H\}, \label{lemma-norm-2-eq-19}
\end{eqnarray*}
defined on $\mathZ$. With similar strategy as the proof of Lemma \ref{lemma-entropy-1}, it easy to check that there exists a constant $A>0$, such that
\begin{eqnarray*}
N_{[]}(\epsilon, \mathF, L_2(\PP)) \lesssim 1/\epsilon^A. \label{lemma-norm-2-eq-20}
\end{eqnarray*}
Furthermore, based on Condition 2, for sufficiently large $n$, every function in this class satisfies $\PP f_{n,h}^2 \lesssim  1/\sqrt{n}$ and $\|f_{n,h}\|_\infty \lesssim 1$. Applying Lemma \ref{lemma-tail-bound-2}, we can derive
\begin{eqnarray}
E\left\| \widetilde f_{n,h}^{(2)} (\bz_1, \bz_2) \right \|_{h\in H; \bz_1 \in \mathZ; \bz_2\in \mathZ} = o(n^{-\alpha}), \label{lemma-norm-2-eq-21}
\end{eqnarray}
for any $0<\alpha<1/4$.

For any given values of $\left\{Z_i^{(2)}\right\}_{i=1,\ldots,n}$, and $\widetilde f_{n,h}^{(2)} (\bz_1, \bz_2) $ defined by (\ref{lemma-norm-2-eq-17}), consider the function class:
\begin{eqnarray}
\widetilde \mathF_n\left(Z_1^{(2)}, \ldots, Z_n^{(2)}\right) = \left \{\widetilde f_{n,h, \bz_2}^{(2)} (\bz_1)\equiv \widetilde f_{n,h}^{(2)} (\bz_1, \bz_2):  \bz_2\in \mathZ, h\in H \right\}. \label{lemma-norm-2-eq-22}
\end{eqnarray}
When $n$ is large, it is a subset of the function class ``$\mathF_n\left(Z_1^{(2)}, \ldots, Z_n^{(2)}\right)$" defined in Lemma \ref{lemma-bracketing-F-n-1}, since $H$ is compact and therefore $\|\theta_{n,h}-\theta_0\|_2\leq \eta_0$ for large $n$.
Note that for every $\widetilde f_{n,h, \bz_2}^{(2)}\in \widetilde \mathF_n\left(Z_1^{(2)}, \ldots, Z_n^{(2)}\right)$, $\widetilde f_{n,h, \bz_2}^{(2)} (Z_i^{(1)})$ for $i=1,\ldots, n$ are i.i.d., conditioning on $\left\{Z_k^{(2)}\right\}_{k=1,\ldots,n}$.  Let
\begin{eqnarray}
 \bar F_n(\bz_1) = \left\|\widetilde f_{n,h, \bz_2}^{(2)} (\bz_1) \right\|_{\bz_2 \in \mathZ; h \in H} \label{lemma-norm-2-eq-23}
\end{eqnarray}
be an envelope function for $\widetilde \mathF_n\left(Z_1^{(2)}, \ldots, Z_n^{(2)}\right)$. Applying Lemma \ref{lemma-tail-bound-1}, for sufficiently large $n$, we have
\begin{eqnarray}
&&E\left( \left\|\GG_n^{(1)}\right\|_{\widetilde \mathF_n\left(Z_1^{(2)}, \ldots, Z_n^{(2)}\right)} \Big| Z_1^{(2)}, \ldots, Z_n^{(2)}  \right) \nonumber \\ &\lesssim& J_{[]}\left(1, \widetilde \mathF_n\left(Z_1^{(2)}, \ldots, Z_n^{(2)}\right), L_2(\PP)\right) \left\|  \bar F_n \right\|_{2, \PP}. \label{lemma-norm-2-eq-24}
\end{eqnarray}
Based on Lemma \ref{lemma-bracketing-F-n-1}, for large $n$,
\begin{eqnarray}
&& J_{[]}\left(1, \widetilde \mathF_n\left(Z_1^{(2)}, \ldots, Z_n^{(2)}\right), L_2(\PP)\right) \nonumber \\ &=& \int_0^1 \sqrt{1+ \log N_{[]}\left(\epsilon\|\bar F_n\|_{2,\PP}, \widetilde \mathF_n\left(Z_1^{(2)}, \ldots, Z_n^{(2)}\right), L_2(\PP)\right) } d\epsilon \nonumber \\
&\lesssim & \int_0^1 \sqrt{ 1 + (p+2)\log n  - 2(p+2) \log \|\bar F_n\|_{2, \PP} - 2(p+2) \log \epsilon } d \epsilon \nonumber \\
&\lesssim & \int_0^1 \sqrt{\log n} d\epsilon + \int_0^1 \sqrt{\left|\log \epsilon\right|} d\epsilon + \int_0^1 \sqrt{\left( - \log \|\bar F_n \|_{2,\PP} \right)^+} d \epsilon  \nonumber \\
&\lesssim & \sqrt{\log n} +  \sqrt{\left( -\log \|\bar F_n \|_{2,\PP} \right)^+}. \label{lemma-norm-2-eq-25}
\end{eqnarray}
Note that there exists a constant $0<c<1$, such that the function $x \sqrt{\log(1/x)}$ is concave when $x\in (0, c)$, and it is bounded when $x\in [c, 1)$. As a consequence, combining (\ref{lemma-norm-2-eq-24}) and (\ref{lemma-norm-2-eq-25}), we have
\begin{eqnarray}
&& E\left\| \GG_n^{(1)} \widetilde f_{n,h}^{(2)}(\cdot, \bz_2) \right\|_{h\in H; \bz_2 \in \mathZ} \nonumber \\ &=& E \left[E\left\{ \left\|\GG_n^{(1)}\right\|_{\widetilde \mathF_n\left(Z_1^{(2)}, \ldots, Z_n^{(2)}\right)} \Big| Z_1^{(2)}, \ldots, Z_n^{(2)} \right\}   \right] \nonumber \\
&\lesssim& E\left[\left\{\sqrt{\log n} +  \sqrt{\left( -\log \| \bar F_n \|_{2,\PP} \right)^+} \right\} \| \bar F_n \|_{2,\PP} \right] \nonumber \\
& = & (\log n) E\left(\|\bar F_n\|_{2, \PP}\right) + E \left\{\| \bar F_n \|_{2,\PP} \sqrt{\left( -\log \| \bar F_n \|_{2,\PP} \right)^+}\right\} \nonumber \\
&=& (\log n) E\left(\|\bar F_n\|_{2, \PP}\right) + E \left\{I (\| \bar F_n \|_{2,\PP}<1) \| \bar F_n \|_{2,\PP} \sqrt{ \log \frac{1}{\| \bar F_n \|_{2,\PP}} }\right\} \nonumber \\
&=& (\log n) E(\|\bar F_n\|_{2,\PP}) + E \left\{I (c\leq \| \bar F_n \|_{2,\PP}<1) \| \bar F_n \|_{2,\PP} \sqrt{ \log \frac{1}{\| \bar F_n \|_{2,\PP}} }\right\} \nonumber \\
&& + E \left\{I (\| \bar F_n \|_{2,\PP}<c) \| \bar F_n \|_{2,\PP} \sqrt{ \log \frac{1}{\| \bar F_n \|_{2,\PP}} }\right\} \nonumber \\
&\lesssim& (\log n) E(\|\bar F_n\|_{2,\PP}) + P(c\leq \| \bar F_n \|_{2,\PP}<1) \nonumber \\
&&+ \frac{E\left\{ I (\| \bar F_n \|_{2,\PP}<c) \| \bar F_n \|_{2,\PP} \right\}}{P(\| \bar F_n \|_{2,\PP}<c)} \sqrt{ \log \frac{P(\| \bar F_n \|_{2,\PP}<c)}{E\left\{ I (\| \bar F_n \|_{2,\PP}<c) \| \bar F_n \|_{2,\PP} \right\}} }, \label{lemma-norm-2-eq-26}
\end{eqnarray}
where the last $\lesssim$ is based on the Jensen's inequality. We need some properties of $\bar F_n$ to further bound (\ref{lemma-norm-2-eq-26}). Recalling the definitions of $\widetilde f_{n,h, \bz_2}^{(2)} (\cdot)$ and $\bar F_n(\cdot)$  in (\ref{lemma-norm-2-eq-22}) and (\ref{lemma-norm-2-eq-23}), we have
\begin{eqnarray*}
\| \bar F_n \|_{2,\PP} &=& \left\|\left\|\widetilde f_{n,h, \bz_2}^{(2)} (\cdot) \right\|_{\bz_2 \in \mathZ; h \in H} \right\|_{2,\PP} \\
&\leq & \left\|\widetilde f_{n,h, \bz_2}^{(2)} (\bz_1) \right\|_{\bz_1 \in \mathZ, \bz_2 \in \mathZ; h \in H}\\
&=& \left\|\widetilde f_{n,h}^{(2)} (\bz_1, \bz_2) \right\|_{\bz_1 \in \mathZ, \bz_2 \in \mathZ; h \in H}. \label{lemma-norm-2-eq-27}
\end{eqnarray*}
Based on Chebyshev's inequality and (\ref{lemma-norm-2-eq-21}), for any $\alpha\in (0, 1/4)$, we have
\begin{eqnarray}
P(c\leq \| \bar F_n \|_{2,\PP}<1) &\leq & P(\| \bar F_n \|_{2,\PP} \geq c) \nonumber \\
&\leq & P\left(\left\|\widetilde f_{n,h}^{(2)} (\bz_1, \bz_2) \right\|_{\bz_1 \in \mathZ, \bz_2 \in \mathZ; h \in H} \geq c\right) = o(n^{-\alpha}), \label{lemma-norm-2-eq-28}
\end{eqnarray}
and
\begin{eqnarray}
E\left\{ I (\| \bar F_n \|_{2,\PP}<c) \| \bar F_n \|_{2,\PP} \right\} &\leq& E\left\{ \| \bar F_n \|_{2,\PP} \right\} \nonumber \\
&\leq& E \left\{ \left\|\widetilde f_{n,h}^{(2)} (\bz_1, \bz_2) \right\|_{\bz_1 \in \mathZ, \bz_2 \in \mathZ; h \in H} \right\} \nonumber\\
&=& o(n^{-\alpha}). \label{lemma-norm-2-eq-29}
\end{eqnarray}
Combining (\ref{lemma-norm-2-eq-26})--(\ref{lemma-norm-2-eq-29}), we conclude
\begin{eqnarray*}
E\left\| \GG_n^{(1)} \widetilde f_{n,h}^{(2)}(\cdot, \bz_2) \right\|_{h\in H; \bz_2 \in \mathZ}  = o(1),
\end{eqnarray*}
which combined with (\ref{lemma-norm-2-eq-18}) leads to (\ref{lemma-norm-2-eq-9}), and therefore (\ref{lemma-norm-2-eq-7}) is verified. We have completed the proof of this lemma. \epf

\begin{lemma}\label{lemma-norm-3}

Assume Conditions 1--3. We have
\begin{eqnarray}
E\left\|\sqrt{n}\GG_n \psi_{n,h}(\bz) - h^T \GG_n \psi(\bz)\right\|_{h\in H} =  o(1),\label{lemma-norm-3-eq-1}
\end{eqnarray}
where
\begin{eqnarray*}
\psi_{n,h}(\bz)
&=& E\left[\frac{F_0(V_{0, 2, 1}) - F_{\theta_{n,h}}(V_{\theta_{n,h}, 2,1})}{F_{\theta_{n,h}}(V_{\theta_{n,h}, 2,1})\left\{1-F_{\theta_{n,h}}(V_{\theta_{n,h}, 2,1}) \right\}}I\left(\bv_{\theta_{n,h}} \leq V_{\theta_{n,h}, 2,1}\right)\right]\label{lemma-norm-3-eq-2}.\\
\psi(\bz) &=& -E\left[\frac{\dot{F}_0(V_{0,2,1}) + F_0'(V_{0,2,1}) \dot{V}_{0,2,1}}{F_{0}(V_{0, 2,1})\left\{1-F_{0}(V_{0, 2,1}) \right\}}I\left(\bv_{0} \leq V_{0, 2,1}\right)\right]. \label{lemma-norm-3-eq-2-added-1}
\end{eqnarray*}
Note that the definition of $\psi(\bz)$ complies with (\ref{def-psi}).

\end{lemma}

%
%
\proof We can decompose
\begin{eqnarray}
&&\sqrt{n}\GG_n \psi_{n,h}(\bz)  \nonumber \\ &=& \sqrt{n}\GG_n\left\{\psi_{n,h}(\bz) - \widetilde \psi_{n,h}(\bz) \right\} + \sqrt{n}\GG_n\left\{\widetilde \psi_{n,h}(\bz) -  \frac{h^T}{\sqrt{n}}\psi(\bz) \right\} + h^T \GG_n \psi(\bz), \nonumber \\ \label{lemma-norm-3-eq-2-added-2}
\end{eqnarray}
where
\begin{eqnarray*}
\widetilde \psi_{n,h}(\bz)
&=& E\left[\frac{F_0(V_{0, 2, 1}) - F_{\theta_{n,h}}(V_{\theta_{n,h}, 2,1})}{F_{0}(V_{0, 2,1})\left\{1-F_{0}(V_{0, 2,1}) \right\}}I\left(\bv_{0} \leq V_{0, 2,1}\right)\right]. \label{lemma-norm-3-eq-3}
\end{eqnarray*}
We shall show this lemma by showing that
\begin{eqnarray}
E\left\|\sqrt{n}\GG_n\left\{\psi_{n,h}(\bz) - \widetilde \psi_{n,h}(\bz) \right\}\right\|_{h\in H} &=& o(1) \label{lemma-norm-3-eq-3-added-1}\\
E\left\|\sqrt{n}\GG_n\left\{\widetilde \psi_{n,h}(\bz) -  \frac{h^T}{\sqrt{n}}\psi(\bz) \right\} \right\|_{h\in H} &=& o(1). \label{lemma-norm-3-eq-3-added-2}
\end{eqnarray}
Then combining (\ref{lemma-norm-3-eq-2-added-2}), (\ref{lemma-norm-3-eq-3-added-1}), and (\ref{lemma-norm-3-eq-3-added-2}) leads to (\ref{lemma-norm-3-eq-1}). We show (\ref{lemma-norm-3-eq-3-added-1}) first. Consider the function class
\begin{eqnarray}
\Psi_n = \left\{\sqrt{n}\left\{\psi_{n,h}(\bz) - \widetilde \psi_{n,h}(\bz)\right\}: h\in H \right\},  \label{lemma-norm-3-eq-4}
\end{eqnarray}
defined on $\mathZ$. For any $h_1, h_2\in H$, because of Condition 2,  when $n$ is sufficiently large,
\begin{eqnarray}
&&\sqrt{n}\left| \left\{\psi_{n,h_1}(\bz) - \widetilde \psi_{n,h_1}(\bz)\right\} - \left\{\psi_{n,h_2}(\bz) - \widetilde \psi_{n,h_2}(\bz)\right\}\right| \\
&\leq& \sqrt{n}\left| \psi_{n,h_1}(\bz) - \psi_{n,h_2}(\bz)\right| + \sqrt{n}\left| \widetilde \psi_{n,h_1}(\bz) - \widetilde \psi_{n,h_2}(\bz) \right| \nonumber\\
&\lesssim & \sqrt{n}E\Bigg| \frac{F_0(V_{0, 2, 1}) - F_{\theta_{n,h_1}}(V_{\theta_{n,h_1}, 2,1})}{F_{\theta_{n,h_1}}(V_{\theta_{n,h_1}, 2,1})\left\{1-F_{\theta_{n,h_1}}(V_{\theta_{n,h_1}, 2,1}) \right\}}     \nonumber \\ && \hspace{0.5in} -      \frac{F_0(V_{0, 2, 1}) - F_{\theta_{n,h_2}}(V_{\theta_{n,h_2}, 2,1})}{F_{\theta_{n,h_2}}(V_{\theta_{n,h_2}, 2,1})\left\{1-F_{\theta_{n,h_2}}(V_{\theta_{n,h_2}, 2,1}) \right\}}  \Bigg| \nonumber \\
&& + \sqrt{n}E\left| I\left(\bv_{\theta_{n,h_2}} \leq V_{\theta_{n,h_2}, 2,1}\right) -   I\left(\bv_{\theta_{n,h_1}} \leq V_{\theta_{n,h_1}, 2,1}\right)\right| \nonumber \\ &&+ \sqrt{n}E\left|\frac{F_{\theta_{n,h_1}}(V_{\theta_{n,h_1}, 2,1}) - F_{\theta_{n,h_2}}(V_{\theta_{n,h_2}, 2,1})}{F_{0}(V_{0, 2,1})\left\{1-F_{0}(V_{0, 2,1}) \right\}}\right| \nonumber\\
&\lesssim& \|h_1 - h_2\|_2 +  \sqrt{n}E\left| I\left(\bv_{\theta_{n,h_2}} \leq V_{\theta_{n,h_2}, 2,1}\right) -   I\left(\bv_{\theta_{n,h_1}} \leq V_{\theta_{n,h_1}, 2,1}\right)\right|. \label{lemma-norm-3-eq-5}
\end{eqnarray}
With Condition 3 and the same derivation as (\ref{lemma-norm-1-eq-9}), we can establish
\begin{eqnarray}
E\left| I\left(\bv_{\theta_{n,h_2}} \leq V_{\theta_{n,h_2}, 2,1}\right) -   I\left(\bv_{\theta_{n,h_1}} \leq V_{\theta_{n,h_1}, 2,1}\right)\right| \lesssim \|h_1 - h_2\|_2/\sqrt{n}. \label{lemma-norm-3-eq-6}
\end{eqnarray}
Combining (\ref{lemma-norm-3-eq-5}) and (\ref{lemma-norm-3-eq-6}), we conclude that the function class $\Psi_n$ defined by (\ref{lemma-norm-3-eq-4}) satisfies (\ref{lemma-bracketing-eq-1}) with ``$\widetilde F(\bz) = C$" where $C$ is a universal constant. Applying Lemma \ref{lemma-bracketing}, we have
\begin{eqnarray}
N_{[]}(\epsilon, \Psi_n, L_2(\PP)) \lesssim 1/\epsilon^{p+1}. \label{lemma-norm-3-eq-7}
\end{eqnarray}
Furthermore, based on Condition 2 and the compactness of $H$, we can derive that every function in $\Psi_n$ satisfies
\begin{eqnarray*}
&&\sqrt{n}\left|\psi_{n,h}(\bz) - \widetilde \psi_{n,h}(\bz)\right|\\
&\leq& \sqrt{n}\sup_{\bz\in \mathZ}\Bigg[|F_0(\bv_0) - F_{\theta_{n,h}}(\bv_{\theta_{n,h}})|\\ && \times E\left|\frac{I\left(\bv_{\theta_{n,h}} \leq V_{\theta_{n,h}, 2,1}\right)}{F_{\theta_{n,h}}(V_{\theta_{n,h}, 2,1})\left\{1-F_{\theta_{n,h}}(V_{\theta_{n,h}, 2,1}) \right\}} - \frac{I\left(\bv_{0} \leq V_{0, 2,1}\right)}{F_{0}(V_{0, 2,1})\left\{1-F_{0}(V_{0, 2,1}) \right\}}\right|\Bigg] \\
&\lesssim& \|h\|_2^2/\sqrt{n} \lesssim 1/\sqrt{n}.
\end{eqnarray*}
Therefore, $C/\sqrt{n}$ is an envelope function for $\Psi_n$, and for this envelope function, based on (\ref{lemma-norm-3-eq-7}),
\begin{eqnarray*}
J_{[]}(1, \Psi_n) &=& \int_0^1 \sqrt{1+ \log N_{[]}(\epsilon \|C/\sqrt{n}\|_{2,\PP}, \Psi_n, L_2(\PP))} d\epsilon\\
&\lesssim & \int_0^1 \sqrt{1+ \log(n^{(p+1)/2}/\epsilon^{p+1})}\\
&\lesssim &\sqrt{\log n}.
\end{eqnarray*}
Applying Lemma \ref{lemma-tail-bound-1}, we immediately have
\begin{eqnarray*}
E(\|\GG_n \|_{\Psi_n}) \lesssim \sqrt{\log n}/\sqrt{n},
\end{eqnarray*}
which proves (\ref{lemma-norm-3-eq-3-added-1}).

We proceed to show (\ref{lemma-norm-3-eq-3-added-2}). Consider the function class
\begin{eqnarray*}
\widetilde \Psi_n = \left\{\sqrt{n}\left\{\widetilde \psi_{n,h}(\bz) -  \frac{h^T}{\sqrt{n}}\psi(\bz) \right\}: h \in H \right\}
\end{eqnarray*}
defined on $\mathZ$. Based on Condition 2, it is straightforward to check that for every $h_1, h_2\in H$,
\begin{eqnarray*}
\left |\sqrt{n}\left\{\widetilde \psi_{n,h_1}(\bz) -  \frac{h_1^T}{\sqrt{n}}\psi(\bz) \right\} - \sqrt{n}\left\{\widetilde \psi_{n,h_2}(\bz) -  \frac{h_2^T}{\sqrt{n}}\psi(\bz) \right\} \right |\lesssim \|h_1 - h_2\|_2.
\end{eqnarray*}
This implies $\widetilde \Psi_n$ satisfies (\ref{lemma-bracketing-eq-1}) with ``$\widetilde F(\bz) = C$". Applying Lemma \ref{lemma-bracketing}, we have
\begin{eqnarray}
N_{[]}(\epsilon, \widetilde\Psi_n, L_2(\PP)) \lesssim 1/\epsilon^{p+1}.
\end{eqnarray}
Furthermore, set
\begin{eqnarray*}
\alpha_n = \sup_{h\in H, \bz\in \mathZ} \left|\sqrt{n}\left\{\widetilde \psi_{n,h}(\bz) -  \frac{h^T}{\sqrt{n}}\psi(\bz) \right\}\right| \to 0,
\end{eqnarray*}
as $n\to \infty$, because of Condition 2 and compactness of $H$ and $\mathZ$.
Clearly $\alpha_n$ can serve as an envelope function for $\widetilde \Psi_n$. Then
\begin{eqnarray*}
J_{[]}(1, \widetilde \Psi_n) &=& \int_0^1 \sqrt{1+ \log N_{[]}(\epsilon \|C \alpha_n\|_{2,\PP}, \widetilde \Psi_n, L_2(\PP))} d\epsilon\\
&\lesssim & \int_0^1 \sqrt{1+ \log(1/(\epsilon\alpha_n)^{p+1})}\\
&\lesssim &\sqrt{\log \alpha_n}.
\end{eqnarray*}
Applying Lemma \ref{lemma-tail-bound-1}, we have
\begin{eqnarray*}
E(\|\GG_n \|_{\widetilde \Psi_n}) \lesssim \alpha_n \sqrt{\log \alpha_n} \to 0.
\end{eqnarray*}
This verifies  (\ref{lemma-norm-3-eq-3-added-2}). We complete the proof of this lemma. \epf

\vspace{3em}
\centerline{\bf\sc References}
\begin{description}


\item Arcones, M.A. and Gin${\acute{e}}$, E. (1993). Limit Theorems for U-processes. {\it The Annals of Probability}, 21, 1494-1542.





\item de la Pe$\tilde{n}$a, V.H. (1992). Decoupling and Khintchine's inequalities for U-statistics. {\it The Annals of Probability}, 20, 1877-1892.






\item  Kosorok, M. R. (2008). {\it Introduction to Empirical Processes and Semiparametric Inference.} New York: Springer.




\item Serfling, R.J. (1980). {\it Approximation Theorems of Mathematrical Statistics}. Wiley: New York.

\item Shao, J. (2003). {\it Mathematical Statistics}, 2nd ed., Springer: New York. 

\item Sherman, R.P. (1994). Maximal Inequalities for Degenerate U-Processes with Applications to Optimization Estimators. {\it The Annals of Probability}, 22, 439-459.





\item  van der Vaart, A. W. and Wellner, J. A. (1996). {\it Weak Convergence and Empirical Processes: With Applications to Statistics.} New York: Springer.

\item Wald, A. (1949). Note on the consistency of the maximum likelihood estimate. {\it Annals of Mathematical Statistics}, 20, 595-601.



\end{description}

\end{document}